\documentclass[prd,showpacs]{revtex4}
\usepackage{epsf}
\newcommand{\vphi}{\varphi}
\newcommand{\DS}{\displaystyle}
\newcommand{\pih}{\frac{\pi}{2}}

\newcommand{\Psiel}{\tilde{\Psi}}
\newcommand{\psiel}{\tilde{\psi}}

\newcounter{fixy}
 \newenvironment{fixy}[1]{\setcounter{figure}{#1}}
{\addtocounter{fixy}{1}}

\begin{document}

\title{
Rotating Dilaton Black Holes with Hair}
\vspace{1.5truecm}
\author{Burkhard Kleihaus}
\affiliation{Department of Mathematical Physics, University College, Dublin,
Belfield, Dublin 4, Ireland}
\author{Jutta Kunz}
\affiliation{Fachbereich Physik, Universit\"at Oldenburg, Postfach 2503
D-26111 Oldenburg, Germany}
\author{Francisco Navarro-L\'erida}
\affiliation{ Dept. de F\'{\i}sica Te\'orica II, 
Ciencias F\'{\i}sicas, Universidad Complutense de Madrid, 
E-28040 Madrid, Spain}
\altaffiliation[Present address:]{Dept. de F\'{\i}sica At\'omica, 
Molecular y Nuclear, Ciencias F\'{\i}sicas, Universidad Complutense de Madrid,
E-28040 Madrid, Spain}

\date{\today}

\begin{abstract}
We consider stationary rotating black holes in SU(2) 
Einstein-Yang-Mills theory,
coupled to a dilaton. The black holes possess non-trivial non-Abelian 
electric and magnetic fields outside their regular event horizon.
While generic solutions carry no non-Abelian magnetic charge, 
but non-Abelian electric charge, the presence of the dilaton field
allows also for rotating solutions with no non-Abelian charge at all.
As a consequence, these special solutions do not exhibit the
generic asymptotic non-integer power fall-off of the non-Abelian gauge 
field functions. The rotating black hole solutions form sequences,
characterized by the winding number $n$ and the node number $k$
of their gauge field functions, tending to embedded Abelian black holes.
The stationary non-Abelian black hole solutions satisfy a mass formula, 
similar to the Smarr formula, where the dilaton charge enters 
instead of the magnetic charge. Introducing a topological charge, we 
conjecture, that black hole solutions in SU(2) Einstein-Yang-Mills-dilaton
 theory
are uniquely characterized by their mass, their angular momentum,
their dilaton charge, their non-Abelian electric charge, and their
topological charge.
\end{abstract}


\maketitle

\section{Introduction}

In Einstein-Maxwell (EM) theory
the unique family of stationary asymptotically flat black holes
comprises the rotating Kerr-Newman (KN) and Kerr black holes
and the static Reissner-Nordstr\o m (RN) and Schwarzschild black holes.
EM black holes are uniquely determined
by their mass $M$, their angular momentum $J$,
their electric charge $Q$, and their magnetic charge $P$,
i.e.~EM black holes have ``no hair'' \cite{nohair1,nohair2}.

The EM ``no-hair'' theorem does not readily generalize to theories with
non-Abelian gauge fields coupled to gravity \cite{su2bh,review}.
The hairy black hole solutions of SU(2) Einstein-Yang-Mills (EYM) theory 
possess non-trivial magnetic fields outside their regular event horizon,
but carry no magnetic charge \cite{su2bh,review,kk,kkrot,kkn}.
Their magnetic fields are characterized by the winding number $n$ 
and by the node number $k$ of the gauge field functions. 
In the static limit, the solutions with winding number $n=1$ are
spherically symmetric, whereas the solutions with winding number $n>1$
possess only axial symmetry, showing that Israel's theorem does not
generalize to non-Abelian theories, either \cite{kk}.

In many unified theories, including string theory, dilatons appear.
When a dilaton is coupled to EM theory, 
this has profound consequences for the black hole solutions.
Although uncharged Einstein-Maxwell-dilaton (EMD) black holes correspond 
to the EM black holes, since the source term for the dilaton vanishes,
charged EMD black hole solutions possess a non-trivial dilaton field,
giving rise to an additional charge, the dilaton charge $D$,
and to qualitatively new features of the black holes.
Charged static EMD black hole solutions, for instance,
exist for arbitrarily small horizon size \cite{emd},
and the surface gravity of `extremal' solutions
depends in an essential way on the dilaton coupling constant $\gamma$.
Rotating EMD black holes, known exactly only for Kaluza-Klein (KK) 
theory with $\gamma = \sqrt{3}$ \cite{FZB,Rasheed}, no longer possess 
the gyromagnetic ratio $g=2$ \cite{HH}, the value of KN black holes.
Extremal charged rotating EMD black holes can possess non-zero angular 
momentum, while their event horizon has zero angular velocity 
\cite{Rasheed}.

Here we consider rotating black hole solutions 
of SU(2) Einstein-Yang-Mills-dilaton (EYMD) theory \cite{eymd,kk,kkrot,kkn}. 
For fixed dilaton coupling constant $\gamma$ and winding number $n$,
the black hole solutions form sequences, which with increasing node 
number $k$, tend to limiting solutions.
The black hole solutions of a given sequence 
carry no non-Abelian magnetic charge, but generically 
they carry a small non-Abelian electric charge \cite{pert1,kkrot,kkn}. 
The limiting solutions, in contrast, correspond 
to embedded Abelian black hole solutions, which carry no electric charge,
but a magnetic charge, equal to the winding number $n$.

We investigate the physical properties of these black hole solutions.
In particular, we consider their global charges, obtained
from an asymptotic expansion of the fields. 
The asymptotic expansion of the gauge field functions generically
involves non-integer powers of the radial coordinate, 
depending on the non-Abelian electric charge of the black hole solutions
\cite{kkrot}. In the presence of the dilaton, there is, however, 
also a special set of solutions, whose non-Abelian electric charge vanishes
\cite{kkn}. For these, the asymptotic expansion involves only integer powers
of the radial coordinate.
We further consider the horizon properties of the non-Abelian black hole 
solutions, such as their horizon area $A$, their surface gravity 
$\kappa_{\rm sg}$, 
their horizon curvature $K$ and horizon topology,
and we introduce a topological horizon charge $N$ \cite{ash,kkn}.

For the rotating non-Abelian black holes the zeroth law of black hole
mechanics holds \cite{kkrot,kkn}, 
as well as a generalized first law \cite{heustrau}.
The non-Abelian black holes further satisfy the mass formula \cite{kkn},
\begin{equation}
M = 2 TS + 2 \Omega J + \frac{D}{\gamma} + 2\psi_{\rm el} Q
\ , \label{namass} \end{equation}
where $T$ denotes the temperature of the black holes, $S$ their entropy,
$\Omega$ their horizon angular velocity, and $\psi_{\rm el}$ their
horizon electrostatic potential.
This non-Abelian mass formula is similar to the Smarr formula \cite{smarr}.
However, instead of the magnetic charge $P$, the dilaton charge $D$ enters.
This is crucial, since the genuinely non-Abelian black hole solutions
carry magnetic fields, but no magnetic charge.
Thus the dilaton charge term takes into account the contribution
to the total mass from the magnetic fields outside the horizon.
This mass formula holds for all non-perturbatively known black hole
solutions of SU(2) EYMD theory \cite{eymd,kk,kkrot}, including the 
rotating generalizations of the static non-spherically symmetric non-Abelian
black hole solutions \cite{kk}.
It also holds for embedded Abelian solutions \cite{kkn,kkn-emd}.

Concerning the uniqueness conjecture for EYMD black holes,
it is not sufficient
to simply replace the magnetic charge $P$ by the dilaton charge $D$.
Black holes in SU(2) EYMD theory are not uniquely characterized by
their mass $M$, their angular momentum $J$,
their dilaton charge $D$, and their non-Abelian electric charge $Q$.
Adding as an additional charge a topological charge $N$, however,
a new uniqueness conjecture can be formulated \cite{kkn}:
{\sl Black holes in SU(2) EYMD theory
are uniquely determined by their mass $M$, their angular momentum $J$,
their dilaton charge $D$, their non-Abelian electric charge $Q$, 
and their topological charge $N$}.

In section {II} we recall the SU(2) EYMD action and the equations of
 motion.
We discuss the stationary ansatz for the metric and the gauge and 
dilaton fields,
and we present the boundary conditions.
In section {III} we address the properties of the black hole solutions.
We briefly present the asymptotic expansion at infinity and at
the horizon, from which we obtain the global charges and the
horizon properties, as well as the proof of the mass formula.
Our numerical results are discussed in section {IV}.
In section {V} we present our conclusions.
Appendices A and B give details of the expansion at infinity
and at the horizon, respectively. In Appendix C we discuss the energy conditions and algebraic type of the stress-energy tensor in SU(2) EYMD theory.

\section{\bf Non-Abelian Black Holes}

After recalling the SU(2) EYMD action and the general
set of equations of motion,
we discuss the ansatz for the stationary non-Abelian black hole
solutions, employed for the metric, the dilaton field and the 
gauge field \cite{kkn}. The ansatz for the metric represents the
stationary axially symmetric Lewis-Papapetrou metric \cite{book}
in isotropic coordinates. The ansatz for the gauge field
includes an arbitrary winding number $n$ \cite{kk,kkn},
and satisfies the Ricci circularity and Frobenius conditions \cite{book}.
As implied by the boundary conditions,
the stationary axially symmetric black hole solutions
are asymptotically flat, and possess a regular event horizon.

\subsection{SU(2) EYMD Action}

We consider the SU(2) Einstein-Yang-Mills-dilaton (EYMD) action
\begin{equation}
S=\int \left ( \frac{R}{16\pi G} + L_M \right ) \sqrt{-g} d^4x \ ,
\ \label{action} \end{equation}
with scalar curvature $R$ and matter Lagrangian $L_M$ given by
\begin{equation}
L_M=-\frac{1}{2}\partial_\mu \Phi \partial^\mu \Phi
 - \frac{1}{2} e^{2 \kappa \Phi } {\rm Tr} (F_{\mu\nu} F^{\mu\nu})
\ , \label{lagm} \end{equation}
with dilaton field $\Phi$,
field strength tensor $ F_{\mu \nu} = 
\partial_\mu A_\nu -\partial_\nu A_\mu + i e \left[A_\mu , A_\nu \right] $,
gauge field $A_\mu =  A_\mu^a \tau_a/2$,
and Newton's constant $G$, dilaton coupling constant $\kappa$,
and Yang-Mills coupling constant $e$.

Variation of the action with respect to the metric and the matter fields
leads, respectively, to the Einstein equations 
\begin{equation}
G_{\mu\nu}= R_{\mu\nu}-\frac{1}{2}g_{\mu\nu}R = 8\pi G T_{\mu\nu}
\  \label{ee} \end{equation}
with stress-energy tensor
\begin{eqnarray}
T_{\mu\nu} &=& g_{\mu\nu}L_M -2 \frac{\partial L_M}{\partial g^{\mu\nu}}
 \nonumber \\
  &=& \partial_\mu \Phi \partial_\nu \Phi
     -\frac{1}{2} g_{\mu\nu} \partial_\alpha \Phi \partial^\alpha \Phi
      + 2 e^{2 \kappa \Phi }{\rm Tr}
    ( F_{\mu\alpha} F_{\nu\beta} g^{\alpha\beta}
   -\frac{1}{4} g_{\mu\nu} F_{\alpha\beta} F^{\alpha\beta})
\ , \label{tmunu}
\end{eqnarray}
and the matter field equations,
\begin{eqnarray}
& &\frac{1}{\sqrt{-g}} \partial_\mu \left(\sqrt{-g} 
 \partial^\mu \Phi \right)=
 \kappa e^{2 \kappa \Phi } 
  {\rm Tr} \left( F_{\mu\nu} F^{\mu\nu} \right)  \ ,
\label{feqD} \end{eqnarray}
\begin{eqnarray}
& &\frac{1}{\sqrt{-g}} D_\mu(\sqrt{-g} e^{2 \kappa \Phi } F^{\mu\nu}) = 0 \ ,
\label{feqA} \end{eqnarray}
where $D_{\mu}=\partial_{\mu} + ie \left[A_\mu , \cdot \right]$.

\subsection{\bf Stationary Axially Symmetric Ansatz}

The system of partial differential equations, 
(\ref{ee}), (\ref{feqD}), and (\ref{feqA}), 
is highly non-linear and complicated. 
In order to generate solutions to these equations,
one profits from the use of symmetries,
simplifying the equations.

Here we consider black hole solutions,
which are both stationary and axially symmetric.
We therefore impose on the spacetime the presence of
two commuting Killing vector fields, 
$\xi$ (asymptotically timelike) and $\eta$ (asymptotically spacelike).
Since the Killing vector fields commute,
we may adopt a system of adapted coordinates, 
say $\{t, r, \theta, \varphi\}$, such that 
\begin{equation}
\xi=\partial_t \ , \ \ \ \eta=\partial_{\varphi}
\ . \label{xieta} \end{equation}
In these coordinates the metric is independent of $t$ and $\varphi$. 
We also assume that the symmetry axis of the spacetime, 
the set of points where $\eta=0$, is regular,
and satisfies the elementary flatness condition \cite{book}
\begin{equation}
\frac{X,_\mu X^{,\mu}}{4X} = 1 \ , \ \ \
X=\eta^\mu \eta_\mu \
\ .  \label{regcond} \end{equation}

Apart from the symmetry requirement on the metric 
(${\cal L}_{\xi} g = {\cal L}_{\eta} g =0$, 
i.e., $g_{\mu \nu}=g_{\mu \nu}(r,\theta)$), 
we impose that the matter fields are also symmetric 
under the spacetime transformations generated by $\xi$ and $\eta$. 

This implies for the dilaton field
\begin{equation}
{\cal L}_{\xi} \Phi = {\cal L}_{\eta} \Phi =0
\ ,  \label{symd} \end{equation}
so $\Phi$ depends on $r$ and $\theta$ only.

For the gauge potential $A=A_\mu dx^\mu$, 
the concept of generalised symmetry \cite{Forgacs,radu2} requires
\begin{eqnarray}
\displaystyle ({\cal L}_{\xi}A)_{\mu} &=&  D_{\mu} W_{\xi}
\ , \nonumber \\
\displaystyle ({\cal L}_{\eta}A)_{\mu} &=&  D_{\mu} W_{\eta}
\ , \label{symA} \end{eqnarray}
where $W_{\xi}$ and $W_{\eta}$ are two compensating su(2)-valued functions 
satisfying
\begin{equation}
{\cal L}_{\xi}W_{\eta}-{\cal L}_{\eta}W_{\xi} 
                + i e \left[W_{\xi},W_{\eta}\right] = 0
\ .  \label{compat} \end{equation}
Performing a gauge transformation to set $W_{\xi}=0$, 
leaves $A$ and $W_{\eta}$ independent of $t$.

To further simplify the system of equations, one can impose 
that the Killing fields generate orthogonal 2-surfaces 
(Frobenius condition)
\begin{equation}
\mbox{\boldmath $\xi$} \wedge \mbox{\boldmath $\eta$} 
 \wedge d\mbox{\boldmath $\xi$} = \mbox{\boldmath $\xi$} 
 \wedge \mbox{\boldmath $\eta$} \wedge d\mbox{\boldmath $\eta$} = 0 
\ ,  \label{Frob} \end{equation}
where $\mbox{\boldmath $\xi$}$ and $\mbox{\boldmath $\eta$}$ 
are considered as 1-forms. 
By virtue of the circularity theorem \cite{trump}
such a condition may be written in terms of the Ricci tensor
\begin{equation}
\mbox{\boldmath $\xi$} \wedge \mbox{\boldmath $\eta$} 
\wedge \mbox{\boldmath $R$}(\xi) = \mbox{\boldmath $\xi$} 
\wedge \mbox{\boldmath $\eta$} \wedge \mbox{\boldmath $R$}(\eta) = 0 
\ ,  \label{circul} \end{equation}
where $(\mbox{\boldmath $R$}(v))_{\mu} = R_{\mu \nu} v^{\nu}$.

The metric can then be written in the Lewis-Papapetrou 
form \cite{lew_pap}, which in isotropic coordinates reads
\begin{equation}
ds^2 = -fdt^2+\frac{m}{f}\left[dr^2+r^2 d\theta^2\right] 
       +\sin^2\theta r^2 \frac{l}{f}
          \left[d\vphi-\frac{\omega}{r}dt\right]^2 \  
\ , \label{metric} \end{equation}
where $f$, $m$, $l$ and $\omega$ are functions of $r$ and $\theta$ only.
Note the change of sign of the function $\omega$ 
with respect to Ref.~\cite{kkrot}.

The $z$-axis represents the symmetry axis.
The regularity condition along the $z$-axis (\ref{regcond}) requires
\begin{equation}
m|_{\theta=0,\pi}=l|_{\theta=0,\pi}
\ . \label{lm} \end{equation}

The event horizon of stationary black hole solutions
resides at a surface of constant radial coordinate, $r=r_{\rm H}$,
and is characterized by the condition $f(r_{\rm H},\theta)=0$ \cite{kkrot}.
The Killing vector field
\begin{equation}
\chi = \xi + \frac{\omega_{\rm H}}{r_{\rm H}} \eta
\ , \label{chi} \end{equation}
is orthogonal to and null on the horizon \cite{wald}.
The ergosphere, defined as the region in which $\xi_\mu \xi^\mu$ is positive,
is bounded by the event horizon and by the surface where
\begin{equation}
 -f +\sin^2\theta \frac{l}{f} \omega^2 = 0 \ .
 \label{ergo}
\end{equation}

Due to the Einstein equations (\ref{ee}), 
the circularity conditions (\ref{circul}) 
have consequences on the matter content, namely,
\begin{equation}
\mbox{\boldmath $\xi$} \wedge \mbox{\boldmath $\eta$} 
\wedge \mbox{\boldmath $T$}(\xi) = \mbox{\boldmath $\xi$} 
\wedge \mbox{\boldmath $\eta$} \wedge \mbox{\boldmath $T$}(\eta) = 0 
\ ,  \label{Riccirc} \end{equation}
with $(\mbox{\boldmath $T$}(v))_{\mu} = T_{\mu \nu} v^{\nu}$.
However, contrary to the case of Abelian fields,
the conditions (\ref{Riccirc}) are not just a consequence 
of the symmetry requirements, but they give rise to additional restrictions 
on the form of the gauge potential.

For the gauge fields we employ a generalized ansatz \cite{kkn}, 
which trivially fulfils both the symmetry constraints 
(\ref{symA}) and (\ref{compat}) 
and the circularity conditions (\ref{Riccirc}),
\begin{equation}
A_\mu dx^\mu
  =   \Psi dt +A_\varphi (d\varphi-\frac{\omega}{r} dt)
+\left(\frac{H_1}{r}dr +(1-H_2)d\theta \right)\frac{\tau_\varphi^n}{2e}
 , \label{a1} \end{equation}
\begin{equation}
\Psi =B_1 \frac{\tau_r^n}{2e} + B_2 \frac{\tau_\theta^n}{2e}
\ , \label{a3} \end{equation}
\begin{equation}
A_\varphi=   -n\sin\theta\left[H_3 \frac{\tau_r^n}{2e}
            +(1-H_4) \frac{\tau_\theta^n}{2e}\right]  
 . \label{a2} \end{equation}
Here the symbols $\tau_r^n$, $\tau_\theta^n$ and $\tau_\vphi^n$
denote the dot products of the Cartesian vector of Pauli matrices,
$\vec \tau = ( \tau_x, \tau_y, \tau_z) $,
with the spatial unit vectors
\begin{eqnarray}
\vec e_r^{\, n}      &=&
(\sin \theta \cos n \varphi, \sin \theta \sin n \varphi, \cos \theta)
\ , \nonumber \\
\vec e_\theta^{\, n} &=&
(\cos \theta \cos n \varphi, \cos \theta \sin n \varphi,-\sin \theta)
\ , \nonumber \\
\vec e_\varphi^{\, n}   &=& (-\sin n \varphi, \cos n \varphi,0)
\ , \label{rtp} \end{eqnarray}
respectively.
Since the gauge fields wind $n$ times around, while the
azimuthal angle $\varphi$ covers the full trigonometric circle once,
we refer to the integer $n$ as the winding number of the solutions.
The gauge field functions $B_i$ and $H_i$ 
depend on the coordinates $r$ and $\theta$ only.
(For $n=1$ and $\kappa=0$, the previously employed stationary ansatz 
\cite{kkrot} is obtained, whereas for $\omega=B_1=B_2=0$
the static axially symmetric ansatz is recovered \cite{kk},
including the static spherically symmetric ansatz \cite{eymd},
for $n=1$ and $H_1=H_3=0$, $H_2=H_4=w(r)$ and $\Phi=\Phi(r)$.)

The ansatz is form-invariant under Abelian gauge transformations $U$
\cite{kk,kkrot}
\begin{equation}
 U= \exp \left({\frac{i}{2} \tau^n_\varphi \Gamma(r,\theta)} \right)
\ .\label{gauge} \end{equation}
With respect to this residual gauge degree of freedom 
we choose the gauge fixing condition 
$r\partial_r H_1-\partial_\theta H_2 =0$
\cite{kk,kkrot}.

For the gauge field ansatz, Eqs.~(\ref{a1})-(\ref{a2}),
the compensating matrix $W_{\eta}$ is given by
\begin{equation}
W_{\eta}=  n\frac{\tau_z}{2 e}
\ . \label{Wcomp} \end{equation}
We note, that by employing the gauge transformation $U$ \cite{foot}
\begin{equation}
U = \exp \left({  \frac{i}{2} \tau_z n \varphi }\right)
\ , \label{Wcompg} \end{equation}
one can choose $W_{\eta}'=0$ \cite{Forgacs,FR},
leading to the gauge field $A_\mu'$,
\begin{equation}
A_\mu' dx^\mu
  =   \Psi' dt +A_\varphi' (d\varphi-\frac{\omega}{r} dt)
+\left(\frac{H_1}{r}dr +(1-H_2)d\theta \right)\frac{\tau_y}{2e}
 , \label{a1'} \end{equation}
\begin{equation}
\Psi'=B_1 \left(
 \sin \theta  \frac{\tau_x}{2e} + \cos \theta  \frac{\tau_z}{2e}
 \right) +
      B_2 \left(
 \cos \theta  \frac{\tau_x}{2e} - \sin \theta  \frac{\tau_z}{2e}
 \right) 
 - n\frac{\tau_z}{2e} \frac{\omega}{r} 
\ , \label{a3'} \end{equation}
\begin{equation}
A_\varphi'=   -n\sin\theta\left[H_3 \left(
 \sin \theta  \frac{\tau_x}{2e} + \cos \theta  \frac{\tau_z}{2e} \right)
            +(1-H_4) \left(
 \cos \theta  \frac{\tau_x}{2e} - \sin \theta  \frac{\tau_z}{2e} \right)
 \right]
 - n\frac{\tau_z}{2e} 
 . \label{a2'} \end{equation}

We further note, that when imposing restricted circularity conditions
by requiring
$F(\mbox{\boldmath $\xi$} , \mbox{\boldmath $\eta$})
={^*}F(\mbox{\boldmath $\xi$} , \mbox{\boldmath $\eta$}) =0$,
only Abelian solutions are possible if asymptotic flatness is assumed \cite{FR}.

\subsection{\bf Boundary Conditions}

{\bf Dimensionless Quantities}

For notational simplicity
we introduce the dimensionless coordinate $x$,
\begin{equation}
x=\frac{e}{\sqrt{4\pi G}} r
\ , \label{dimless} \end{equation}
the dimensionless electric gauge field functions
${\bar B}_1$ and ${\bar B}_2$,
\begin{equation}
{\bar B}_1 = \frac{\sqrt{4 \pi G}}{e}  B_1 \ , \ \ \
{\bar B}_2 = \frac{\sqrt{4 \pi G}}{e}  B_2 \ ,
\label{barb} \end{equation}
the dimensionless dilaton function $\phi$,
\begin{equation}
\phi = \sqrt{4\pi G} \Phi
\ , \label{dimp} \end{equation}
and the dimensionless dilaton coupling constant $\gamma$,
\begin{equation}
\gamma =\frac{1}{\sqrt{4\pi G}} \kappa
\ . \label{dimg} \end{equation}
For $\gamma = 1$ contact with the low energy effective action
of string theory is made,
whereas in the limit $\gamma \rightarrow 0$
the dilaton decouples and EYM theory is obtained. 
\\
\\
\noindent {\sl \bf Boundary conditions at infinity}

To obtain asymptotically flat solutions, we impose
on the metric functions
the boundary conditions at infinity 
\begin{equation}
f|_{x=\infty}= m|_{x=\infty}= l|_{x=\infty}=1 \ , \ \ \
\omega|_{x=\infty}= 0
\ . \label{bc1a} \end{equation}

For the dilaton function we choose
\begin{equation}
\phi|_{x=\infty}=0
\ , \label{bc1b} \end{equation}
since any finite value of the dilaton field at infinity
can always be transformed to zero via
$\phi \rightarrow \phi - \phi(\infty)$,
$x \rightarrow x e^{-\gamma \phi(\infty)} $.

We further impose, that the two electric gauge field functions 
$\bar B_i$ vanish asymptotically,
$$
\bar B_1|_{x=\infty}= \bar B_2|_{x=\infty}= 0
 \ . $$
For magnetically neutral solutions,
the gauge field functions $H_i$ have to satisfy
\begin{equation}
H_1|_{x=\infty}= H_3|_{x=\infty}= 0 \ , \ \ \
H_2|_{x=\infty}= H_4|_{x=\infty}= (-1)^k
\ , \label{bc1c} \end{equation}
where the node number $k$ is defined as the number of nodes of
the functions $H_2$ and $H_4$ along 
the positive (or negative) $z$-axis \cite{kkreg,kk}.
Note, that for each node number there is a  degenerate solution
with $H_2|_{x=\infty}= H_4|_{x=\infty}= -(-1)^k$,
related by the large gauge transformation $U=i\tau_r^n$.
Under this transformation the functions transform according to
$$H_1 \rightarrow -H_1 \ , \ \ \
  H_2 \rightarrow -H_2 \ , \ \ \
  H_3 \rightarrow +H_3 \ , \ \ \
  H_4 \rightarrow -H_4  \ , $$
$$
  \bar B_1 \rightarrow +\bar B_1 \ , \ \ \
  \bar B_2 \rightarrow -\bar B_2 -2 n \sin\theta \frac{\omega}{x} \ .
$$

\noindent {\sl \bf  Boundary conditions at the horizon}

The event horizon of stationary black hole solutions
resides at a surface of constant radial coordinate, $x=x_{\rm H}$,
and is characterized by the condition $f(x_{\rm H},\theta)=0$ \cite{kkrot}.

Regularity at the horizon then requires the following boundary conditions
for the metric functions
\begin{equation}
f|_{x=x_{\rm H}}=
m|_{x=x_{\rm H}}=
l|_{x=x_{\rm H}}=0
\ , \ \ \ \omega|_{x=x_{\rm H}}=\omega_{\rm H}= {\rm const.}
\ , \label{bh2a} \end{equation}
for the dilaton function 
\begin{equation}
\partial_x \phi|_{x=x_{\rm H}} =0
\ , \label{bh2b} \end{equation}
and for the `magnetic' gauge field functions
\begin{equation}
           H_1 |_{x=x_{\rm H}}= 0 \ , \ \ \
\partial_x H_2 |_{x=x_{\rm H}}= 0 \ , \ \ \
\partial_x H_3 |_{x=x_{\rm H}}= 0 \ , \ \ \
\partial_x H_4 |_{x=x_{\rm H}}= 0 \  
\ , \label{bh2c} \end{equation}
with the gauge condition $\partial_\theta H_1=0$ taken into account
\cite{kkrot}.

As discussed in \cite{kkrot},
the boundary conditions for the `electric' part of the 
gauge potential are obtained from the requirement that
the electro-static potential $\Psiel=\chi^\mu A_\mu$
is constant at the horizon \cite{isorot},
\begin{equation}
\Psiel(r_{\rm H}) = \chi^\mu A_\mu |_{r=r_{\rm H}}= \Psi_{\rm H}
\ . \label{esp0} \end{equation}
Defining the dimensionless electrostatic potential $\psiel$,
\begin{equation}
{\psiel} = {\sqrt{4 \pi G}}  \Psiel \ , 
\label{psi} \end{equation}
and the dimensionless horizon angular velocity $\Omega$,
\begin{equation}
 \Omega = \frac{\omega_{\rm H}}{x_{\rm H}}
\ , \label{Omega} \end{equation}
this yields the boundary conditions 
\begin{equation}
\bar B_1 |_{x=x_{\rm H}}  =   n \Omega \cos \theta  \ , \ \ \
\bar B_2 |_{x=x_{\rm H}}  =  -n \Omega \sin \theta  \ .
\label{bh2d}
\end{equation}
With these boundary conditions, the horizon electrostatic
potential reads
\begin{equation}
\psiel_{\rm H} = n \Omega \frac{\tau_z}{2}
= \psi_{\rm el} \frac{\tau_z}{2}
\ , \label{esp} \end{equation}
defining $\psi_{\rm el}$ for the non-Abelian mass formula.
\\
\\
\noindent {\sl \bf Boundary conditions along the axes}

The boundary conditions along the $\rho$- and $z$-axis
($\theta=\pi/2$ and $\theta=0$) are determined by the
symmetries.
They are given by
\begin{eqnarray}
& &\partial_\theta f|_{\theta=0} =
   \partial_\theta m|_{\theta=0} =
   \partial_\theta l|_{\theta=0} =
   \partial_\theta \omega|_{\theta=0} = 0 \ ,
\nonumber \\
& &\partial_\theta f|_{\theta=\pih} =
   \partial_\theta m|_{\theta=\pih} =
   \partial_\theta l|_{\theta=\pih} =
   \partial_\theta \omega|_{\theta=\pih} = 0 \ ,
\label{bc4a} \end{eqnarray}
\begin{eqnarray}
& &\partial_\theta \phi|_{\theta=0} = 0 \ ,
\nonumber \\
& &\partial_\theta \phi|_{\theta=\pih} = 0 \ ,
\label{bc4b} \end{eqnarray}
$$
    \bar B_2|_{\theta=0}=0 \ , \ \ \ \partial_\theta \bar B_1|_{\theta=0}=0 \ ,
$$
$$
 H_1|_{\theta=0}=H_3|_{\theta=0}=0 \ , \ \ \
   \partial_\theta H_2|_{\theta=0} =
   \partial_\theta H_4|_{\theta=0}  = 0 \ ,
$$
$$
    \bar B_1|_{\theta=\pih}=0 \ , \ \ \ 
    \partial_\theta \bar B_2|_{\theta=\pih}=0 \ ,
$$
\begin{equation}
 H_1|_{\theta=\pih}=H_3|_{\theta=\pih}=0 \ , \ \ \
    \partial_\theta H_2|_{\theta=\pih} =
    \partial_\theta H_4|_{\theta=\pih} = 0 \ .
\   \label{bc4c} \end{equation}
In addition, regularity on the $z$-axis requires condition (\ref{lm})
for the metric functions to be satisfied,
and regularity of the energy density on the $z$-axis requires
\begin{equation}
H_2|_{\theta=0}=H_4|_{\theta=0}
\ . \label{h2h4} \end{equation}

\section{\bf Black Hole Properties}

We derive the properties of the stationary axially symmetric
black holes from the expansions of their gauge field and matter functions
at infinity and at the horizon. The expansion at infinity yields
the global charges of the black holes and their magnetic moments.
Generically the gauge field
functions of the black hole solutions show a non-integer power
fall-off asymptotically, with the exponents determined by the 
non-Abelian electric charge $Q$. 
The expansion at the horizon yields the horizon properties,
such as the area parameter, the surface gravity, and the horizon deformation.
We also introduce a topological charge of the horizon.
We then give a detailed account of the non-Abelian mass formula.

\subsection{\bf Global Charges}

The mass ${\cal M}$ and the angular momentum ${\cal J}$
of the black hole solutions are obtained
from the metric components $g_{tt}$ and $g_{t\varphi}$, respectively.
The asymptotic expansion for the metric function $f$,
$$
f = 1-\frac{2 M}{x} +  O \left(\frac{1}{x^2}\right)\ ,
$$
yields for the dimensionless mass $M$,
\begin{equation}
 M= \frac{1}{2}\lim_{x\rightarrow\infty}x^2\partial_x f \ ,
\  \label{MJ1} \end{equation}
where
\begin{equation}
{\cal M} =
\frac{\sqrt{4\pi G}}{eG} M \ .
\label{dimlessM} \end{equation}
Likewise,
the asymptotic expansions for the metric function $\omega$,
$$
\omega = \frac{2 J}{x^2} + O\left(\frac{1}{x^3}\right)\ ,
$$
yields for the dimensionless angular momentum $J$,
\begin{equation}
 J = \frac{1}{2}\lim_{x\rightarrow\infty}x^2\omega
\ , \label{MJ2} \end{equation}
where
\begin{equation}
{\cal J} = \frac{4 \pi}{e^2} J \ .
\label{dimlessJ} \end{equation}

The asymptotic expansion for the dilaton function $\phi$,
$$
\phi = -\frac{D}{x}-\frac{\gamma Q^2}{2 x^2} + O\left(\frac{1}{x^3}\right)
\ ,
$$
yields the dimensionless dilaton charge $D$,
\begin{equation}
 D =  \lim_{x\rightarrow\infty}x^2\partial_x\phi
\ , \label{D} \end{equation}
related to the dilaton charge ${\cal D}$ via
\begin{equation}
{\cal D} = \frac{1}{e} D \ .
\label{dimlessD} \end{equation}

The asymptotic expansion of the gauge field yields the global
non-Abelian electromagnetic charges, $\cal Q$ and $\cal P$.
The asymptotic expansion of the electric gauge field functions
$\bar B_1$ and $\bar B_2$,
$$
{\bar B}_1 = \frac{Q \cos{\theta}}{x}
+ O\left(\frac{1}{x^2}\right)\ , \ \ \ \
{\bar B}_2 = -(-1)^k \frac{Q \sin{\theta}}{x}
 + O\left(\frac{1}{x^2}\right)\ ,
$$
yield the dimensionless non-Abelian
electric charge $Q$,
\begin{equation}
  Q =- \lim_{x\rightarrow\infty}x^2\partial_x \left( \cos \theta \bar B_1
                                -(-1)^k \sin \theta \bar B_2 \right)
\ , \label{Q} \end{equation}
where
\begin{equation}
 {\cal Q} =\frac{Q}{e}
\ . \label{dimlessQ} \end{equation}
The boundary conditions of the magnetic gauge field functions
guarantee that the dimensionless non-Abelian magnetic charge $P$ vanishes.

The non-Abelian global charges $Q$ and $P$ appear to be gauge dependent.
In particular, the definition of $Q$ corresponds to rotating to a gauge,
where the gauge field component $\psi=\sqrt{4 \pi G} \Psi$ asymptotically
only has a $\tau_z$ component \cite{kkrot},
$$
 \psi \rightarrow \frac{Q}{x} \frac{\tau_z}{2} \ .
$$
Identifying the global non-Abelian electric charge $Q$ in this way,
corresponds to the usual choice \cite{kkrot,radu2,ad}.

We note, that the modulus of the non-Abelian electric charge, $|Q|$,
corresponds to a gauge invariant definition for the
non-Abelian electric charge given in Ref.~\cite{iso1},
\begin{equation}
 {\cal Q}^{\rm YM} = \frac{1}{4\pi}
 \oint \sqrt{\sum_i{\left({^*}F^i_{\theta\varphi}\right)^2}}
 d\theta d\varphi = \frac{|Q|}{e}
\ , \label{Qdelta} \end{equation}
where ${^*}F$ represents the dual field strength tensor,
and the integral is evaluated at spatial infinity.

Likewise, a gauge invariant definition of the non-Abelian magnetic charge
is given in Ref.~\cite{iso1},
\begin{equation}
 {\cal P}^{\rm YM} = \frac{1}{4\pi}
 \oint \sqrt{\sum_i{\left(F^i_{\theta\varphi}\right)^2}} d\theta d\varphi
 =  \frac{|P|}{e}
\ , \label{Pdelta} \end{equation}
where again the integral is evaluated at spatial infinity,
yielding $P=0$.

Note that the lowest order terms in the expansions, as needed for
the expressions of the global charges $M$, $J$, $D$, $Q$ and $P$,
do not involve non-integer powers.

\subsection{Non-integer power fall-off}
Here we present the lowest order terms of the expansion of
the `magnetic' gauge field functions $H_1$ -- $H_4$ for winding
number $n=1$ and $n=3$. (For details see Appendix A).
Since for $n=2$ the lowest order expansion contains non-analytic terms
like $\log(x)/x^2$ already in the static limit \cite{addendum}, we refrained from
an analysis of the rotating solutions in this case.

The asymptotic expansion yields for $n=1$
\begin{eqnarray}
H_1&=&\left[ \frac{2 C_5}{x^2}
+ \frac{8 C_4}{\beta-1} x^{-\frac{1}{2} (\beta-1)}
-\frac{2 C_2 C_3 (\alpha+3)}{(\alpha+5)Q^2}
 x^{-\frac{1}{2}(\alpha+1)}\right] \sin{\theta} \cos{\theta}
+ o\left(\frac{1}{x^2}\right)\ ,\nonumber\\
H_2&=&(-1)^k+C_3 x^{-\frac{1}{2}(\alpha-1)}
+o\bigg(x^{-\frac{1}{2}(\alpha-1)}\bigg)\ ,\nonumber\\
H_3&=&\bigg(\frac{C_2}{x} - (-1)^k C_3 x^{-\frac{1}{2}(\alpha-1)} \bigg)
 \sin{\theta} \cos{\theta}
 + o\left(\frac{1}{x}\right)\ , \nonumber \\
H_4&=&(-1)^k+C_3 x^{-\frac{1}{2}(\alpha-1)} + \bigg((-1)^k\frac{C_2}{x}
- C_3 x^{-\frac{1}{2}(\alpha-1)} \bigg) \sin^2{\theta}
+ o\left(\frac{1}{x}\right)
\ , \label{asymp_short_n1}
\end{eqnarray}
where $C_i$ are dimensionless constants.

Here $\alpha$ and $\beta$ determine the non-integer fall-off
of the gauge field functions,
\begin{equation}
\alpha = \sqrt{9-4 Q^2} \ , \ \ \
\beta = \sqrt{25-4 Q^2} \ .
\label{alpha-beta} \end{equation}

For $n=3$ the lowest order terms of the expansion involve only
integer powers,
\begin{eqnarray}
H_1 &=& \frac{C_5}{x^2}\sin{2 \theta}
 + o\left(\frac{1}{x^2}\right) \ , \nonumber \\
H_2 &=& {\DS (-1)^k + \frac{C_5}{x^2} \cos{2 \theta}
 + o\left(\frac{1}{x^2}\right) }\ , \nonumber \\
H_3 &=& \frac{C_2}{x} \sin\theta \cos\theta + \frac{(M+\gamma D) C_2
 -2 (-1)^k C_5 -a M Q}{2 x^2} \sin\theta \cos\theta
+ o\left(\frac{1}{x^2}\right) \ , \nonumber \\
H_4 &=& (-1)^k + (-1)^k \frac{C_2}{x}\sin^2\theta + \frac{C_5}{x^2}
 + (-1)^k \frac{(M+\gamma D)C_2 - 2 (-1)^k C_5 - aMQ}{2 x^2} \sin^2\theta
\nonumber \\
& &  + o\left(\frac{1}{x^2}\right) \ , \label{asymp_short_n3}
\end{eqnarray}
but higher order terms contain non-integer powers involving
${\displaystyle \epsilon= \sqrt{49-4Q^2}}$.

To determine, whether the generic non-integer fall-off is a physical
property of the solutions or a gauge artefact, we consider the
asymptotic behaviour of the gauge invariant quantities
${\rm Tr} \left(F_{\mu\nu} F^{\mu\nu}\right)$ and
${\rm Tr} \left( F_{\mu\nu} {^*}F^{\mu\nu}\right)$.
Inserting the expansions Eqs.~(\ref{asymp_short_n1})
and (\ref{asymp_short_n3})
we find
\begin{equation}
{\rm Tr} \left(F_{\mu\nu} F^{\mu\nu}\right) = -\frac{Q^2}{x^4}
 + \frac{4(M-\gamma D)Q^2}{x^5} + o\left(\frac{1}{x^5}\right) \ ,
\ \ \ n=1 \ {\rm and} \ n=3  \ ,
\end{equation}
and
\begin{eqnarray}
{\rm Tr} \left(F_{\mu\nu} {^*}F^{\mu\nu}\right)  & = &
\frac{4 C_2 Q}{x^5}\cos\theta
 + o\left(\frac{1}{x^5}\right)
 \ , \ \ \ n=1 \ , \nonumber \\
{\rm Tr} \left(F_{\mu\nu} {^*}F^{\mu\nu}\right) & = &
 \frac{12 C_2 Q}{x^5}\cos\theta
 + o\left(\frac{1}{x^5}\right)  \ , \ \ \ n=3 \ .
\nonumber
\end{eqnarray}
Although lowest order terms do not contain the non-integer powers,
they occur in the higher order terms, indicating that the
non-integer power decay cannot be removed by a gauge transformation.

\subsection{\bf Magnetic Moment}

In Abelian gauge theory,
the dimensionless magnetic field $\vec {\cal B}_{\cal A}$
of a magnetic dipole with dipole moment
$\vec{\mu}^{\cal A} = \mu_z^{\cal A} \vec e_z$ is given by
$$
\vec {\cal B}_{\cal A} =
\frac{\mu_z^{\cal A}}{x^3} \left(2 \cos \theta \vec e_r
                                    + \sin \theta \vec e_\theta \right) \ .
$$
To investigate the magnetic moment for the non-Abelian black holes
we work in the gauge where asymptotically
$A_0 = \frac{Q}{x} \frac{\tau_z}{2} + o(1/x)$.
From the asymptotic expansion of the gauge field functions,
given in  Appendix A,
we find for the magnetic field for winding number $n=1$ solutions
\begin{eqnarray}
{\cal B}_{\hat r} &=& - \frac{C_2}{x^3} 2 \cos \theta \frac{\tau_z}{2}
        + b_{\hat r} \frac{\tau_\rho^1}{2}
 \ , \nonumber \\
{\cal B}_{\hat \theta} &=& - \frac{C_2}{x^3} \sin \theta \frac{\tau_z}{2}
        + b_{\hat \theta} \frac{\tau_\rho^1}{2}
 \ , \nonumber \\
{\cal B}_{\hat \vphi} &=&
          b_{\hat \vphi} \frac{\tau_\vphi^1}{2}
        \  ,
\label{B-Field} \end{eqnarray}
where the functions $b_{\hat r},b_{\hat \theta},b_{\hat \vphi}$
involve non-integer powers of $x$, and contain the leading terms
in the expansion for $Q\neq 0$. Since the leading terms decay slower
than $x^{-3}$ for $Q\neq 0$ we cannot extract the magnetic moment
in general.

In the following we give a suggestion for the definition of the
magnetic moment.
The asymptotic form of $A_0$ suggests to
interpret $\frac{\tau_z}{2}$ as `electric charge operator'
$\hat{Q}\in {\rm su(2)}$ in the given gauge.
We then observe that the projection of
${\cal B}_{\hat \mu}$
in $\hat{Q}$-direction,
\begin{eqnarray}
{\cal B}^{(\hat{Q})}_{\hat r} &=& - \frac{C_2}{x^3} 2 \cos \theta
+o(1/x^3)
 \ , \nonumber \\
{\cal B}^{(\hat{Q})}_{\hat \theta} &=& - \frac{C_2}{x^3} \sin \theta
+o(1/x^3)
 \ , \nonumber \\
{\cal B}^{(\hat{Q})}_{\hat \vphi} &=& o(1/x^3)
\label{Bz-Field} \end{eqnarray}
corresponds asymptotically to the magnetic field of a magnetic
dipole $\vec{\mu}= -C_2 \vec{e}_z$ with magnitude
\begin{equation}|\vec{\mu}| = |C_2| \ . \label{magmom} \end{equation}
We note that $\vec{\mu}$ is invariant under time independent gauge
transformations, since such
a gauge transformation rotates
the `electric charge operator' $\hat{Q}$
and the magnetic field ${\cal B}_{\hat \mu}$
in the same way and leaves the projection
${\cal B}^{(\hat{Q})}_{\hat\mu}$ invariant.

A special case arises for $n=1$ and $Q=0$, when the
functions $b_{\hat r},b_{\hat \theta},b_{\hat \vphi}$ are of order
$O(x^{-3})$. In this case the asymptotic form of the magnetic field reads
\begin{eqnarray}
{\cal B}_{\hat r} &=& - \frac{C_2}{x^3} 2 \cos \theta \frac{\tau_z}{2}
        +\frac{ C_3}{x^3}2\sin\theta \frac{\tau_\rho^1}{2} + O(1/x^4)
 \ , \nonumber \\
{\cal B}_{\hat \theta} &=& - \frac{C_2}{x^3} \sin \theta \frac{\tau_z}{2}
        - \frac{C_3}{x^3}\cos\theta \frac{\tau_\rho^1}{2} + O(1/x^4)
 \ , \nonumber \\
{\cal B}_{\hat \vphi} &=&
          -\frac{C_3}{x^3} \frac{\tau_\vphi^1}{2} + O(1/x^4)
        \  ,
\label{Bapp-Field} \end{eqnarray}
Comparison with the magnetic field of a magnetic dipole
$\vec{\mu}^{\cal A}$ in Abelian gauge theory,
\begin{eqnarray}
\vec{B}_{\cal A} & = &
-\left( \vec{\mu}^{\cal A}\cdot \vec{\nabla}\right) \frac{\vec{x}}{x^3}
\nonumber \\
 & = &
\frac{1}{x^3} \left(
 \mu_z^{\cal A}(2 \cos\theta \vec{e}_r + \sin\theta  \vec{e}_\theta)
+\mu_\rho^{\cal A}(2\sin\theta \vec{e}_r - \cos\theta \vec{e}_\theta)
+\mu_\vphi^{\cal A}\vec{e}_\vphi \right) \ ,
\nonumber
\end{eqnarray}
suggests the identification of the su(2) valued magnetic moment
\begin{eqnarray}
\mu_x & = & C_3 \frac{\tau_x}{2} \ ,
\nonumber\\
\mu_y & = & C_3 \frac{\tau_y}{2} \ ,
\nonumber\\
\mu_z & = & -C_2 \frac{\tau_z}{2} \ .
\label{muc2c3}
\end{eqnarray}
In the static case $C_3=-C_2$, reflecting the spherical symmetry of
the solution. For the stationary rotating solutions the spherical
symmetry is broken to axial symmetry, corresponding to
$C_3 \neq -C_2$.

For winding number $n=3$ the asymptotic expansion of the
gauge field functions yields for the magnetic field
\begin{eqnarray}
{\cal B}_{\hat r} &=& - \frac{C_2}{x^3} 2 \cos \theta \frac{\tau_z}{2}
        + O(1/x^4)
 \ , \nonumber \\
{\cal B}_{\hat \theta} &=& - \frac{C_2}{x^3} \sin \theta \frac{\tau_z}{2}
        + O(1/x^4)
 \ , \nonumber \\
{\cal B}_{\hat \vphi} &=&          O(1/x^4)
        \  ,
\label{Bn3-Field} \end{eqnarray}
leading again to $\vec{\mu} =-C_2 \vec{e}_z$.

\subsection{\bf Expansion at the Horizon}

Expanding the metric and gauge field functions at the horizon
in powers of
\begin{equation}
\delta=\frac{x}{x_{\rm H}}-1
\end{equation}
yields to lowest order
\begin{eqnarray}
f(\delta,\theta)&=&\delta^2 f_2 (1 -\delta) + O(\delta^4) 
\ , \nonumber\\
m(\delta,\theta)&=&\delta^2 m_2 (1 -3\delta) + O(\delta^4) 
\ , \nonumber\\
l(\delta,\theta)&=&\delta^2 l_2 (1 -3\delta) + O(\delta^4) 
\ , \nonumber\\
\omega(\delta,\theta)&=&\omega_{\rm H} (1 + \delta) + O(\delta^2) 
\ , \nonumber\\
{\phi}(\delta,\theta)&=&\phi_0 + O(\delta^2) 
\ , \nonumber\\
{\bar B}_1(\delta,\theta)&=&n\frac{\omega_{\rm H} }{x_{\rm H}}\cos{\theta}
+ O(\delta^2) \ , \nonumber\\
{\bar B}_2(\delta,\theta)&=&-n\frac{\omega_{\rm H} }{x_{\rm H}}\sin{\theta}
+ O(\delta^2) \ , \nonumber\\
H_1(\delta,\theta)&=&\delta \left(1 -\frac{1}{2}\delta \right) H_{11}
+ O(\delta^3) \ , \nonumber\\
H_2(\delta,\theta)&=&H_{20}+O(\delta^2) 
\ , \nonumber\\
H_3(\delta,\theta)&=&H_{30} + O(\delta^2)
\ , \nonumber\\
H_4(\delta,\theta)&=&H_{40}+ O(\delta^2) \ ,\label{H4_hor_red}
\end{eqnarray}
The expansion coefficients $f_2$, $m_2$, $l_2$, $\phi_0$, $H_{11}$,
$H_{20}$, $H_{30}$, $H_{40}$ are functions of the variable $\theta$.
Among these coefficients the following relations hold,
\begin{equation}
0=\frac{\partial_{\theta}m_2}{m_2}
-2 \frac{\partial_{\theta}f_2}{f_2}
\ , \label{relation_hor_1} \end{equation}
\begin{equation}
H_{11}=\partial_{\theta} H_{20}
\ . \label{relation_hor_2} \end{equation}
Further details of the expansion at the horizon
are given in Appendix B.

\subsection{Horizon Properties}

Let us now discuss the horizon properties of the SU(2) EYMD black hole solutions.
The first quantity of interest is the area of the black hole horizon. 
The dimensionless area $A$, given by
\begin{equation}
A = 2 \pi \int_0^\pi  d\theta \sin \theta
\frac{\sqrt{l_2 m_2}}{f_2} x_{\rm H}^2
\ . \label{area} \end{equation}
defines the area parameter $x_\Delta$ via \cite{iso1} 
\begin{equation}
A = 4 \pi x_\Delta^2 \ ,
\label{xDelta} \end{equation}
and the dimensionless entropy $S$ of the black hole
\begin{equation}
S = \frac{A}{4} \ .
\label{entro} \end{equation}

The surface gravity of the black hole solutions is obtained from \cite{wald}
\begin{equation}
\kappa_{\rm sg}^2 =
 - \frac{1}{4} (\nabla_\mu \chi_\nu)(\nabla^\mu \chi^\nu)
\ , \label{sgwald} \end{equation}
with Killing vector
$\chi$, Eq.~(\ref{chi}).
Inserting the expansion at the horizon,
Eqs.~(\ref{H4_hor_red}),
yields the dimensionless surface gravity
\begin{equation}
\kappa_{\rm sg} = \frac{f_2(\theta)}{x_{\rm H} \sqrt{m_2(\theta)}}
\ . \label{temp} \end{equation}
As seen from Eq.~(\ref{relation_hor_1}),
$\kappa_{\rm sg}$ is indeed constant on the horizon,
as required by the zeroth law of black hole mechanics.
The dimensionless temperature $T$ 
of the black hole is proportional to the surface gravity,
\begin{equation}
T = \frac{\kappa_{\rm sg}}{2 \pi} 
\ . \label{tempt} \end{equation}

To obtain a measure for the deformation of the horizon we
compare the dimensionless circumference of the horizon along the equator, 
$L_{\rm e}$,
with the dimensionless circumference of the horizon along 
a great circle passing through the poles, 
$L_{\rm p}$,
\begin{equation}
L_{\rm e} = \int_0^{2 \pi} { d \vphi \left.
 \sqrt{ \frac{l}{f}} x \sin\theta
 \right|_{x=x_{\rm H}, \theta=\pi/2} } \ , \ \ \
L_{\rm p} = 2 \int_0^{ \pi} { d \theta \left.
 \sqrt{ \frac{m  }{f}} x
 \right|_{x=x_{\rm H}, \vphi={\rm const.}} }
\ . \label{lelp} \end{equation}
We obtain further information about the horizon deformation
by considering its Gaussian curvature $K$,
\begin{equation}
K(\theta) =  \frac{R_{\theta\vphi\theta\vphi}}{g_2} \ , \ 
g_2 =   g_{\theta\theta}g_{\vphi\vphi}-g_{\theta\vphi}^2
\ . \label{Gauss} \end{equation}
We determine the topology of the horizon, 
by computing its Euler characteristic $\chi_{\rm E}$,
\begin{equation}
\chi_{\rm E}= \frac{1}{2\pi} \int K \sqrt{g_2} d\theta d\vphi
\ . \label{topol} \end{equation}
Using the expansion Eqs.~(\ref{H4_hor_red}) in the integrand,
\begin{equation}
K\sqrt{g_2}|_{x=x_{\rm H}} = -\frac{\partial}{\partial \theta}
             \left[ \cos\theta \sqrt{\frac{l_2}{m_2}}
                   +\frac{\sin\theta}{2\sqrt{l_2}}
                   \frac{\partial}{\partial \theta}
                     \sqrt{\frac{l_2^2}{m_2}} \right]
\ , \label{topolint} \end{equation}
and the fact that $l_2=m_2$ on the $z$-axis,
this yields for the Euler number $\chi_{\rm E} = 2$,
indicating that the horizon has the topology of a 2-sphere.

To find the horizon electric charge $Q_\Delta$ 
and the horizon magnetic charge $P_\Delta$, one can evaluate the
integrals, Eqs.~(\ref{Qdelta}) and (\ref{Pdelta}),
representing dimensionless gauge invariant quantities \cite{iso1}.
Whereas such a definition of the non-Abelian horizon charges 
appears adequate in the static case, it appears problematic
in the stationary case.
For embedded Abelian solutions, for instance, one obtains
a horizon electric charge, which differs from the global electric charge.
The reason is that when evaluating the horizon electric charge 
according to this prescription, one is
taking the absolute value of the dual field strength tensor.
Thus one does not allow for the cancellation,
present in a purely Abelian theory,
when the integral involves the dual field strength tensor.

We finally introduce 
a topological horizon charge $N$, suggested by Ashtekar \cite{ash}.
For that purpose we consider the su(2)-algebra 
valued 2-form $F_{\rm H}$,
corresponding to the pull-back of the Yang-Mills field strength to 
the horizon $H$,
\begin{equation}
F_{\rm H} = F_{\theta\vphi}|_{\rm H} d\theta \wedge d\vphi 
\ .  \end{equation}
Its dual ${^*}F_{\rm H}$ on the horizon
\begin{equation}
{^*}F_{\rm H} =  \left. g_2^{-1/2} F_{\theta\vphi}
\right|_{\rm H}
\ ,  \end{equation}
represents an su(2)-algebra valued function on the horizon.
We decompose ${^*}F_{\rm H}$ as
\begin{equation}
{^*}F_{\rm H} =  {^*}F_{\rm H}^{\rho} \left(
   \cos n \varphi  \frac{\tau_x}{2} +
   \sin n \varphi  \frac{\tau_y}{2} \right)
       +         {^*}F_{\rm H}^{z} \frac{\tau_z}{2}
\ ,  \end{equation}
with norm $\left|\,{^*}F_{\rm H}\right|$,
\begin{equation}
\left|\,{^*}F_{\rm H}\right| = \sqrt{ \left({^*}F_{\rm H}^{\rho}\right)^2
 + \left({^*}F_{\rm H}^{z}\right)^2 }
\ .  \end{equation}
Next we define a normalised su(2) valued function on the horizon by
\begin{equation}
\sigma =  
 \frac{ {^*}F_{\rm H} }{ \left|\,{^*}F_{\rm H}\right| }
 =  \sin \Theta \left(
   \cos n \varphi  \frac{\tau_x}{2} +
   \sin n \varphi  \frac{\tau_y}{2} \right)
       +         \cos \Theta \frac{\tau_z}{2}
\ ,  \label{sigma} \end{equation}
where 
$$ \sin \Theta = {^*}F_{\rm H}^{\rho}/\left|\,{^*}F_{\rm H}\right|
\ , \ \ \ \ 
\cos \Theta = {^*}F_{\rm H}^{z}/\left|\,{^*}F_{\rm H}\right|
\ . 
$$ 
$\sigma$ represents a map from the 2-sphere of the horizon
to the 2-sphere in the Lie-algebra su(2).
The degree of the map,
\begin{equation}
 N = \frac{1}{4 \pi} \int_{\rm H}
 \frac{1}{2} \varepsilon_{abc} \sigma^a d \sigma^b \wedge d\sigma^c
\ , \label{topN} \end{equation}
defines the topological horizon charge $N$, 
\begin{equation}
N = \frac{1}{4\pi} \int_{\rm H}
 \left( \sin \Theta \, n \partial_\theta \Theta \right) 
 d\theta d\varphi
  =  - \frac{n}{2} \left( \cos \Theta |_{\theta=\pi}
  - \cos \Theta |_{\theta=0} \right)
\ . \end{equation}

To evaluate the topological horizon charge $N$ for the 
stationary non-Abelian black hole solutions,
we express $\cos \Theta$ and $\sin \Theta$ in terms of 
the functions $H_i$. This yields 
\begin{equation}
\sin \Theta = \frac{\sin \theta G_r + \cos \theta G_\theta}
 {\sqrt{G_r^2 + G_\theta^2}} \ , \ \ \
\cos \Theta = \frac{\cos \theta G_r - \sin \theta G_\theta}
 {\sqrt{G_r^2 + G_\theta^2}}
\ , \end{equation}
where
\begin{eqnarray}
G_r  & = & 
- \left[ H_{3,\theta}+H_3{\rm cot}\theta +H_2 H_4 -1 \right]
\frac{1}{e} \ , 
\nonumber \\
G_\theta & = & 
\left[ H_{4,\theta}+{\rm cot}\theta (H_4-H_2)-H_2 H_3 \right]
\frac{1}{e} \ .
\nonumber
\end{eqnarray}
Symmetry and boundary conditions imply
$G_r|_{\theta=\pi}=G_r|_{\theta=0}$ and
$G_\theta|_{\theta=\pi}=G_\theta|_{\theta=0}=0$, therefore
\begin{equation}
 N = n \left. \frac{G_r}{|G_r|}\right|_{\theta=0} 
   = n \, {\rm sign}\left( G_r|_{\theta=0}\right) 
\ , \label{Nn} \end{equation}
relating the winding number $n$ to the
topological horizon charge $N$.
We note that solutions with winding number $-n$ are gauge equivalent
to solutions with winding number $n$, but carry opposite electric charge 
\cite{foot2}.

\subsection{\bf Non-Abelian Mass Formula}

The non-Abelian mass formula, Eq.~(\ref{namass}),
$$
M = 2 TS + 2 \Omega J + \frac{D}{\gamma} + 2 \psi_H Q
$$
is a generalization of the non-Abelian mass formula
$M = 2 TS + D/\gamma$, obtained previously 
for static axially symmetric non-Abelian solutions \cite{kkreg,kk},
which generically carry no non-Abelian charges. 
Rotating non-Abelian black hole solutions in general do
carry non-Abelian electric charge, 
but they do not carry non-Abelian magnetic charge.
Since they nevertheless carry non-trivial non-Abelian magnetic fields,
the Smarr formula \cite{smarr},
\begin{equation}
M = 2 TS + 2 \Omega J + \psi_{\rm el} Q + \psi_{\rm mag} P
\ , \label{smarr} \end{equation}
(where $\psi_{\rm mag}$ represents the horizon magnetic potential)
is bound to fail: The contribution to the mass
from the non-Abelian magnetic fields 
outside the horizon would not be included. 
In constrast, in the non-Abelian mass formula (\ref{namass}),
this magnetic field contribution to the mass is contained in
the dilaton term $\DS \frac{D}{\gamma}$.

To derive the non-Abelian mass formula,
let us begin by considering the general definition of the local mass
$\cal M_{\rm local}$ and local angular momentum $\cal J_{\rm local}$,
(see e.~g.~\cite{wald})
\begin{equation}
{\cal M_{\rm local}} = - \frac{1}{{8\pi G}} \int_{{\rm S}^2} \frac{1}{2}
 \varepsilon_{\mu\nu\rho\sigma}\nabla^\rho \xi^\sigma dx^\mu dx^\nu
\ , \label{Mloc} \end{equation}
\begin{equation}
{\cal J_{\rm local}} =   \frac{1}{{16\pi G}} \int_{{\rm S}^2} \frac{1}{2}
 \varepsilon_{\mu\nu\rho\sigma}\nabla^\rho \eta^\sigma dx^\mu dx^\nu
\ , \label{Jloc} \end{equation}
for a given 2-sphere $S^2$.
By taking the 2-sphere to infinity, and employing Stokes theorem,
the global charges $\cal M$ and $\cal J$ are obtained \cite{wald},
\begin{eqnarray}
{\cal M} = \frac{1}{{4\pi G}} \int_{\Sigma}
 R_{\mu\nu}      n^\mu \xi^\nu dV
 - \frac{1}{{8\pi G}} \int_{\rm H} \frac{1}{2}
 \varepsilon_{\mu\nu\rho\sigma}\nabla^\rho \xi^\sigma dx^\mu dx^\nu
\ , \label{M} \end{eqnarray}
\begin{eqnarray}
 {\cal J} = - \frac{1}{8 \pi G}    \int_\Sigma 
 R_{\mu\nu}      n^\mu \eta^\nu dV
 +\frac{1}{16\pi G} \int_{\rm H} \frac{1}{2}
 \varepsilon_{\mu\nu\rho\sigma} \nabla^\rho \eta^\sigma dx^\mu dx^\nu
\ . \label{J} \end{eqnarray}
Here $\Sigma$ denotes an asymptotically flat hypersurface 
bounded by the horizon ${\rm H}$,
$n^\mu = (1, 0, 0, -\omega/r)/\sqrt{f}$
is normal to $\Sigma$ with $n_\mu n^\mu = -1$,
and $dV$ is the natural volume element on $\Sigma$
($dV=m \sqrt{l/f^3}\, r^2 \sin \theta dr d\theta d\varphi$),
and consequently,
\begin{equation}
\frac{1}{{4\pi G}} \int_{\Sigma}
 R_{\mu\nu}      n^\mu \xi^\nu dV =
 -\frac{1}{{4\pi G}} \int_{\Sigma}
 R^0_{\ 0}      \sqrt{-g} dr d\theta d\varphi
\ . \label{relmassz} \end{equation}

Defining the horizon mass of the black hole
${\cal M}_{\rm H} = {\cal M}_{\rm local}(r_{\rm H})$ 
and its horizon angular momentum
${\cal J}_{\rm H} = {\cal J}_{\rm local}(r_{\rm H})$ 
one obtains with 
$\chi = \xi + \left(\omega_{\rm H}/r_{\rm H}\right) \eta$
the horizon mass formula \cite{wald}
\begin{equation}
{\cal M}_{\rm H} = 2  {\cal T} {\cal S} 
 + 2 \frac{\omega_{\rm H}}{r_{\rm H}} {\cal J}_{\rm H}
\ , \label{relmassh} \end{equation}
where ${\cal T}$ and ${\cal S}$ denote temperature and entropy.
The global mass is then given by the relation
\begin{equation}
{\cal M} = 2  {\cal T} {\cal S}
 + 2 \frac{\omega_{\rm H}}{r_{\rm H}} {\cal J}_{\rm H}
 -\frac{1}{{4\pi G}} \int_{\Sigma}
 R^0_{\ 0}      \sqrt{-g} dr d\theta d\varphi
\ . \label{relmassy} \end{equation}

Let us next replace the integrand in (\ref{relmassy})
by first making use of the Einstein equations, 
\begin{equation}
 R^0_{\ 0} = 16 \pi G e^{2 \kappa \Phi}
 {\rm Tr} \left( F_{0\mu}F^{0\mu} - \frac{1}{4} F_{\mu\nu} F^{\mu\nu} \right)
\end{equation}
and then of the dilaton field equation,
\begin{equation}
 R^0_{\ 0} = 8 \pi G \left[ -\frac{1}{2\kappa} \frac{1}{\sqrt{-g}}
 \partial_\mu \left( \sqrt{-g} \partial^\mu \Phi \right)
  + 2  e^{2 \kappa \Phi}
  {\rm Tr} \left( F_{0\mu}F^{0\mu} \right) \right]
\ . \end{equation}

We now evaluate the integral involving the dilaton d'Alembertian,
\begin{equation}
\int_{\Sigma}
 \left[ \frac{1}{\sqrt{-g}}
 \partial_\mu \left( \sqrt{-g} \partial^\mu \Phi \right) \right]
 \sqrt{-g} dr d\theta d\varphi
= \int_{\Sigma} \left[
 \partial_r \left( \sqrt{-g} \partial^r \Phi \right)
 +\partial_\theta \left( \sqrt{-g} \partial^\theta \Phi \right) \right]
 dr d\theta d\varphi
\ , \label{relmassx} \end{equation}
with the divergence theorem.
Making use of the boundary conditions at the horizon
($\partial_r \Phi =0$) and at infinity
($\lim_{r\rightarrow \infty} \sqrt{-g} \partial^r \Phi = {\cal D}
\sin \theta$), yields for the first term in (\ref{relmassx})
\begin{equation}
 2 \pi \int_0^\pi 
 \sqrt{-g} \partial^r \Phi |_{r=r_{\rm H}}^{r=\infty} d\theta 
 = 4 \pi {\cal D}
\ , \end{equation}
with the dilaton charge ${\cal D}$.
The second term vanishes, since $ \sqrt{-g}$ vanishes at $\theta=0$
and $\theta=\pi$.
The relation for the global mass then reads
\begin{equation}
{\cal M} = 2  {\cal T} {\cal S}
 + 2 \frac{\omega_{\rm H}}{r_{\rm H}} {\cal J}_{\rm H}
 + \frac{4\pi}{\kappa} {\cal D}
 - 4 \int_\Sigma  e^{2 \kappa \Phi}  {\rm Tr} [ F_{0\mu} F^{0\mu} ]
 \sqrt{-g} dr d\theta d\varphi
\ . \label{relmassw} \end{equation}

Making again use of the Einstein equations,
we next replace the horizon angular momentum ${\cal J}_{\rm H}$
by the global angular momentum ${\cal J}$ \cite{wald},
\begin{equation}
{\cal J} = {\cal J} _{\rm H} + 
 2  \int_\Sigma  e^{2 \kappa \Phi}
 {\rm Tr}  (F_{ \varphi \mu} F^{0\mu})  \sqrt{-g} dr d\theta d\varphi
\ , \end{equation} 
and obtain
\begin{equation}
{\DS
{\cal M} - 2  {\cal T} {\cal S}
 - 2 \frac{\omega_{\rm H}}{r_{\rm H}} {\cal J}
 - \frac{4\pi}{\kappa} {\cal D}
 = {\cal I} \equiv 
 - 4 \int_\Sigma  e^{2 \kappa \Phi} {\rm Tr}
 \left[
  \left( F_{0\mu}+ \frac{\omega_{\rm H}}{r_{\rm H}}
         F_{\varphi\mu} \right) F^{0\mu} \right]
 \sqrt{-g} dr d\theta d\varphi}
\ , 
\label{intb}
\end{equation}
defining the integral ${\cal I}$.

To evaluate ${\cal I}$, Eq.~(\ref{intb}), we make use of the
symmetry relations, Eqs.~(\ref{symA}) \cite{radu2},
\begin{equation}
F_{\mu 0} = D_\mu A_0 \ , \ \ \
F_{\mu \varphi} = D_\mu \left( A_\varphi- W_\eta \right)
\ . \label{fsymA} \end{equation}
The integral then reads
\begin{equation}
 {\cal I} =
   4 \int_\Sigma  e^{2 \kappa \Phi} {\rm Tr} \left[ \left\{ D_\mu
  \left( A_0 + \frac{\omega_{\rm H}}{r_{\rm H}} \left(
         A_{\varphi} - W_\eta \right) \right) \right\} F^{0\mu} \right]
 \sqrt{-g} dr d\theta d\varphi
\ . \label{intc} \end{equation}

Adding zero to the above integral, in the form of the 
gauge field equation of motion for the zero component, 
the covariant derivative then acts on all functions in the integrand,
\begin{equation}
 {\cal I} =
  \displaystyle 4 \int_\Sigma
 {\rm Tr}  
   \left[ D_\mu \left\{ \left( 
    A_0 + \frac{\omega_{\rm H}}{r_{\rm H}} 
   \left( A_{\varphi} - W_\eta \right) 
   \right) e^{2 \kappa \Phi} F^{0\mu} \sqrt{-g} \right\} \right]  
   dr d\theta d\varphi
\ . \label{I} \end{equation}
Since the trace of a commutator vanishes,
we replace the gauge covariant derivative by the partial derivative,
\begin{equation}
 {\cal I} =
  {\displaystyle 4 \int_\Sigma
 {\rm Tr}  \left[ \partial_\mu \left\{ \left(
    A_0 + \frac{\omega_{\rm H}}{r_{\rm H}} 
    \left( A_{\varphi} - W_\eta \right) 
   \right) e^{2 \kappa \Phi} F^{0\mu} \sqrt{-g} \right\} \right]  
   dr d\theta d\varphi}
\ , \label{II} \end{equation}
and employ the divergence theorem.
The $\theta$-term vanishes,
since $\sqrt{-g}$ vanishes at $\theta=0$ and $\theta=\pi$,
and the $\varphi$-term vanishes,
since the integrands at $\varphi=0$ and $\varphi=2\pi$ coincide,
thus we are left with
\begin{equation}
{\cal I}=
 {\displaystyle 4 \int \left.
 {\rm Tr}  \left[  \left( A_0 + \frac{\omega_{\rm H}}{r_{\rm H}}
   \left( A_{\varphi} - W_\eta \right)  \right)
 e^{2 \kappa \Phi} F^{0r} \sqrt{-g} \right]  
 \right|^\infty_{r_{\rm H}} d\theta d\varphi}
\ . \label{int} \end{equation}

We next make use of the fact 
that the electrostatic potential $\Psiel|_{\rm H}$
is constant at the horizon (see Eqs.~(\ref{esp0}) and (\ref{esp})),
\begin{equation}
\left. \Psiel \right|_{\rm H} 
= \left. \left( A_0 
 + \frac{\omega_{\rm H}}{r_{\rm H}} A_{\varphi} \right) \right|_{\rm H}
= \frac{\omega_{\rm H}}{r_{\rm H}} W_\eta 
=\Psi_{\rm el} \frac{\tau_z}{2}
\ . \label{esp2} \end{equation}
Hence the integrand vanishes at the horizon,
and the only contribution to ${\cal I}$ comes from infinity.

At infinity the asymptotic expansion yields 
to lowest order
\begin{eqnarray}
F^{0r} \sqrt{-g} & = & - \frac{Q}{e} \sin \theta 
 \left( \cos \theta \frac{\tau_r^n}{2}
 - (-1)^k \sin \theta \frac{\tau_\theta^n}{2} \right) + o(1)
\ , \nonumber \\
A_0 & = & o(1)
\ , \nonumber \\
A_\varphi & = & - \frac{n}{e}(1-(-1)^k) \sin \theta \frac{\tau_\theta^n}{2} 
  + o(1)
\ . \label{A_asym} \end{eqnarray}
Hence $A_0$ does not contribute to the integral,
and $A_\varphi$ contributes for odd node number $k$. 
The integral ${\cal I}$,
\begin{equation}
{\cal I} =  \frac{nQ}{e^2}  \frac{\omega_{\rm H}}{r_{\rm H}}
  \int {\rm Tr} \left( \cos \theta \tau_r^n - (-1)^k
                       \sin \theta \tau_\theta^n \right)^2 
   \sin \theta d\theta d \varphi
  = \frac{8 \pi n Q}{e^2} \frac{\omega_{\rm H}}{r_{\rm H}}
\ , \label{IIfinal} \end{equation}
is therefore independent of $k$, and the mass formula becomes
\begin{equation}
{\cal M} - 2  {\cal T} {\cal S}
 - 2 \frac{\omega_{\rm H}}{r_{\rm H}} {\cal J}
 - \frac{4\pi}{\kappa} {\cal D} =
   {8 \pi } \Psi_{\rm el} {\cal Q}
\ . \label{Ifinal} \end{equation}

To complete the proof of the mass formula (\ref{namass})
for genuinely non-Abelian black holes,
subject to the above ansatz and boundary conditions,
we now return to dimensionless variables. Noting that
\begin{equation}
{\cal T} {\cal S} =
\frac{\sqrt{4\pi G}}{eG} TS \ ,
\end{equation}
\begin{equation}
\frac{\omega_{\rm H}}{r_{\rm H}} =
\frac{e}{\sqrt{4\pi G}}  \Omega \ , 
\end{equation}
and recalling the definition of $\psi_{\rm el}$, Eq.~(\ref{esp}),
the non-Abelian mass formula is obtained,
$$
M = 2 TS + 2 \Omega J +\frac{D}{\gamma} + 2\psi_{\rm el} Q \ .
$$
The mass formula also holds for embedded Abelian black holes
\cite{kkn,kkn-emd}.

\section{Numerical Results}

We solve the set of eleven coupled non-linear
elliptic partial differential equations numerically \cite{schoen},
subject to the above boundary conditions,
employing compactified dimensionless coordinates,
$\bar x = 1-(x_{\rm H}/x)$. 
The numerical calculations, based on the Newton-Raphson method,
are performed with help of the program FIDISOL \cite{schoen}.
The equations are discretized on a non-equidistant grid in 
$\bar x$ and  $\theta$.
Typical grids used have sizes $100 \times 20$, 
covering the integration region
$0\leq \bar x\leq 1$ and $0\leq\theta\leq\pi/2$.
(See \cite{kk,kkreg,kkrot} and \cite{schoen} 
for further details on the numerical procedure.)

For a given dilaton coupling constant $\gamma$,
the rotating non-Abelian black hole solutions 
then depend on the continuous parameters $x_{\rm H}$
and $\omega_{\rm H}$, representing
the isotropic horizon radius and the metric function
$\omega$ at the horizon, respectively,
while their ratio $\omega_{\rm H}/x_{\rm H}$ corresponds to the 
rotational velocity of the horizon.
The solutions further depend on the discrete parameters $n$ and $k$,
representing the winding number and the node number, respectively.

To construct a rotating non-Abelian black hole solution 
of SU(2) EYMD theory with dilaton coupling constant $\gamma$
and parameters $x_{\rm H}$, $\omega_{\rm H}$, $n$, and $k$,
we either start from the static SU(2) EYMD black hole solution 
possessing the same set of parameters except for $\omega_{\rm H}=0$, 
and then increase $\omega_{\rm H}$ \cite{kk},
or we start from the corresponding rotating SU(2) EYM black hole solution
and then increase $\gamma$ \cite{kkrot}.

\boldmath
\subsection{$n=1$ Black Holes}
\unboldmath
 
Rotating non-Abelian black hole solutions with winding number
$n=1$ have been studied before in EYM theory \cite{kkrot}.
Here we study the influence of the presence of the dilaton
on the properties of these black hole solutions,
with particular emphasis on the $Q=0$ black hole solutions.
\\
\\
\noindent {\sl \bf Global charges}

Many features of the $n=1$ EYMD black hole solutions agree with those
of the $n=1$ EYM solutions, studied before \cite{kkrot}. In particular,
when increasing $\omega_{\rm H}$ from zero, while keeping $x_{\rm H}$ fixed,
a lower branch of black hole solutions forms
and extends up to a maximal value $\omega^{\rm max}_{\rm H}(x_{\rm H},\gamma)$,
where an upper branch bends backwards towards $\omega_{\rm H}=0$.
This is seen in Figs.~1a-d, where the mass $M$,
the angular momentum per unit mass $a=J/M$,
the relative dilaton charge $D/\gamma$,
and the non-Abelian electric charge $Q$
are shown as functions of the parameter $\omega_{\rm H}$
for black holes with horizon radius $x_{\rm H}=0.1$, 
winding number $n=1$, and node number $k=1$
for five values of the dilaton coupling constant,
$\gamma=0$,$0.5$, $1$, $\sqrt{3}$ and $3$.
For $\gamma=0$ the relative dilaton charge $D/\gamma$
is extracted from the mass formula Eq.~(\ref{namass}).

The mass $M$ (Fig.~1a) and the angular momentum per unit mass $a$
(Fig.~1b)
of the non-Abelian solutions increase monotonically along both branches,
and diverge with $\omega_{\rm H}^{-1}$ 
in the limit $\omega_{\rm H} \rightarrow 0$ along the upper branch.
Both mass and angular momentum become (almost) independent of
the dilaton coupling constant along the upper branch.
The relative dilaton charge $D/\gamma$  (Fig.~1c) decreases monotonically 
along 
both branches, approaching zero in the limit $\omega_{\rm H} \rightarrow 0$.
(The curves are discontinued along the upper branch 
when the numerical procedure no longer provides high accuracy.)
Also the dilaton charge approaches its limiting value 
along the upper branch with a slope, (almost) independent of 
the dilaton coupling constant.

The non-Abelian electric charge $Q$ (Fig.~1d) increases monotonically 
along both branches, when $\gamma < 1.15$,
whereas when $\gamma > 1.15$ (and $x_{\rm H}$ sufficiently small), it first decreases
on the lower branch until it reaches a minimum and then increases along
both branches. 
The magnitude of $Q$ remains always small, however.
For $\gamma=0$, $Q$ approaches the finite limiting value
$Q_{\rm lim} \approx 0.124$ on the upper branch,
as $\omega_{\rm H}$ tends to zero,
independent of the isotropic horizon radius $x_{\rm H}$ \cite{kkrot}.
Apparently, this remains true when the dilaton is coupled.

For comparison, we exhibit in Figs.~1a-c also the mass, 
the specific angular momentum, and the relative dilaton charge of embedded Abelian solutions
with the same horizon radius (dotted curves) with $Q=0$ and $P=1$.
(Mass, angular momentum and dilaton charge of embedded Abelian solutions,
possessing the same charge $Q$ as the non-Abelian solutions and $P=1$
are graphically indistinguishable from mass, angular momentum
and dilaton charge of the $Q=0$ solutions, shown.
Since the Abelian $\gamma=0$ and $\gamma=\sqrt{3}$ solutions are known
analytically, their global charges
along the upper branch are shown up to $\omega_{\rm H}=0$.)
As expected \cite{kkrot},
mass, angular momentum per unit mass, and relative dilaton charge of the non-Abelian EYMD solutions,
are close to mass, specific angular momentum, and dilaton charge of the embedded
EMD solutions with $Q=0$ and $P=1$.
This is true, in particular,
for solutions with small horizon radii $x_{\rm H}$,
where large deviations of these global properties from 
those of the corresponding Kerr solutions arise \cite{kkrot}.
The dilaton charge of the Abelian solutions, however, approach
their limiting value $D/\gamma=0$ on the upper branch
with a common slope, different from the common slope of the non-Abelian
solutions.

The $\gamma$-dependence of the global charges is demonstrated
further in Figs.~2a-b. Here we exhibit the mass $M$,
the angular momentum per unit mass $a=J/M$,
the relative dilaton charge $D/\gamma$,
and the non-Abelian electric charge $Q$
for non-Abelian black hole solutions with
winding number $n=1$, node number $k=1$, horizon radius $x_{\rm H}=0.1$, 
and $\omega_{\rm H}=0.020$ on the lower branch (Fig.~2a)
and on the upper branch (Fig.~2b)
as functions of $\gamma$ for $0\leq \gamma \leq 8$.
On the lower branch $M$, $a=J/M$, $D/\gamma$, and $Q$ decrease with 
increasing $\gamma$. 
In particular, we note that the non-Abelian electric charge $Q$ passes
zero at $\gamma \approx 1.5$.
On the upper branch only the relative dilaton charge $D/\gamma$ decreases
with increasing $\gamma$, whereas the mass $M$ and the
angular momentum per unit mass $a$ increase with increasing $\gamma$.
The  non-Abelian electric charge $Q$ exhibits a minimum on the upper
branch.

It is remarkable that the non-Abelian electric charge $Q$
can change sign in the presence of the dilaton field \cite{kkn}.
In contrast, in EYM theory the non-Abelian electric charge $Q$
always has the same sign, depending on the direction of rotation
\cite{kkrot,foot3}.

Thus we observe the new feature of rotating EYMD black holes that,
for certain values of the dilaton coupling constant $\gamma$
and the parameters $x_{\rm H}$ and $\omega_{\rm H}$,
the non-Abelian electric charge $Q$ of rotating black hole solutions
with $n=1$ and $k=1$ can vanish.
Cuts through the parameter space of solutions with vanishing $Q$
are exhibited in Fig.~3a, where we show $\omega_{\rm H}$ 
as a function of $x_{\rm H}$ for $Q=0$ solutions 
with dilaton coupling constants $\gamma=1.5$, $\sqrt{3}$, and $2$.
For fixed $\gamma$ these curves
divide the parameter space into a region 
with solutions possessing negative $Q$ and 
a region with solutions with positive $Q$.
Negative $Q$ solutions correspond to parameter values below
the curves.

The maximum of a curve coincides with 
$\omega^{\rm max}_{\rm H}(x_{\rm H},\gamma)$.
We note that, for fixed $\gamma$,
the values of $\omega_{\rm H}$ to the right of the 
maximum correspond to solutions on the lower branches whereas the values to
the left correspond to solutions on the upper branches. 
In the limit $\gamma \rightarrow \gamma_{\rm min} \approx 1.15$ the 
curves degenerate to a point, implying that rotating $Q=0$ solutions 
exist only above $\gamma_{\rm min}$.
This is demonstrated in Fig.~3b,
where the dependence of the maximal value of $\omega_{\rm H}$
together with its location  $x_{\rm H}$ on $\gamma$ are shown,
clearly exhibiting the minimal value of $\gamma$ for rotating $Q=0$ solutions.

These $Q=0$ EYMD black holes represent
the first black hole solutions, which carry
non-trivial non-Abelian electric and magnetic fields
and no non-Abelian charge.
Perturbative studies \cite{pert2} previously suggested 
the existence of rotating non-Abelian $Q=0$ black hole solutions
in EYM theory, satisfying, however, a different set of boundary conditions.
Such EYM black hole solutions could not be obtained numerically 
\cite{radu2,kkrot}.

As a consequence of their vanishing non-Abelian electric charge,
these black hole solutions do not exhibit the
generic asymptotic non-integer power fall-off of the
non-Abelian gauge field functions \cite{kkrot}.
For these solutions, the magnetic moment can be identified
according to Eq.~(\ref{muc2c3}), where $\mu_x \sim C_3$
and $\mu_z \sim -C_2$. The coefficient $-C_2=|C_2|$ is exhibited in Fig.~3c 
for the sets of solutions shown in Fig.~3a.
Note that the right end points of these curves correspond to the
static limit.

The corresponding curves for $C_3$ are graphically almost indistinguishable. 
The difference between $-C_2$ and $C_3$ is due to the rotation,
since in the static case $C_2+C_3=0$. 
In Fig.~3d we show this difference for the same set of solutions.

Having considered the rotating hairy black hole solutions
for fixed horizon radius $x_{\rm H}$ as a function of $\omega_{\rm H}$,
we now keep $\omega_{\rm H}$ fixed and vary the horizon radius.
In Figs.~4a-d we show the mass $M$, 
the angular momentum per unit mass $a=J/M$,
the relative dilaton charge $D/\gamma$,
and the non-Abelian electric charge $Q$
of the non-Abelian black hole solutions
as functions of the isotropic horizon radius $x_{\rm H}$
for $\omega_{\rm H}=0.040$ and $\gamma=0$, $0.5$, $1$, $\sqrt{3}$ and $3$.
(Solutions along the steeper branch are numerically difficult
to obtain, therefore the $D/\gamma$ and $Q$ curves are discontinued
at $x_{\rm H} \approx 0.43$.)
For EYMD black hole solutions, as for EYM black hole solutions,
there is a minimal value of the horizon radius 
$x_{\rm H}^{\rm min}(\omega_{\rm H})$,
for a given value of $\omega_{\rm H}$.
In particular, the limit $x_{\rm H}^{\rm min} \rightarrow 0$
is only reached for $\omega_{\rm H} \rightarrow 0$.
We note that for KK black holes with $Q=0$ and $P=1$ the family of 
solutions characterised by $[\omega_{\rm H}$, 
$x_{\rm H}^{\rm min}(\omega_{\rm H})]$
tends to the static `extremal' solution as $x_{\rm H} \to 0$.
Anticipating that the non-Abelian black holes
behave essentially analogously to the Abelian solutions,
we expect to find the static regular EYMD solutions in this limit
\cite{kkrot}.
We have no indication that rotating regular solutions could be 
reached in this limit \cite{pert2,foot1}.
\\
\\

\noindent {\sl \bf Horizon Properties}

Let us next turn to the horizon properties of the
rotating non-Abelian black holes.
In Fig.~5a-d we show the area parameter $x_\Delta$,
the surface gravity $\kappa_{\rm sg}$,
the deformation of the horizon as quantified by the ratio
of equatorial to polar circumferences, $L_{\rm e}/L_{\rm p}$,
and the Gaussian curvature at the poles $K(0)$
as functions of $\omega_{\rm H}$, 
for black hole solutions with horizon radius $x_{\rm H} =0.1$, 
winding number $n=1$ and node number $k=1$,
and for dilaton coupling constants
$\gamma=0$, 0.5, 1, $\sqrt{3}$ and 3.
Also shown are the horizon properties of the corresponding
embedded Abelian black hole solutions with $Q=0$ and $P=1$.

The horizon size is quantified by the area parameter $x_\Delta$,
shown in Fig.~5a.
Starting from the value of the corresponding 
static non-Abelian black hole solution,
the area parameter grows monotonically along both branches,
and diverges along the upper branch.
The horizon size of the corresponding Abelian black hole solutions
is slightly larger, in particular along the lower branch.
As illustrated, the difference decreases with increasing $\gamma$.
The difference also decreases with increasing horizon radius $x_{\rm H}$.

The surface gravity $\kappa_{\rm sg}$ of the non-Abelian black holes
is shown in Fig.~5b.
Starting from the value of the corresponding 
static non-Abelian black hole solution
on the lower branch in the limit $\omega_{\rm H} \rightarrow 0$,
it decreases monotonically along both branches,
and reaches zero on the upper branch
in the limit $\omega_{\rm H} \rightarrow 0$, 
corresponding to the value assumed by extremal black hole solutions.
The surface gravity of the corresponding Abelian black hole solutions
is slightly smaller, in particular along the lower branch.
The difference decreases with 
increasing horizon radius $x_{\rm H}$.
We recall, that for static non-Abelian black holes
the surface gravity diverges in the limit $x_{\rm H}\rightarrow 0$
independent of $\gamma$ \cite{kks},
whereas for static Abelian black holes
the limiting value depends on $\gamma$,
and diverges only for $\gamma > 1$, whereas it is finite for $\gamma = 1$,
and zero for $\gamma < 1$.

The deformation of the horizon is revealed when measuring
the circumference of the horizon along the equator, $L_{\rm e}$,
and the circumference of the horizon along 
a great circle passing through the poles, $L_{\rm p}$,
Eqs.~(\ref{lelp}).
The ratio $L_{\rm e}/L_{\rm p}$ grows monotonically along both branches,
as shown in Fig.~5c.
As $\omega_{\rm H}$ tends to zero on the upper branch, the ratio
tends to the value  
$L_{\rm e}/L_{\rm p} = \pi/(\sqrt{2} E(1/2)) \approx 1.645$,
for both non-Abelian and Abelian black holes, independent of $\gamma$.
Clearly, the matter fields lose their influence on the black hole properties,
in the limit $M \to \infty$.
On the lower branch the ratio $L_{\rm e}/L_{\rm p}$ assumes the value one
in the limit $\omega_{\rm H} \to 0$,
corresponding to the value of
a static spherically symmetric black hole.
Overall, the deformation of the horizon of the non-Abelian black holes
is close to the deformation of the horizon of the corresponding EMD
black holes with $Q=0$ and $P=1$.

For rapidly rotating EYMD and EMD black holes, the Gaussian curvature of the horizon 
can become negative, as observed before for EM black holes \cite{smarr}.
In Fig.~5d we show the Gaussian curvature of the horizon at the pole $K(0)$
as a function of $\omega_{\rm H}$ for non-Abelian and Abelian black holes. 
Starting with a positive curvature at the pole of the static solution, 
the curvature at the pole decreases with increasing angular momentum,
crosses zero and becomes negative at some point along the upper branch.
It then reaches a minimum value and starts to increase again,
to become zero in the infinite mass limit,
when $\omega_{\rm H} \rightarrow 0$.

In Figs.~6 we consider the deformation of the horizon of the non-Abelian
black holes in more detail.
We exhibit in  Fig.~6a the angular dependence of the Gaussian curvature 
at the horizon for $\gamma=1$ black hole solutions with 
horizon radius $x_{\rm H}=0.1$, winding number $n=1$, node number $k=1$, 
for several values of the angular momentum per unit mass $a$. 
When $a>0$, the Gaussian curvature increases monotonically from the pole 
to the equator. 
For larger values of $a$, the Gaussian curvature is negative in the vicinity
of the pole.

Fig.~6b shows the shape of the horizon of the non-Abelian black holes,
obtained from isometric embeddings of the horizon in Euclidean space
\cite{smarr}.
As pointed out in \cite{smarr}, 
when the Gaussian curvature becomes negative,
the embedding cannot be performed completely in Euclidean space
with metric $ds^2=dx^2+dy^2+dz^2$,
but the region with negative curvature
must be embedded in pseudo-Euclidean space 
with metric $ds^2=dx^2+dy^2-dz^2$,
represented by dashed lines in the figure.
\\
\\

\noindent {\sl \bf Limiting Abelian Solutions}

As seen above, the global charges and the horizon properties
of the non-Abelian black hole solutions with $n=1$ and $k=1$
are rather close to those of the embedded EMD solutions
with $Q=0$ and $P=1$.
But also their metric and dilaton functions as well as 
their gauge field functions are impressively close to their
EMD counterparts,
except for those gauge field functions
which do not vanish in the static limit.

With increasing node number $k$, the non-Abelian solutions
get increasingly closer to the Abelian solutions, 
converging pointwise in the limit $k \rightarrow \infty$.
In particular, the gauge field functions
$H_2$ and $H_4$ approach zero 
with increasing node number $k$ in an increasing interval,
while their boundary conditions $H_2(\infty)=H_4(\infty)=(-1)^k$
enforce a value, differing from the value of the limiting Abelian solutions,
$H_2^{\cal A}(\infty)=H_4^{\cal A}(\infty)=0$.

Convergence of the solutions with increasing node number $k$ 
towards the corresponding EMD solutions is demonstrated in Figs.~7a-e
for the metric function $f$, the dilaton function $\phi$,
the gauge field functions $H_2$, $H_3$ and for the function
${x\bar{B}_1}$
for non-Abelian black hole solutions with $x_{\rm H}=0.1$,
$\omega_{\rm H}=0.020$, $n=1$, $k=1-3$,
and dilaton coupling constants $\gamma=0$ and 1.
From Fig.~7e we conclude that the limiting 
solution has vanishing electric charge.

\boldmath
\subsection{$n>1$ Black Holes}
\unboldmath

Here we give for the first time a detailed account
of rotating $n>1$ black hole solutions \cite{kkn}.
Whereas the rotating $n=1$ black hole solutions reduce to
spherically symmetric black hole solutions in the static limit,
the rotating $n>1$ black hole solutions remain
axially symmetric in the static limit \cite{kk,kkn}.
\\
\\
\noindent {\sl \bf Global charges}

Many features of the $n>1$ EYMD black hole solutions,
agree with those of the $n=1$ EYMD solutions. 
In Figs.~8a-d we exhibit the mass $M$,
the angular momentum per unit mass $a=J/M$,
the relative dilaton charge $D/\gamma$,
and the non-Abelian electric charge $Q$
as functions of the parameter $\omega_{\rm H}$
for black holes with horizon radius $x_{\rm H}=0.1$, 
winding numbers $n=1-3$, and node number $k=1$
for the dilaton coupling constants $\gamma=0$, $1$, and $\sqrt{3}$.
Again, for $\gamma=0$ the relative dilaton charge $D/\gamma$
is extracted from the mass formula Eq.~(\ref{namass}), discussed below.

The mass $M$ shown in Fig.~8a 
and the angular momentum per unit mass $a$ shown in Fig.~8b,
increase monotonically and diverge on the upper branch in the limit  
$\omega_{\rm H} \rightarrow 0$, as in the $n=1$ case.
Whereas the mass on the upper branch becomes (almost) independent
both of the coupling constant $\gamma$ and the winding number $n$
already for moderately large values for $\omega_{\rm H}$,
the specific angular momentum becomes independent only of 
the coupling constant $\gamma$, but retains its dependence
on the winding number $n$ up to considerably smaller values
of  $\omega_{\rm H}$.

The relative dilaton charge shown in Fig.~8c,
apparently approaches zero in the limit
$\omega_{\rm H} \rightarrow 0$, 
as in the case of the $n=1$ black hole solutions. 
We note, however, that for winding number $n=3$, 
the dilaton curve developes a minimum with a negative value of $D$, 
and only then bends upwards towards zero. 
The non-Abelian electric charge exhibited in Fig.~8d, increases with 
increasing $n$. 
As in the $n=1$ case, $Q$ apparently reaches a limiting value
on the upper branch as $\omega_{\rm H} \rightarrow 0$. 
This limiting value increases with increasing winding number $n$
and appears to be independent of the coupling constant $\gamma$.
(For small values of $\omega_{\rm H}$ on the upper branch,
the numerical error becomes too large
to extract the electric charge and thus the limiting values of $Q$.)
For $n=3$, the electric charge developes a maximum on the upper branch 
before tending to the limiting value.

Comparing the global charges of these non-Abelian black hole solutions
with those of the corresponding $Q=0$ and $P=n$ Abelian black hole solutions,
we observe
that the discrepancies between Abelian and non-Abelian solutions 
become larger with increasing winding number $n$.
As in the $n=1$ case, the discrepancies decrease 
with increasing dilaton coupling constant $\gamma$,
and with increasing horizon radius $x_{\rm H}$.

Trying to find $n>1$ black holes 
with vanishing non-Abelian electric charge,
we constructed solutions for a wide range of parameters.
However, in contrast to the $n=1$ case,
we did not observe a change of sign of $Q$ for any set of solutions
constructed.
\\
\\
\noindent {\sl \bf Horizon Properties}

Turning next to the horizon properties of the
rotating non-Abelian $n>1$ black holes,
we exhibit in Figs.~9a-d the area parameter $x_\Delta$,
the surface gravity $\kappa_{\rm sg}$,
the deformation of the horizon as quantified by the ratio
of equatorial to polar circumferences, $L_{\rm e}/L_{\rm p}$,
and the Gaussian curvature at the poles $K(0)$
as functions of $\omega_{\rm H}$, 
for black hole solutions with $x_{\rm H} =0.1$, $n=1-3$ and $k=1$,
and for dilaton coupling constants
$\gamma=0$, $1$,and $\sqrt{3}$. 

As shown in Fig.~9a,  
the horizon size is monotonically increasing along both branches,
and diverges as $\omega_{\rm H} \rightarrow 0$ on the upper branch. 
Along the upper upper branch, the horizon size shows
a distinct dependence on the winding number $n$,
but little dependence on the coupling constant $\gamma$.
The $\gamma$ dependence increases though with increasing $n$.
The surface gravity decreases monotonically along both branches, 
and reaches a vanishing value on the upper branch 
as $\omega_{\rm H} \rightarrow 0$, as demonstrated in Fig.~9b.  
For black hole solutions with the same value of $\gamma$ but different
values of $n$, we observe crossings of the lower branches.
This is in contrast to the Abelian black hole solutions, where
analogous crossings do not occur. 
It is thus a distinct feature of
the non-Abelian nature of the black holes.

The deformation parameter $L_{\rm e}/L_{\rm p}$ is shown in Fig.~9c.
For small values of $\omega_{\rm H}$ along the upper branch,
we observe, that the curves become (rather) independent of $\gamma$,
but their slope keeps a distinct $n$-dependence.
All curves appear to have the same limiting value though,
when $\omega_{\rm H} \rightarrow 0$, independent of 
$\gamma$ and $n$.
(Unfortunately, the numerical error for small $\omega_{\rm H}$ along the 
upper branches is too large to extract the limiting value as 
$\omega_{\rm H}\to 0$.) 
We expect this common limiting value to be 
$L_{\rm e}/L_{\rm p}=1.645$,
which is the value of the extremal Kerr black holes, Kerr-Newman
black holes and Kaluza-Klein black holes.

The Gaussian curvature of the horizon at the poles,
exhibited in Fig.~9d, 
possesses negative values for fast rotating black holes,
also for $n>1$ black holes.
The topology of the horizon of these $n>1$ black holes is also 
that of a 2-sphere, Eq.~(\ref{topolint}).

As expected, these properties of the non-Abelian black hole solutions 
are rather similar to those of their Abelian counterparts, 
though the differences become more pronounced as $n$ increases.
\\
\\
\noindent {\sl \bf Limiting Abelian Solutions}

Considering the limit of node number $k\to \infty$ 
and fixed winding number $n$ as well as fixed $\gamma$,
we observe that the black hole solutions with winding number $n$ 
tend to the corresponding Abelian black hole solutions 
with vanishing electric charge $Q=0$ and with magnetic charge $P=n$.
The convergence is observed for the global charges,
and for the horizon properties,
and pointwise convergence is seen for the
metric, dilaton and gauge field functions.

\subsection{Mass Formula and Uniqueness}

The numerical stationary axially symmetric EYMD black hole solutions
satisfy the mass formula, Eq.~(\ref{namass}), with an accuracy of $10^{-3}$.
So do the numerical EMD black hole solutions.
EYM black holes are included in the limit $\gamma \rightarrow 0$,
since $D/\gamma$ remains finite.

We demonstrate the mass formula for the non-Abelian black hole
solutions in Figs.~10a-c, where we show the four terms in the mass formula,
$2TS$, $2\Omega J$, $D/\gamma$, and $2\psi_{\rm el}Q$,
as well as their sum (denoted by $M$ in the figures) as functions of the mass $M$.
Black hole solutions with horizon radius $x_{\rm H}=0.1$, 
winding number $n=1$, and node number $k=1$,
and dilaton coupling constant $\gamma=1$
are shown in Fig.~10a, $n=1$ and $k=2$ solutions in Fig.~10b,
and $n=2$ and $k=1$ solutions in Fig.~10c.

Reconsidering the uniqueness conjecture for EYMD black holes,
one may try to obtain a new uniqueness conjecture
by only replacing the magnetic charge $P$ 
by the dilaton charge $D$ or the relative dilaton charge $D/\gamma$.
After all, such a replacement gave rise to the new non-Abelian mass
formula, Eq.~(\ref{namass}).
However, in the case of the uniqueness conjecture, such a 
replacement is not sufficient. Indeed,
black holes in SU(2) EYMD theory are not uniquely characterized by
their mass $M$, their angular momentum $J$,
their non-Abelian electric charge $Q$, and their dilaton charge $D$.
This becomes evident already in the static case, 
where genuine EYMD black holes have $J=Q=0$, and thus their
only charges are the mass $M$ and the dilaton charge $D$.
In Fig.~11 we show the relative dilaton charge $D/\gamma$
as a function of the mass $M$,
for the static non-Abelian black hole solutions
with $n=1-3$ and $k=1-3$ for $\gamma=1$.
Also shown are the corresponding limiting Abelian solutions with $P=n$.
As seen in the figures, solutions with the same winding number $n$ 
and different node number $k$ do not intersect.
However, solutions with different winding numbers $n$ can intersect.
The curve of $n=3$, $k=1$ solutions, for instance, intersects
the curves of all $n=2$, $k>2$ solutions.

To remedy this deficiency, we make use of
the topological charge $N$, Eq.~(\ref{topN}).
Recalling that for the non-Abelian black holes
$N=n$, Eq.~(\ref{Nn}),
and setting $N=0$ for embedded Abelian black holes,
we can attach to each black hole solution a topological charge.
We then find, that all static SU(2) EYMD black hole solutions are 
uniquely determined by their mass $M$, 
their dilaton charge $D$ (or the relative dilaton charge $D/\gamma$),
and their topological charge $N$.
We note, that for a given winding number $N=n$,
the dilaton charge $D$ (or the relative dilaton charge $D/\gamma$)
of the non-Abelian solutions increases monotonically with increasing $k$, 
converging to the dilaton charge of the embedded Abelian solution with
magnetic charge $P=n$.

This observation has led us to suggest a new uniqueness conjecture \cite{kkn},
stating that
{\sl black holes in SU(2) EYMD theory
are uniquely determined by their mass $M$, their angular momentum $J$,
their non-Abelian electric charge $Q$, their dilaton charge $D$,
and their topological charge $N$}.
Since the relative dilaton charge $D/\gamma$ remains finite in the
limit $\gamma \rightarrow 0$, this conjecture may formally be
extended to the EYM case, by replacing the dilaton charge $D$
by the relative dilaton charge $D/\gamma$.

The uniqueness conjecture is supported by all our numerical data. 
To illustrate the validity of the conjecture also for stationary black holes,
we show in Figs.~12a-c the relative dilaton charge $D/\gamma$
for several values of the specific angular momentum $a$
for black holes with $n=1-3$, $k=1-3$ ($\gamma=1$).
Also shown are the corresponding limiting Abelian solutions with $P=n$.
For instance, we observe in Fig.~12a that the 
curve with $n=1$, $k=2$, and $a=0.25$ crosses 
the other $k=2$ curves and also the $k=3$ curves
with $a=0$ and $a=0.10$. 
However, since it does not cross the $k=3$ curve with the same
specific angular momentum $a=0.25$, the conjecture is not violated.

\section{Conclusions}

We have constructed non-perturbative rotating hairy black hole solutions 
of Einstein-Yang-Mills(-dilaton) theory and investigated their properties.
The rotating black hole solutions emerge from the corresponding
static hairy black hole solutions, when a small horizon angular
velocity is imposed via the boundary conditions.
The hairy black hole solutions are labeled by the node number $k$
and the winding number $n$ of their gauge field functions.
We have extended our previous studies \cite{kkrot},
by considering black holes with higher winding number $n$, 
and by including a dilaton field with coupling constant $\gamma$.
We have also provided a detailed analysis of the asymptotic expansion 
of black hole solutions with $n=1$ and $n=3$. 

The generic hairy black hole solutions possess a small non-Abelian 
electric charge, but vanishing magnetic charge.
For fixed winding number $n$, the non-Abelian black hole solutions 
form sequences, labeled by the node number $k$. 
In the limit $k \rightarrow \infty$,
they tend to embedded Abelian black hole solutions 
with magnetic charge $P=n$ and electric charge $Q=0$.
For $\gamma=0$ the limiting solutions are Kerr-Newman solutions,
for $\gamma\ne0$ they are rotating EMD solutions 
\cite{FZB,Rasheed,kkn-emd}.
A comprehensive study of the black hole solutions in EMD theory for 
general dilaton coupling constant will be presented elsewhere
\cite{kkn-emd}.

The presence of the dilaton field has a number of significant
consequences for the hairy black hole solutions.
First, the presence of the dilaton field allows for
solutions with vanishing electric charge \cite{kkn}.
Thus there are rotating black hole solutions with no non-Abelian charge at all.
These $Q=0$ solutions exist in a certain range of parameters
for black holes with winding number $n=1$, and node number $k=1$.
The asymptotic expansion of these $Q=0$ solutions 
involves only integer powers, in constrast to the generic
non-integer power fall-off for solutions with $Q \neq 0$.
We did not find corresponding $Q=0$
solutions with $n>1$ or $k>1$, when searching for them
in a wide range of parameters.

Second, the presence of the dilaton is crucial in deriving the
mass formula \cite{kkn} for rotating black holes, involving only
global charges and horizon properties of the black holes.
Here the derivation of the mass formula was presented in detail.
The mass formula is similar to the Smarr formula for Abelian
black holes, but instead of the magnetic charge $P$
the dilaton charge $D$ enters. This is crucial, since
the hairy black holes have $P=0$, but non-vanishing magnetic fields
and thus a magnetic contribution to the mass, which is taken into
account by the dilaton charge term.
The mass formula also holds for EMD black holes.
Involving the relative dilaton charge $D/\gamma$,
the mass formula may formally be extended to the EYM and EM case,
since the relative dilaton charge $D/\gamma$ remains finite in the
limit $\gamma \rightarrow 0$.

Third, the presence of the dilaton allows for a new uniqueness
conjecture for hairy black holes \cite{kkn}.
Including the horizon topological charge $N=n$,
our non-perturbative set of solutions supports the conjecture, that
{\sl black holes in SU(2) EYMD theory
are uniquely determined by their mass $M$, their angular momentum $J$,
their non-Abelian electric charge $Q$, their dilaton charge $D$,
and their topological charge $N$}.
Again, since the relative dilaton charge $D/\gamma$ remains finite in the
limit $\gamma \rightarrow 0$, this conjecture may formally be
extended to the EYM case, by replacing the dilaton charge $D$
by the relative dilaton charge $D/\gamma$.

We have not yet fully addressed a number of further
issues for these rotating black holes.
For instance, in the numerical part of the paper 
we have restricted the calculations to small values of the 
dilaton coupling constant $\gamma$. 
For larger  values of $\gamma$, new phenomena might arise.
We have neither considered in any detail the existence
of stationary rotating `extremal' EYMD solutions \cite{kkrot}.
Both these issues deserve further investigation.

Most importantly, we have not addressed the existence of
rotating regular solutions in EYMD theory.
While perturbative studies indicated the existence 
of slowly rotating regular solutions \cite{pert2},
non-perturbative numerical studies did not support their
existence \cite{radu2,kkrot}.
Independent of the (non-)existence of slowly rotating solutions,
there might however exist rapidly rotating branches 
of regular as well as black holes solutions,
not connected to the static solutions.
The investigation of such rapidly rotating solutions
remains a major challenge.

\begin{acknowledgments}
FNL has been supported in part by a FPI Predoctoral Scholarship from
Ministerio de Educaci\'on (Spain) and by DGICYT Project PB98-0772.
\end{acknowledgments}

\appendix

\section{Expansion at Infinity}

The asymptotic expansion depends on two integers, labelling
the black hole solutions, the winding number $n$ and the node
number $k$. We here explicitly show the dependence on $k$,
by choosing the boundary conditions according to Eq.~(\ref{bc1c}).
Note, however, that all solutions can be brought to satisfy the
same asymptotic boundary conditions by a gauge transformation
(see the discussion in section {II.C}).
Here we restrict to winding number $n=1$ and $n=3$.
The analysis for $n=2$ seems to be `prohibitively complicated'
\cite{radu2} because already in the static case the leading order 
terms involve non-analytic terms $\sim \log(x)/x^2$ \cite{addendum}.
\\
\\
{\sl $n=1$ Solutions}

Generalizing our previous expansion for $n=1$ \cite{kkrot}
including the dilaton, we obtain
\begin{eqnarray}
f&=&1-\frac{2 M}{x} + \frac{2 M^2 + Q^2}{x^2}
+  o \left(\frac{1}{x^2}\right)\ , \nonumber\\
m&=&1 + \frac{C_1}{x^2} + \frac{Q^2-M^2-D^2-2 C_1}{x^2}\sin^2{\theta}
+ o\left( \frac{1}{x^2}\right)\ , \nonumber\\
l&=&1+\frac{C_1}{x^2} + o\left(\frac{1}{x^2}\right)\ , \nonumber \\
\omega&=&\frac{2 a M}{x^2} - \frac{6 a  M^2 + C_2 Q}{x^3}
+ o\left(\frac{1}{x^3}\right)\ , \nonumber\\
\phi &=& -\frac{D}{x}-\frac{\gamma Q^2}{2 x^2} + \frac{C_1 D - 4 C_7 + 2\gamma (M - \gamma D) Q^2}{6 x^3} + \frac{C_7}{x^3} \sin^2{\theta} + 
 o\left(\frac{1}{x^3}\right)
\ , \nonumber\\
{\bar B}_1 &=& \frac{Q \cos{\theta}}{x} -\frac{(M-\gamma D) Q \cos{\theta}}{x^2}
+ \bigg[-\frac{2 (-1)^k C_4 (\beta-5)}{ (\beta-1)Q}x^{-\frac{1}{2} (\beta+1)}\nonumber
\\&&+\frac{4 (-1)^k C_2 C_3 Q}{(\alpha-3) (\alpha+5)} x^{-\frac{1}{2}(\alpha+3)} 
+\frac{C^2_3 (\alpha^2-\alpha-8) Q}
{\alpha (\alpha-1) (\alpha+2) (\alpha-3)} x^{-\alpha}\nonumber\\
&&+\frac{2 (1+\gamma^2)Q^3-C_1 Q+4 (M-\gamma D)^2 Q - 4 (-1)^k C_5 Q -2 C_6}{6 x^3}\bigg]
 \cos{\theta} \nonumber\\
&&+ \bigg[\frac{C_6}{x^3} + \frac{C^2_3 Q}{(\alpha+2)(\alpha-3)}x^{-\alpha}
- \frac{8 (-1)^k C_4 Q}{(\beta-1)(\beta+5)} x^{-\frac{1}{2}(\beta+1)}\nonumber \\
&& - \frac{4 (-1)^k C_2 C_3 Q}{(\alpha-3)(\alpha+5)}
 x^{-\frac{1}{2}(\alpha+3)} \bigg] \cos^3{\theta}
+ o\left(\frac{1}{x^3}\right)\ , \nonumber \\
{\bar B}_2 
&=& -(-1)^k \frac{Q \sin{\theta}}{x} 
+ (-1)^k \frac{(M-\gamma D) Q \sin{\theta}}{x^2}
-(-1)^k \bigg[\frac{C^2_3  (\alpha^2-\alpha- 8)Q}{\alpha (\alpha-1)(\alpha+2)(\alpha-3)} x^{-\alpha}\nonumber\\
&&+\frac{2 (1+\gamma^2)Q^3-C_1 Q + 4 (M-\gamma D)^2Q + 2 (-1)^k C_5 Q -2 C_6}{6 x^3}\nonumber\\ 
&&+\frac{2 ((-1)^k -1) a M}{ x^3}\bigg] \sin{\theta}
+ \bigg[-(-1)^k \frac{C_6}{x^3}
-\frac{(-1)^k C^2_3 Q}{(\alpha+2)(\alpha-3)}x^{-\alpha}\nonumber\\
&&+\frac{4 C_2 C_3 Q}{(\alpha-3)(\alpha+5)} x^{-\frac{1}{2}(\alpha+3)}
-\frac{2 C_4 (\beta-5)}{(\beta-1)Q} x^{-\frac{1}{2}(\beta+1)}\bigg]
\sin{\theta} \cos^2{\theta} + o\left(\frac{1}{x^3}\right),\nonumber\\
H_1&=&\left[ \frac{2 C_5}{x^2}
+ \frac{8 C_4}{\beta-1} x^{-\frac{1}{2} (\beta-1)}
-\frac{2 C_2 C_3 (\alpha+3)}{(\alpha+5)Q^2}
 x^{-\frac{1}{2}(\alpha+1)}\right] \sin{\theta} \cos{\theta}
+ o\left(\frac{1}{x^2}\right)\ ,\nonumber\\
H_2&=&(-1)^k+C_3 x^{-\frac{1}{2}(\alpha-1)} + \bigg[\frac{C_5}{x^2}
+ C_4 x^{-\frac{1}{2}(\beta-1)}
+ \frac{C_3 (\alpha^2+2 \alpha -11)}{2 (\alpha+1)} \nonumber\\
&&\bigg( M+\gamma D- \frac{2 C_2 (\alpha+3)}{ (\alpha+5)Q^2} \bigg)
 x^{-\frac{1}{2} (\alpha+1)} \bigg]
+ \bigg[-\frac{2 C_5}{x^2} - 2 C_4 x^{-\frac{1}{2}(\beta-1)} \nonumber\\
&&+ \frac{C_2 C_3 (\alpha+1) (\alpha+3)}{2 (\alpha+5) Q^2}
 x^{-\frac{1}{2}(\alpha+1)}\bigg] \sin^2{\theta}
+o\bigg(\frac{1}{x^2}\bigg)\ ,\nonumber\\
H_3&=&\bigg(\frac{C_2}{x} - (-1)^k C_3 x^{-\frac{1}{2}(\alpha-1)} \bigg)
 \sin{\theta} \cos{\theta}
+ \bigg[-(-1)^k\frac{C_3 (\alpha^2+2 \alpha-11)}{2 (\alpha+1)} \nonumber\\
&&\bigg(M+\gamma D -\frac{2 C_2 (\alpha+3)}{ (\alpha+5)Q^2}\bigg)
 x^{-\frac{1}{2}(\alpha+1)}-(-1)^k C_4 x^{-\frac{1}{2}(\beta-1)}
+\frac{3 C^2_3}{(\alpha+1)(\alpha-2)}x^{-(\alpha-1)} \nonumber\\
&&-\frac{2 (-1)^k C_5 - C_2 (M+\gamma D) + 3 a M Q}{2 x^2}\bigg]
 \sin{\theta} \cos{\theta} + o\left(\frac{1}{x^2}\right)\ , \nonumber \\
H_4&=&(-1)^k+C_3 x^{-\frac{1}{2}(\alpha-1)} + \bigg((-1)^k\frac{C_2}{x}
- C_3 x^{-\frac{1}{2}(\alpha-1)} \bigg) \sin^2{\theta}
+\bigg[\frac{C_3 (\alpha^2+2 \alpha -11)}{2 (\alpha+1)} \nonumber\\
&&\bigg(M+\gamma D-\frac{2 C_2 (\alpha+3)}{(\alpha+5)Q^2}\bigg)x^{-\frac{1}{2}(\alpha+1)}
+C_4 x^{-\frac{1}{2} (\beta-1)} + \frac{C_5}{x^2}\bigg] \nonumber\\
&&- \bigg[\frac{C_3 (\alpha^2+2 \alpha -11)}{2 (\alpha+1)}
\bigg(M+\gamma D-\frac{2 C_2 (\alpha+3)}{(\alpha+5)Q^2}\bigg)x^{-\frac{1}{2}(\alpha+1)}
+C_4 x^{-\frac{1}{2} (\beta-1)} \nonumber \\
&&- \frac{3 (-1)^k C^2_3}{(\alpha+1)(\alpha-2)} x^{-(\alpha-1)}
+ \frac{2 C_5 -(-1)^k C_2 (M+\gamma D) +(-1)^k 3 a M Q}{2 x^2}\bigg]\sin^2{\theta}\nonumber\\
&&+ o\left(\frac{1}{x^2}\right)
\ , \label{asymp_general}
\end{eqnarray}
where $\alpha=\sqrt{9-4 Q^2}$ and $\beta=\sqrt{25-4 Q^2}$. 
\\
\\
{\sl $Q=0$ solutions ($n=1$, $k=1$)}

Let us now consider the asymptotic expansion for the case $Q=0$,
and expand the constants $C_i$ in powers of $Q$,
\begin{eqnarray}
C_1&=&C_{10}+C_{11}Q+O(Q^2) \ , \nonumber\\
C_2&=&C_{20}+C_{21}Q+O(Q^2) \ , \nonumber\\
C_3&=&C_{30}+C_{31}Q+O(Q^2) \ , \nonumber\\ 
C_4&=&\frac{5}{4}\frac{C_{20}C_{30}}{Q^2} 
 + \frac{C^{*}_{41}}{Q} + O(Q^0) \ , \nonumber\\
C_5&=&-\frac{1}{2}\frac{C_{20}C_{30}}{Q^2}
 + \frac{1}{4} \frac{3(C_{20}C_{31}+C_{21}C_{30})-4C^{*}_{41}}{Q} 
 + O(Q^0) \ , \nonumber\\
C_6&=&\frac{1}{10}\frac{(5C_{20}+3C_{30})C_{30}}{Q}+O(Q) \ .
\end{eqnarray}

The $Q$-dependence of $C_4$ and $C_5$ is indeed observed
numerically, as seen in Figs.~13a-b. For this figure
we first obtained $C_{20}$ and $C_{30}$ from the $Q=0$ solution, 
yielding $C_{20}=-25.014$ and $C_{30}=25.019$.
The dashed lines correspond to the approximation, that
only the first term in $C_4$ and $C_5$ is taken, using these
values for $C_{20}$ and $C_{30}$.
Then we extracted $C_4$ and $C_5$ from the function $H_1$
for several values of $\gamma$ 
(equivalent to several values of $Q$,
since $x_{\rm H}$ and $\omega_{\rm H}$ are fixed),
where we employed $C_2$ and $C_3$,
extracted from the functions $H_2$ and $H_3$
for the same values of $\gamma$. 
In the figure we observe good agreement 
between these points and the approximate theoretical curve.

We also observe, that unlike the pure EYM case \cite{kkrot}, 
we may have $C_{20}+C_{30} \ne 0$, when the dilaton is included. Consequently,
$H_3$ can approach zero asymptotically with an integer power fall-off.
Setting $C_{10}=-1/2 M^2_0$, $C_{11}=0$, 
$C_{20}=-b$, $C_{21}=0$, $C_{30}=b$, $C_{31}=0$, and $C^{*}_{41}=0$, 
we obtain the relations for the EYM case, reported previously
\cite{kkrot}.
\\
\\
{\sl  $n=3$ Solutions}

\begin{eqnarray}
f&=&1-\frac{2 M}{x} + \frac{2 M^2 + Q^2}{x^2}
 +  o \left(\frac{1}{x^2}\right)\ , \nonumber\\
m&=&1 + \frac{C_1}{x^2} + \frac{Q^2-M^2-D^2-2 C_1}{x^2}\sin^2{\theta}
 + o\left( \frac{1}{x^2}\right)\ , \nonumber\\
l&=&1+\frac{C_1}{x^2} + o\left(\frac{1}{x^2}\right)\ , \nonumber \\
\omega&=&\frac{2 a M}{x^2} - \frac{6 a  M^2 + 3 C_2 Q}{x^3}
 + o\left(\frac{1}{x^3}\right)\ , \nonumber\\
\phi &=& -\frac{D}{x}-\frac{\gamma Q^2}{2 x^2} + \frac{C_1 D - 4 C_7 
 + 2\gamma (M - \gamma D) Q^2}{6 x^3} + \frac{C_7}{x^3} \sin^2{\theta} + 
 o\left(\frac{1}{x^3}\right)
\ , \nonumber\\
{\bar B}_1 &=& \frac{Q \cos{\theta}}{x} 
 -\frac{(M-\gamma D) Q \cos{\theta}}{x^2} \nonumber\\
 &&+\frac{[2(1+\gamma^2)Q^2+4(M-\gamma D)^2
 -C_1-4 (-1)^k C_5]Q-2 C_6}{6x^3}\cos\theta \nonumber \\
 && + \frac{C_6}{x^3}\cos^3\theta + o\left(\frac{1}{x^3}\right) \ , \nonumber \\
{\bar B}_2 &=& -(-1)^k\frac{Q \sin{\theta}}{x} 
 +(-1)^k\frac{(M-\gamma D) Q \sin{\theta}}{x^2} \nonumber\\ 
 &&-(-1)^k \Bigg[\frac{[2(1+\gamma^2)Q^2+4(M-\gamma D)^2-C_1
 +2 (-1)^k C_5]Q-2 C_6}{6x^3} \nonumber \\
 &&-\frac{6 (1-(-1)^k) a M}{x^3}+\frac{C_6}{x^3}\cos^2\theta\Bigg] \sin\theta 
 + o\left(\frac{1}{x^3}\right) \ , \nonumber \\
H_1 &=& \frac{C_5}{x^2}\sin{2 \theta}  
 - \frac{1}{2}(-1)^k C_9 (\epsilon -1) x^{-\frac{1}{2}(\epsilon-1)} \sin\theta \cos\theta 
 + o\left(\frac{1}{x^3}\right) \ , \nonumber \\
H_2 &=& {\DS (-1)^k + \frac{C_5}{x^2} \cos{2 \theta} 
 - (-1)^k C_9 \left(1 - \frac{1}{8} (\epsilon-1)^2 \sin^2\theta \right) 
 x^{-\frac{1}{2}(\epsilon-1)} + o\left(\frac{1}{x^3}\right) }\ , \nonumber \\
H_3 &=& \frac{C_2}{x} \sin\theta \cos\theta + \frac{(M+\gamma D) C_2 
 -2 (-1)^k C_5 -a M Q}{2 x^2} \sin\theta \cos\theta + \Bigg[\frac{C_8}{x^3}  \nonumber \\
 &&+ C_9 x^{-\frac{1}{2}(\epsilon-1)}+ \frac{1}{24}\Bigg(\frac{3 C_2 (2 (1+\gamma^2)Q^2 
 + 2 (M+\gamma D)^2 + C_1-8 (-1)^k C_5)}{x^3} \nonumber \\
 &&+\frac{6(M-3\gamma D)aMQ - 30 C_8}{x^3} -C_9(\epsilon+3)(\epsilon-5) 
 x^{-\frac{1}{2}(\epsilon-1)}\Bigg)\sin^2\theta\Bigg]\sin\theta \cos\theta \nonumber \\
 && + o\left(\frac{1}{x^3}\right) \ , \nonumber \\
H_4 &=& (-1)^k + (-1)^k \frac{C_2}{x}\sin^2\theta + \frac{C_5}{x^2} 
 + (-1)^k \frac{(M+\gamma D)C_2 - 2 (-1)^k C_5 - aMQ}{2 x^2} \sin^2\theta \nonumber \\
 && + (-1)^k \Bigg[ -C_9 x^{-\frac{1}{2}(\epsilon-1)} 
 + \frac{1}{24}\Bigg[ C_9 ((\epsilon-1)^2+8) x^{-\frac{1}{2}(\epsilon-1)} 
 + \frac{24 (C_8 +(-1)^k C_2 C_5)}{x^3} \nonumber \\
 &&+\Bigg(\frac{3C_2(2(1+\gamma^2)Q^2 + 2 (M+\gamma D)^2 + C_1 - 8(-1)^k C_5) 
 + 6(M-3\gamma D)aMQ -30 C_8}{x^3} \nonumber \\
 &&- C_9(\epsilon+3)(\epsilon-5)  
 x^{-\frac{1}{2}(\epsilon-1)}\Bigg)\sin^2\theta\Bigg]\sin^2\theta\Bigg] 
 + o\left(\frac{1}{x^3}\right) \ ,
\end{eqnarray}
where $\epsilon=\sqrt{49-4Q^2}$.

\section{ Expansion at the Horizon}

Here we present the expansion of the functions
of the stationary axially symmetric black hole solutions
at the horizon $x_{\rm H}$ in powers of $\delta$.
These expansions can be obtained from the regularity conditions
imposed on the Einstein equations and the matter field equations:
\begin{eqnarray}
f(\delta,\theta)&=&\delta^2 f_2 \bigg\{1 -\delta
+  \frac{\delta^2}{24}\bigg[\left(\frac{n}{x_{\rm H}}\right)^2\frac{f_2}{l_2} e^{2 \gamma \phi_0}
\bigg[24 \cot{\theta}[-(-H_{30,\theta} +1 -{H_{20}}^2) H_{30} \nonumber\\
&&- H_{20} H_{40,\theta} + H_{40} H_{40, \theta}]
+ 12 \bigg(H^2_{20}(H^2_{30}+H^2_{40}-1)+\frac{(H_{20}-H_{40})^2
+H^2_{30}}{\sin^2{\theta}} \nonumber\\
&&-(H^2_{30}+H^2_{40})+2 H_{20} (-H_{30} H_{40,\theta}
+H_{40} H_{30,\theta})+1 -2 H_{30,\theta} +H^2_{30,\theta}\nonumber\\
&&+H^2_{40,\theta} \bigg) \bigg]
-2 \cot{\theta} \bigg(3\frac{f_{2,\theta}}{f_2}
- 2 \frac{l_{2,\theta}}{l_2}\bigg)
- \bigg(3 \frac{f_{2,\theta}}{f_2}\frac{l_{2,\theta}}{l_2}
+ 6\frac{f_{2,\theta \theta}}{f_2}
+ \left(\frac{l_{2,\theta}}{l_2}\right)^2  \nonumber \\
&& -2 \frac{l_{2,\theta\theta}}{l_2}
-18-6\left(\frac{f_{2,\theta}}{f_2}\right)^2 \bigg)
- \frac{24}{f_2} \bigg[- 4 n \sin{\theta} \frac{\omega_2}{x_{\rm H}}e^{2 \gamma \phi_0}
 \bigg( H_{30}B_{12} + (1-H_{40}) B_{22}\bigg)   \nonumber \\
&&-2 e^{2 \gamma \phi_0} (B^2_{12} + B^2_{22})
-\sin^2{\theta}\frac{\omega^2_2}{x^2_{\rm H} f_2} \bigg( x^2_{\rm H} l_2
+ 2 n^2 f_2 e^{2 \gamma \phi_0} \left(H^2_{30}+(1-H_{40})^2\right) \bigg) \bigg] \bigg] \bigg\}\nonumber\\
&&+O(\delta^5)\ , \nonumber\\
m(\delta,\theta)&=&\delta^2 m_2 \bigg\{1-3 \delta
+ \frac{\delta^2}{24} \bigg[150 -4 \frac{l_{2,\theta \theta}}{l_2}
+ 2 \bigg(\frac{l_{2,\theta}}{l_2}\bigg)^2
+ 3 \frac{l_{2,\theta}}{l_2}\frac{m_{2,\theta}}{m_2}
-6 \frac{m_{2,\theta \theta}}{m_2}  \nonumber\\
&&+6 \bigg( \frac{m_{2,\theta }}{m_2} \bigg)^2
-6 \bigg( \frac{f_{2, \theta}}{f_2} \bigg)^2
+ 2 \cot{\theta} \bigg( 3  \frac{m_{2, \theta}}{m_2}
-4  \frac{l_{2,\theta}}{l_2}\bigg)-24 \phi^2_{0,\theta} \nonumber \\
&&+ 24 \sin^2{\theta}\frac{l_2 \omega^2_2}{f^2_2} \bigg] \bigg\}
+ O(\delta^5)\ , \nonumber \\
l(\delta,\theta)&=&\delta^2 l_2 \bigg\{1 -3 \delta
+ \frac{\delta^2}{12} \bigg[ \bigg( \frac{l_{2,\theta}}{l_2}\bigg)^2
- 2 \frac{l_{2,\theta \theta}}{l_2}
+75- 4 \cot{\theta}\frac{l_{2,\theta}}{l_2}\bigg]\bigg\}
+ O(\delta^5)\ ,\nonumber\\
\omega(\delta,\theta)&=&\omega_{\rm H} (1 + \delta)
+ \delta^2\omega_2 + O(\delta^4)\ ,\nonumber\\
\phi(\delta,\theta)&=&\phi_0-\frac{\delta^2}{8}\bigg\{2  \cot{\theta}\phi_{0,\theta} + \phi_{0,\theta}\frac{l_{2,\theta}}{l_2}+2\phi_{0,\theta \theta} \nonumber \\
&&-\bigg(\frac{n}{x_{\rm H}}\bigg)^2\frac{f_2}{l_2}\gamma e^{2 \gamma \phi_0} \bigg[ 4 \cot{\theta}[-H_{30} (-H_{30,\theta}+1)-H_{20}(-H_{20} H_{30}+H_{40,\theta})\nonumber \\
&&+H_{40} H_{40,\theta}] +2 \bigg((H^2_{30}+H^2_{40}-1)H^2_{20}+2H_{20}(-H_{30}H_{40,\theta}+H_{40}H_{30,\theta})\nonumber \\
&&+H^2_{30,\theta}-2H_{30,\theta}+H^2_{40,\theta}+1+\frac{(H_{20}-H_{40})^2}{\sin^2{\theta}}+\frac{H^2_{30}}{\sin^2{\theta}}-(H^2_{30}+H^2_{40})\bigg) \bigg] \nonumber \\
&&+8\frac{1}{x^2_{\rm H} f_2}\gamma e^{2 \gamma \phi_0} \bigg[2 n \sin{\theta} x_{\rm H} \omega_2 (H_{30} B_{12}+(1-H_{40}) B_{22})+x^2_{\rm H} (B^2_{12}+B^2_{22})\nonumber \\
&&+n^2\sin^2{\theta}\omega^2_2 (H^2_{30}+(1-H^2_{40}))\bigg] 
\bigg\} \ , \nonumber\\
{\bar B}_1(\delta,\theta)&=&n\frac{\omega_{\rm H} }{x_{\rm H}}\cos{\theta}
+\delta^2 (1-\delta) B_{12} + O(\delta^4)\ ,\nonumber\\
{\bar B}_2(\delta,\theta)&=&-n\frac{\omega_{\rm H} }{x_{\rm H}}\sin{\theta}
+\delta^2 (1-\delta) B_{22} + O(\delta^4)\ ,\nonumber\\
H_1(\delta,\theta)&=&\delta \left(1 -\frac{1}{2}\delta \right) H_{11}
+ O(\delta^3)\ , \nonumber\\
H_2(\delta,\theta)&=&H_{20}
+\frac{\delta^2}{4} \bigg[n^2\frac{m_2}{l_2}\bigg(H_{20} (H^2_{30}
+H^2_{40}-1)-H_{30}H_{40,\theta}+H_{40} H_{30,\theta}
+\frac{H_{20}-H_{40}}{\sin^2{\theta}} \nonumber\\
&&-\cot{\theta} (-2 H_{20} H_{30} + H_{40,\theta})\bigg)-(H_{11,\theta}
+ H_{20,\theta \theta})\bigg] + O(\delta^3)\ , \nonumber \\
H_3(\delta,\theta)&=&H_{30}
-\frac{\delta^2}{8} \bigg[-\bigg(4 \gamma \phi_{0,\theta}+ 2\frac{f_{2,\theta}}{f_2}
-\frac{l_{2,\theta}}{l_2}\bigg) (1-H_{40}H_{20}-H_{30,\theta}
-\cot{\theta} H_{30}) \nonumber\\
&& -2\cot{\theta} H_{20}(H_{20}-H_{40})+2H_{30,\theta \theta}
+ 4 H_{20} H_{40,\theta} -2\bigg(\frac{H_{30}}{\sin^2{\theta}}
- \cot{\theta} H_{30,\theta}\bigg)\nonumber\\
&&-2 H_{30}H^2_{20} - 2 H_{40}(2H_{11}-H_{20,\theta})
+8\sin{\theta}\frac{l_2 \omega_2}{f^2_2}(\frac{x_{\rm H}}{n} B_{12}
+\sin{\theta} \omega_2 H_{30}) \bigg] + O(\delta^3)\ , \nonumber\\
H_4(\delta,\theta)&=&H_{40}
-\frac{\delta^2}{8}\bigg[ \bigg(4 \gamma \phi_{0,\theta} + 2\frac{f_{2,	\theta}}{f_2}
- \frac{l_{2,\theta}}{l_2}\bigg) [H_{40,\theta}-H_{20} H_{30}
-\cot{\theta} (H_{20}-H_{40})]\nonumber\\
&&+H_{20}(-4 H_{30,\theta} +2) + 2 [H_{30}(2 H_{11}-H_{20,\theta})
+H_{40,\theta \theta}-H_{40} H^2_{20}]\nonumber\\
&&+ 2 \frac{H_{20}-H_{40}}{\sin^2{\theta}} - 2 \cot{\theta}(-2 H_{11}
-H_{40,\theta} + H_{20} H_{30} + H_{20,\theta})\nonumber\\
&&-8 \sin{\theta}\frac{l_2 \omega_2}{f^2_2}[ \frac{x_{\rm H}}{n} B_{22}
+\sin{\theta}\omega_2 (1-H_{40})] \bigg] + O(\delta^3)\ ,
\label{hor_general}
\end{eqnarray}
$H_{20}$, $H_{30}$, $H_{40}$, $H_{11}$, $f_2$, $m_2$, $l_2$, 
$\omega_2$, $\phi_0$, $B_{12}$, and $B_{22}$ are functions of $\theta$.
Relations (\ref{relation_hor_1}) and (\ref{relation_hor_2}) also hold.

\section{Energy conditions and Segr\'e type}

In this section we revisit the energy conditions for SU(2) EYMD theory, the
results being valid for general gauge group SU(N). We also consider the Segr\'e type
of this type of matter.

The {\it dominant energy condition} requires that ``for every timelike vector
 $V$, the stress-energy tensor satisfies $T_{\mu \nu} V^{\mu} V^{\nu} \ge 0$ 
and $T^{\mu \nu} V_{\nu}$ is a non-spacelike vector''. Let us show that
 Eq.~(\ref{tmunu}) satisfies this condition. 

Let $V$ be an arbitrary timelike vector. We may define its associated unit 
vector $E_0$ via 
\begin{equation}
V=\sqrt{-V_{\mu} V^{\mu}} E_0 \ .  \label{def_E_0}
\end{equation}
By constructing an orthonormal basis with $E_0$ as its timelike element, $\{E_0, E_1, E_2, E_3\}$, one can reformulate the dominant energy condition as
\begin{equation}
T_{(0) (0)} \ge |T_{(a) (b)}| \ , \,\,\, \forall a,b=0, 1, 2, 3, \label{dec}
\end{equation}
where the $(\cdot)$ index indicates the component in the orthonormal basis.

In order to simplify the proof we separate Eq.~(\ref{tmunu}) into two parts, namely,
\begin{equation}
T_{\mu\nu} = T^{\rm dil}_{\mu\nu} + T^{\rm YM}_{\mu\nu} \ , \label{T_shift}
\end{equation}
with
\begin{eqnarray}
T^{\rm dil}_{\mu\nu}&=& \partial_\mu \Phi \partial_\nu \Phi
     -\frac{1}{2} g_{\mu\nu} \partial_\alpha \Phi \partial^\alpha \Phi \equiv
H_{\mu} H_{\nu}-\frac{1}{2} g_{\mu\nu} H^2 \ , \label{T_dil} \\
T^{\rm YM}_{\mu\nu} &=& 2 e^{2 \kappa \Phi }{\rm Tr}
    ( F_{\mu\alpha} F_{\nu\beta} g^{\alpha\beta}
   -\frac{1}{4} g_{\mu\nu} F_{\alpha\beta} F^{\alpha\beta}) \ , \label{T_YM}
\end{eqnarray}
where we have defined $H_{\mu}\equiv \partial_{\mu} \Phi$.

A straightforward calculation performed in the orthonormal basis then yields
\begin{equation}
T^{\rm dil}_{(0)(0)} \ge |T^{\rm dil}_{(a)(b)}|, \, \, \,T^{\rm YM}_{(0)(0)} \ge |T^{\rm YM}_{(a)(b)}| \ , \, \, \, \forall a,b=0, 1, 2, 3, \label{dec_shift} 
\end{equation}
and from Eqs.~(\ref{T_shift}) and (\ref{dec_shift}) we conclude that 
condition (\ref{dec}) holds.

Obviously, the {\it weak energy condition} is also satisfied, $T_{(0) (0)} \ge 0 \ , \,\,\, \forall a,b=0, 1, 2, 3$. 

As for the {\it strong energy condition}, it can be formulated as follows: ``A stress-energy tensor is said to satisfy the strong energy condition if it obeys the inequality
\begin{equation}
T_{\mu \nu} V^{\mu} V^{\nu} \ge \frac{1}{2} T V_{\mu} V^{\mu} \ , \label{sec} 
\end{equation}
for any timelike vector $V$ ($T$ stands for the trace of the stress-energy 
tensor, $T=T_{\mu}^{\mu}$)''.

The Yang-Mills part of the stress-energy tensor, $T^{\rm YM}_{\mu \nu}$, 
fulfils the strong energy condition trivially as it is a traceless tensor 
that satisfies the weak energy condition. The dilaton part,  
$T^{\rm dil}_{\mu \nu}$, also satisfies the strong energy condition
 as can be easily shown in an 
orthonormal basis
\begin{equation}
T^{\rm dil}_{\mu \nu} V^{\mu} V^{\nu} 
\ge \frac{1}{2} T^{\rm dil} V_{\mu} V^{\mu}\ . \label{sec_dil} 
\end{equation}
Note that $T^{\rm dil}=T=-H^2$. Thus, Eq.~(\ref{sec}) is satisfied by the full
EYMD stress-energy tensor.

Related to the energy conditions is the Segr\'e type (or algebraic type) of the stress-energy tensor \cite{book}. It consists of a study of the eigenvalue problem of that tensor, namely,
\begin{equation}
{T^{\mu}}_{\nu} u^{\nu} = \lambda u^{\mu} \ . \label{eigenv}
\end{equation}

It is well known that a general EM system has an algebraic type A1[(1 1) (1, 1)], with two double eigenvalues, opposite in sign. (We are not considering the null case). For an arbitrary EYM system the Segr\'e type is that of a diagonalizable general type, A1[1 1 1, 1]. The non-Abelian nature breaks the degeneracy of the Abelian case.

But the inclusion of the dilaton field also destroys that degeneracy  A1[(1 1) (1, 1)] in the Abelian case, since the eigenvalues for a general EMD solution satisfy
\begin{eqnarray}
\lambda_1 + \lambda_2 &=& -H^2 \ , \nonumber\\
\lambda_3 + \lambda_4 &=& 0 \ ,
\end{eqnarray} 
with $H$ defined as before. It is no surprise then to find a generic type for the
EYMD solutions.

If we restrict ourselves to Eqs. (\ref{metric}) and (\ref{a1})-(\ref{a2}), we may write the eigenvalues in the form
\begin{eqnarray}
\lambda_1 &=& \frac{1}{2} \left[({T^t}_t + {T^{\varphi}}_{\varphi}) + \sqrt{({T^t}_t - {T^{\varphi}}_{\varphi})^2 + 4 {T^{\varphi}}_t {T^t}_{\varphi}} \right]\ ,\nonumber \\
\lambda_2 &=& \frac{1}{2} \left[({T^t}_t + {T^{\varphi}}_{\varphi}) - \sqrt{({T^t}_t - {T^{\varphi}}_{\varphi})^2 + 4 {T^{\varphi}}_t {T^t}_{\varphi}} \right]\ ,\nonumber\\
\lambda_3 &=& \frac{1}{2} \left[({T^r}_r + {T^{\theta}}_{\theta}) + \sqrt{({T^r}_r - {T^{\theta}}_{\theta})^2 + 4 {T^{\theta}}_r {T^r}_{\theta}} \right] \ ,\nonumber \\
\lambda_4 &=& \frac{1}{2} \left[({T^r}_r + {T^{\theta}}_{\theta}) - \sqrt{({T^r}_r - {T^{\theta}}_{\theta})^2 + 4 {T^{\theta}}_r {T^r}_{\theta}} \right] \ .
\end{eqnarray}
Note that this is just a consequence of the Lewis-Papapetrou form of the metric and the Einstein equations. By using an orthonormal basis it is easy to show that all the eigenvalues are real for EMYD solutions.

The explicit expressions of these eigenvalues are rather complicated. To conclude this section we show the asymptotic behavior of the dimensionless version of these eigenvalues ($n=1$)
\begin{eqnarray}
\lambda_1 &=& -\frac{D^2-Q^2}{2 x^4} + \frac{M (D^2 - 2 Q^2)}{x^5} + o\left(\frac{1}{x^5}\right) \ ,\nonumber\\
\lambda_2 &=& -\frac{D^2+Q^2}{2 x^4} + \frac{M (D^2 + 2 Q^2) - 2 \gamma D Q^2}{x^5} + o\left(\frac{1}{x^5}\right) \ ,\nonumber\\ 
\lambda_3 &=& \frac{D^2-Q^2}{2 x^4} - \frac{M (D^2 - 2 Q^2)}{x^5} + o\left(\frac{1}{x^5}\right) \ ,\nonumber\\
\lambda_4 &=& -\frac{D^2-Q^2}{2 x^4} + \frac{M (D^2 - 2 Q^2)}{x^5} + o\left(\frac{1}{x^5}\right) \ .
\end{eqnarray}
Obviously, we observe that for charged EYM solutions ($D=0$), the algebraic type of the stress-energy tensor behaves asymptotically as that of a pure EM system.


\newpage
\setcounter{fixy}{1}

   \begin{fixy} {0}
\begin{figure}\centering
{\large Fig. 1a} \vspace{0.0cm}\\
\epsfysize=8cm
\mbox{\epsffile{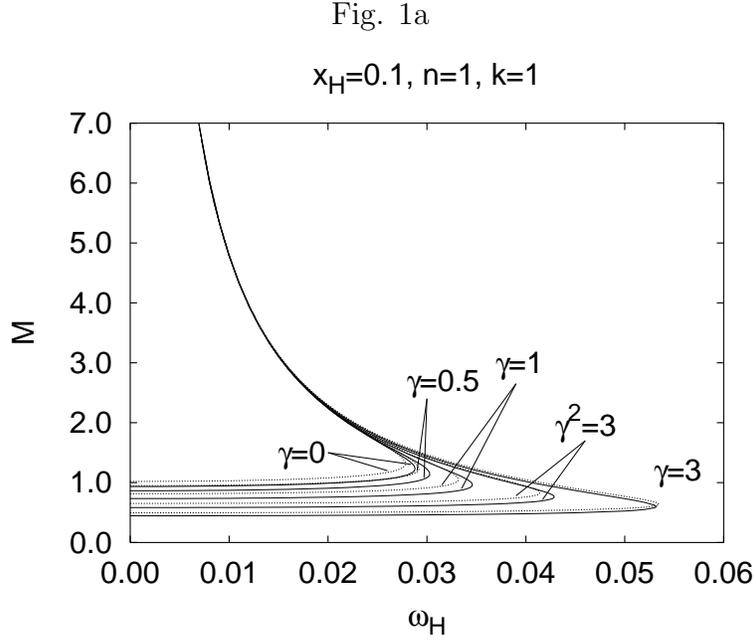}}
\caption{
The dimensionless mass $M$ is shown as a function of $\omega_{\rm H}$ 
for EYMD black holes with
winding number $n=1$, node number $k=1$, horizon radius $x_{\rm H}=0.1$, 
for dilaton coupling constant $\gamma=0$, 0.5, 1, $\sqrt{3}$, and 3 
(solid lines). 
The dimensionless mass of the corresponding EMD solutions with
electric charge $Q=0$ and magnetic charge $P=1$ is also shown 
(dotted lines).
} \end{figure}

\begin{figure}\centering
{\large Fig. 1b} \vspace{0.0cm}\\
\epsfysize=8cm
\mbox{\epsffile{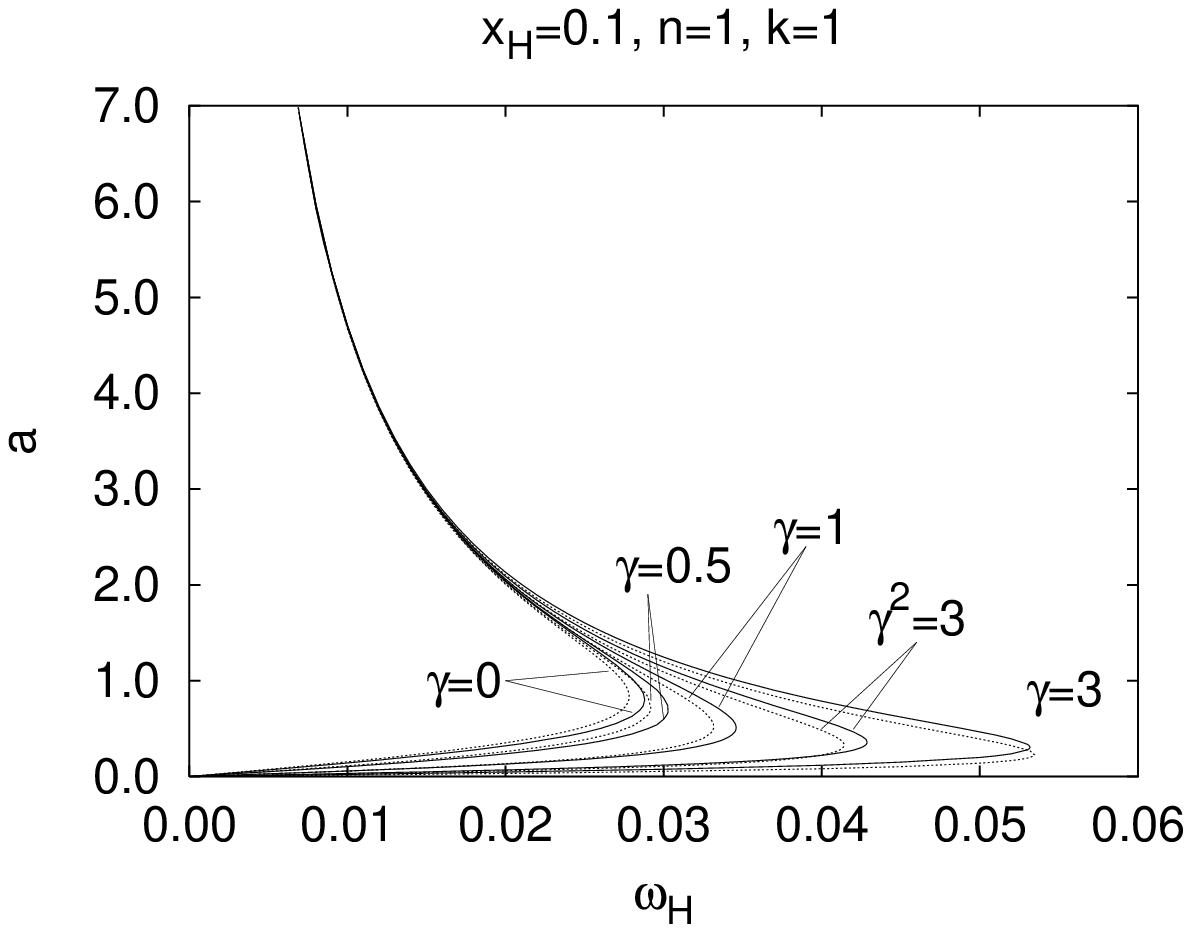}}
\caption{
Same as Fig.~1a for the specific angular momentum $a=J/M$. 
} \end{figure}
\clearpage

\newpage
\begin{figure}\centering
{\large Fig. 1c} \vspace{0.0cm}\\
\epsfysize=8cm
\mbox{\epsffile{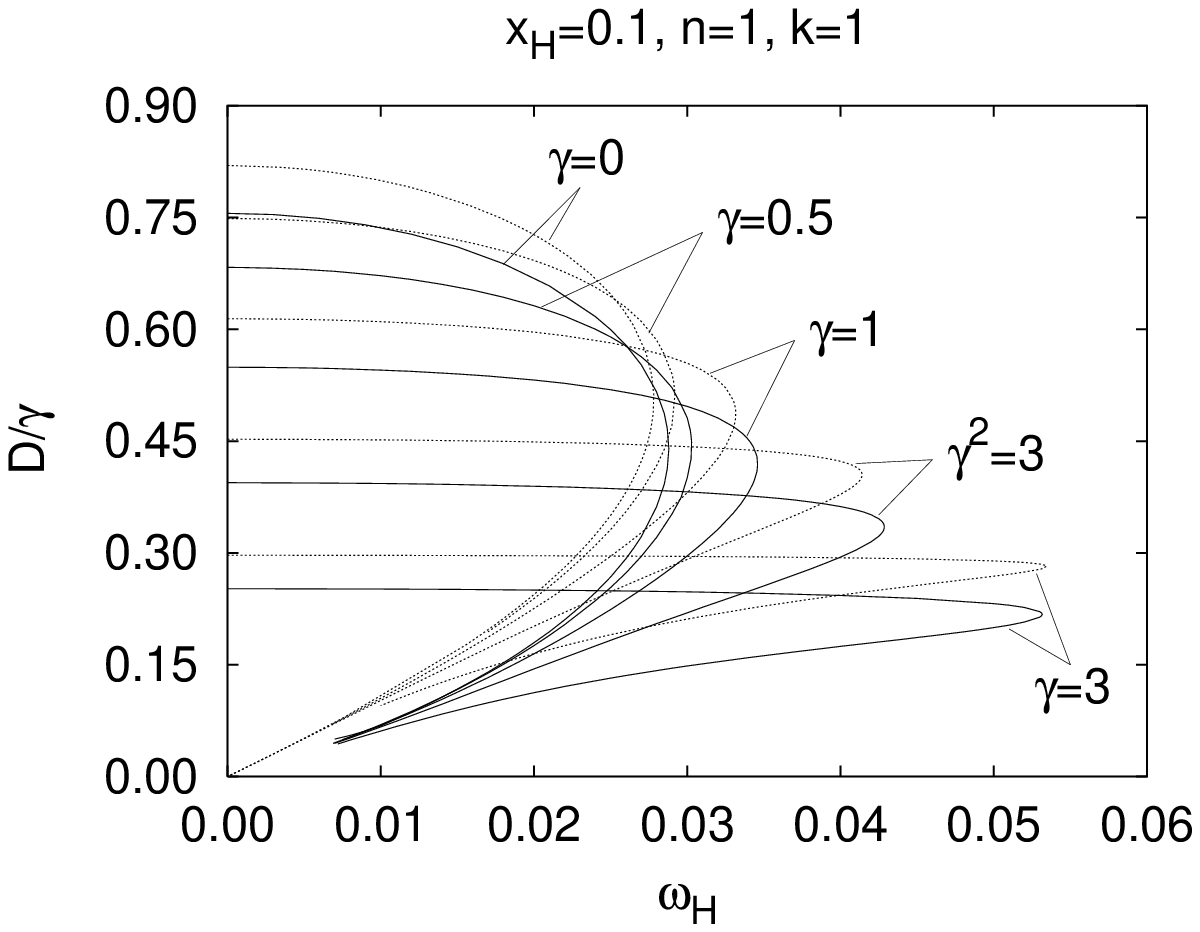}}
\caption{
Same as Fig.~1a for the relative dilaton charge $D/\gamma$. 
} \end{figure}
\begin{figure}\centering
{\large Fig. 1d} \vspace{0.0cm}\\
\epsfysize=8cm
\mbox{\epsffile{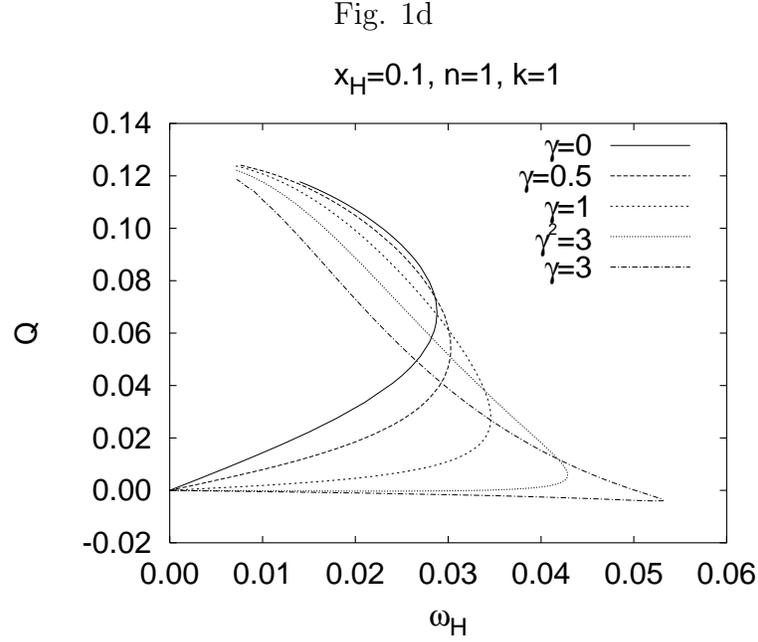}}
\caption{
The electric charge $Q$ is shown as a function of $\omega_{\rm H}$ 
for the same set of EYMD black holes shown in Fig.~1a.
} \end{figure}
\end{fixy}

\clearpage

\newpage
   \begin{fixy} {0}
\begin{figure}\centering
{\large Fig. 2a} \vspace{0.0cm}\\
\epsfysize=8cm
\mbox{\epsffile{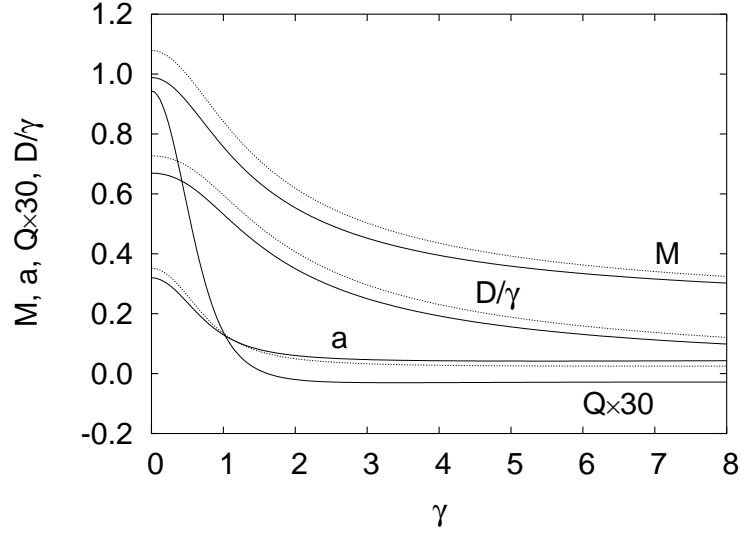}}
\caption{
The global charges, $M$, $a$, $D/\gamma$, and $Q$, 
are shown as functions of $\gamma$ for 
EYMD black holes with
winding number $n=1$, node number $k=1$, horizon radius $x_{\rm H}=0.1$, 
and $\omega_{\rm H}=0.020$ on the lower branch (solid lines). 
The global charges of the corresponding EMD solutions with
electric charge $Q=0$ and magnetic charge $P=1$ are also shown 
(dotted lines).   
} \end{figure}

\begin{figure}\centering
{\large Fig. 2b} \vspace{0.0cm}\\
\epsfysize=8cm
\mbox{\epsffile{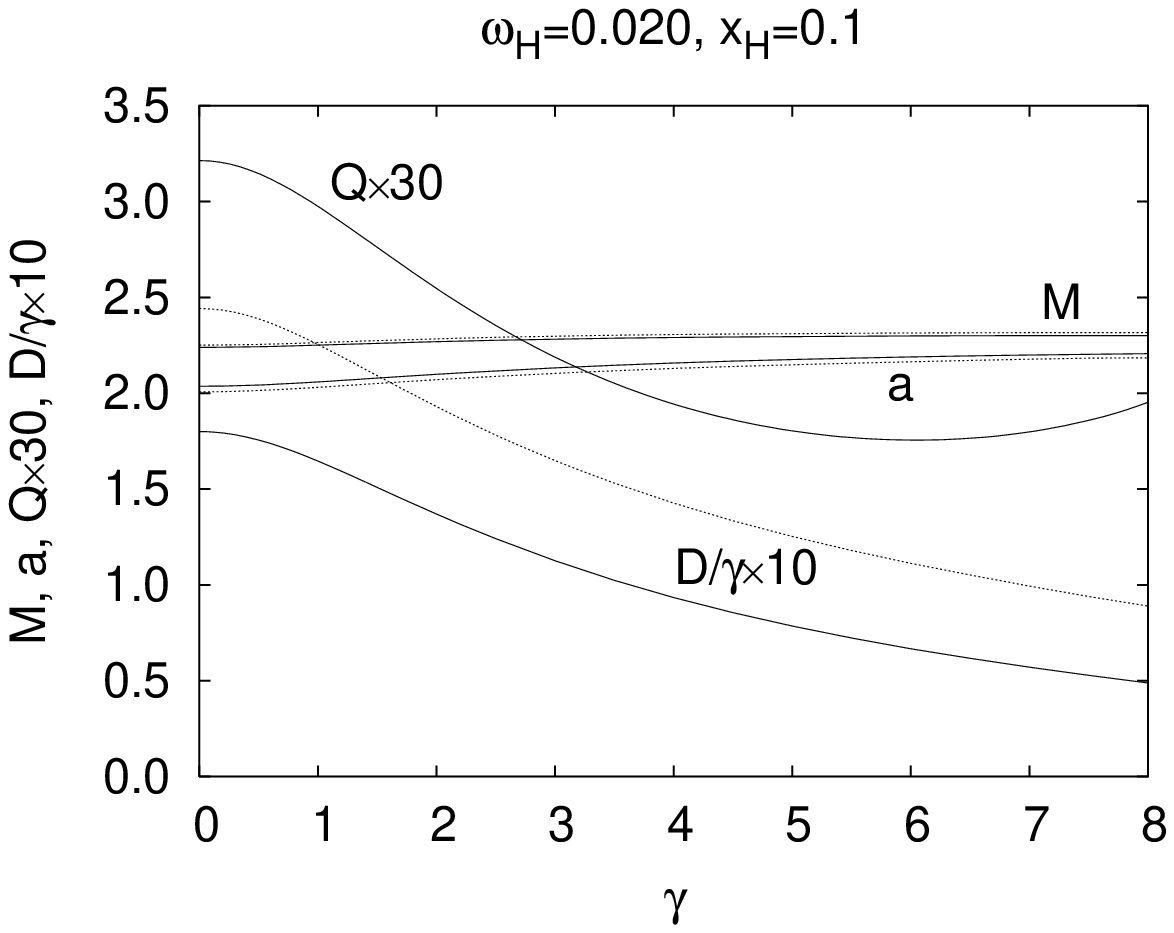}}
\caption{
Same as Fig.~2a for black hole solutions on the upper branch. 
} \end{figure}
\end{fixy}
\clearpage

\newpage
   \begin{fixy} {0}
\begin{figure}\centering
{\large Fig. 3a} \vspace{0.0cm}\\
\epsfysize=8cm
\mbox{\epsffile{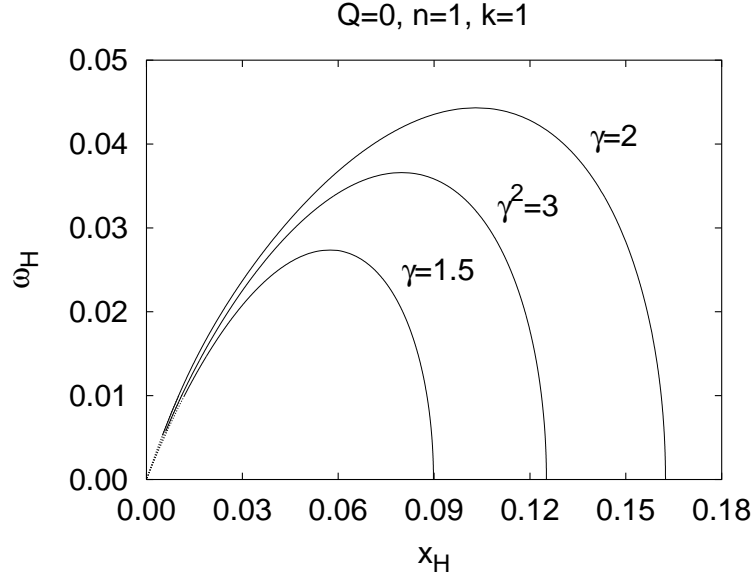}}
\caption{
Cuts through the parameter space of $Q=0$ black hole solutions 
with winding number $n=1$, node number $k=1$, 
and dilaton coupling constant $\gamma=1.5$, $\sqrt{3}$, and 2.  
The curves are extrapolated to $\omega_{\rm H}=x_{\rm H}=0$,
as indicated by the dots.
} \end{figure}

\begin{figure}\centering
{\large Fig. 3b} \vspace{0.0cm}\\
\epsfysize=8cm
\mbox{\epsffile{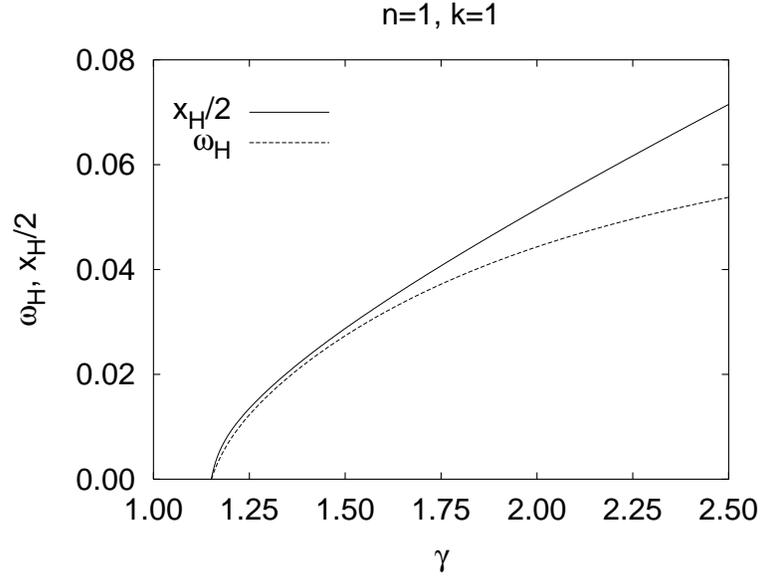}}
\caption{
The  value of the maximum of  $\omega_{\rm H}$ and its location $x_{\rm H}$ 
are shown as functions of $\gamma$ for the $Q=0$ 
EYMD black hole solutions with winding number $n=1$ and node number $k=1$.
} \end{figure}
\clearpage

\newpage

\begin{figure}\centering
{\large Fig. 3c} \vspace{0.0cm}\\
\epsfysize=8cm
\mbox{\epsffile{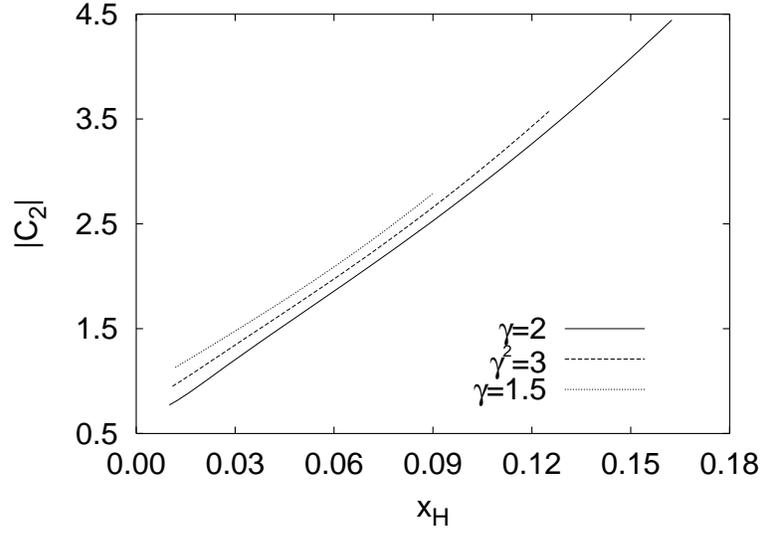}}
\caption{
The coefficient $-C_2=|C_2|$ of the dimensionless magnetic moment $\mu$ is shown
as a function of $x_{\rm H}$
for the sets of EYMD back holes shown in Fig.~3a.
} \end{figure}

\begin{figure}\centering
{\large Fig. 3d} \vspace{0.0cm}\\
\epsfysize=8cm
\mbox{\epsffile{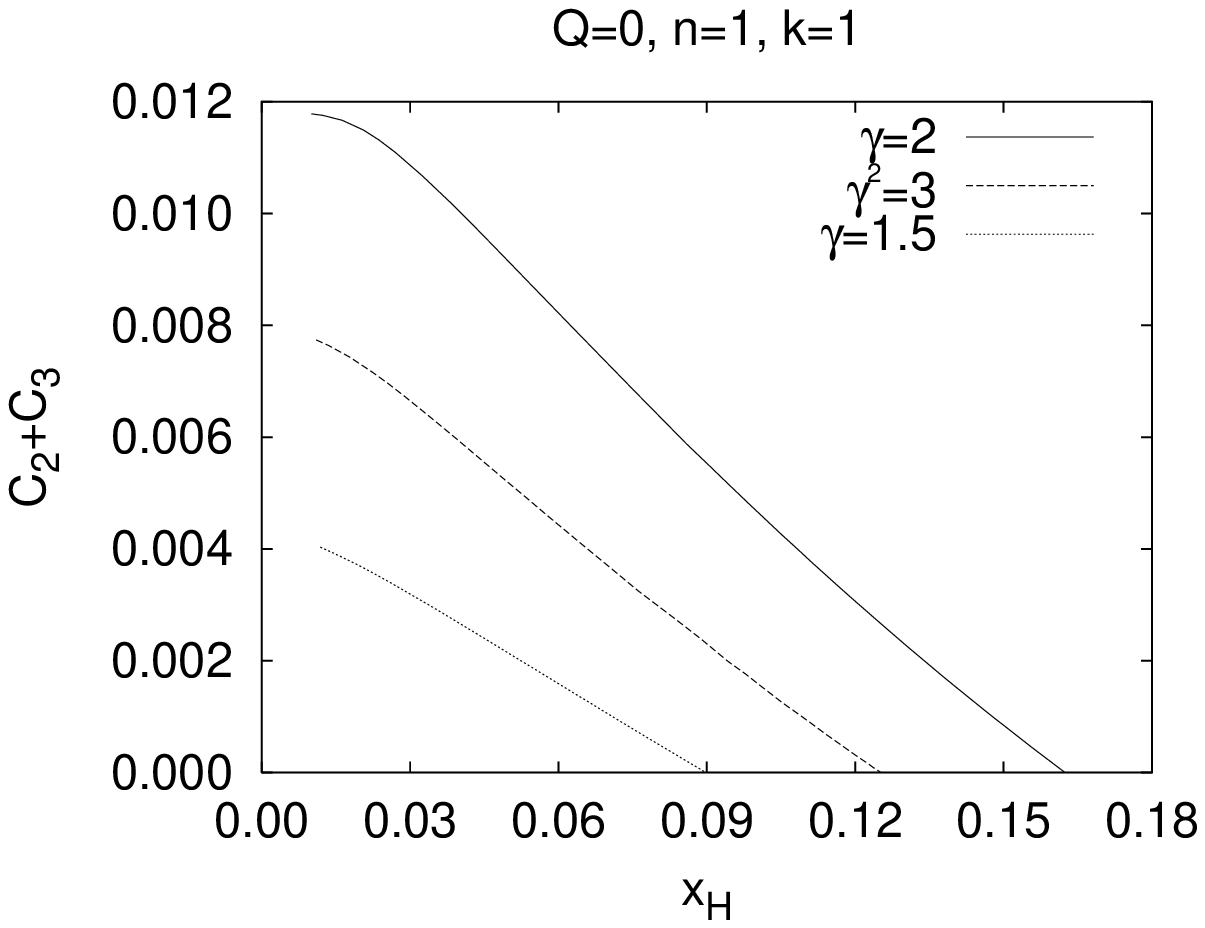}}
\caption{
Same as Fig.~3c for $C_2+C_3$.
} \end{figure}
\end{fixy}

\clearpage
\newpage

   \begin{fixy} {0}
\begin{figure}\centering
{\large Fig. 4a} \vspace{0.0cm}\\
\epsfysize=8cm
\mbox{\epsffile{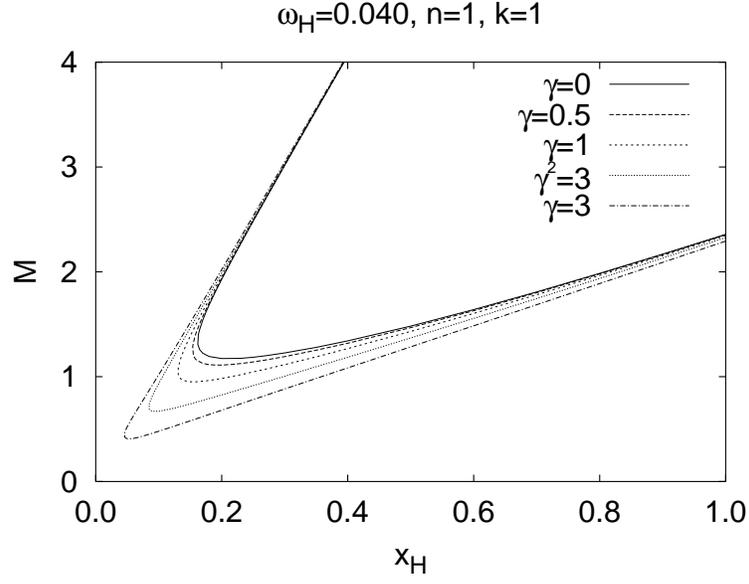}}
\caption{
The dimensionless mass $M$ is shown as a function of $x_{\rm H}$ for EYMD black holes with winding number $n=1$, node number $k=1$, $\omega_{\rm H}=0.040$, for dilaton coupling constant $\gamma=0$, 0.5, 1, $\sqrt{3}$, and 3.
} \end{figure}

\begin{figure}\centering
{\large Fig. 4b} \vspace{0.0cm}\\
\epsfysize=8cm
\mbox{\epsffile{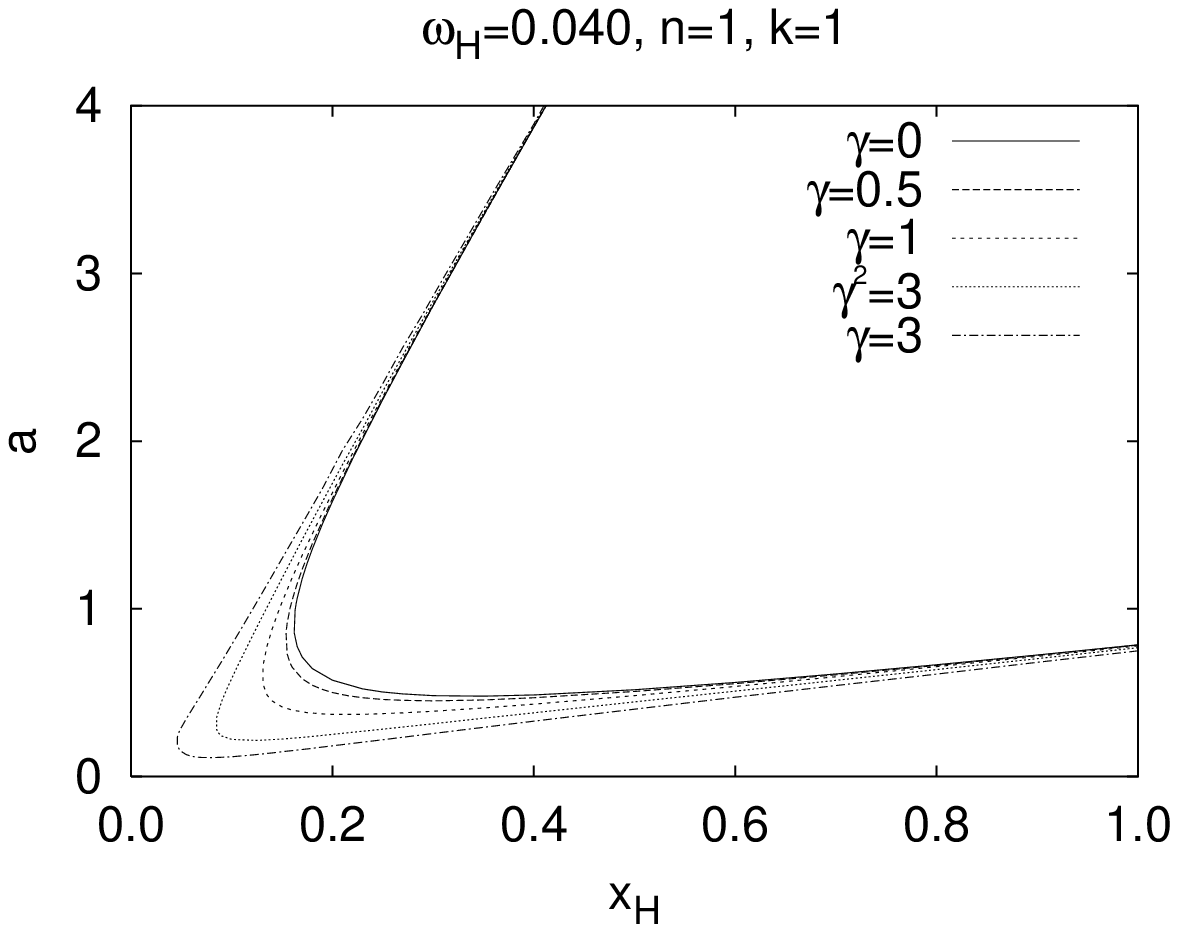}}
\caption{
Same as Fig.~4a for the specific angular momentum $a=J/M$.
} \end{figure}

\clearpage
\newpage

\begin{figure}\centering
{\large Fig. 4c} \vspace{0.0cm}\\
\epsfysize=8cm
\mbox{\epsffile{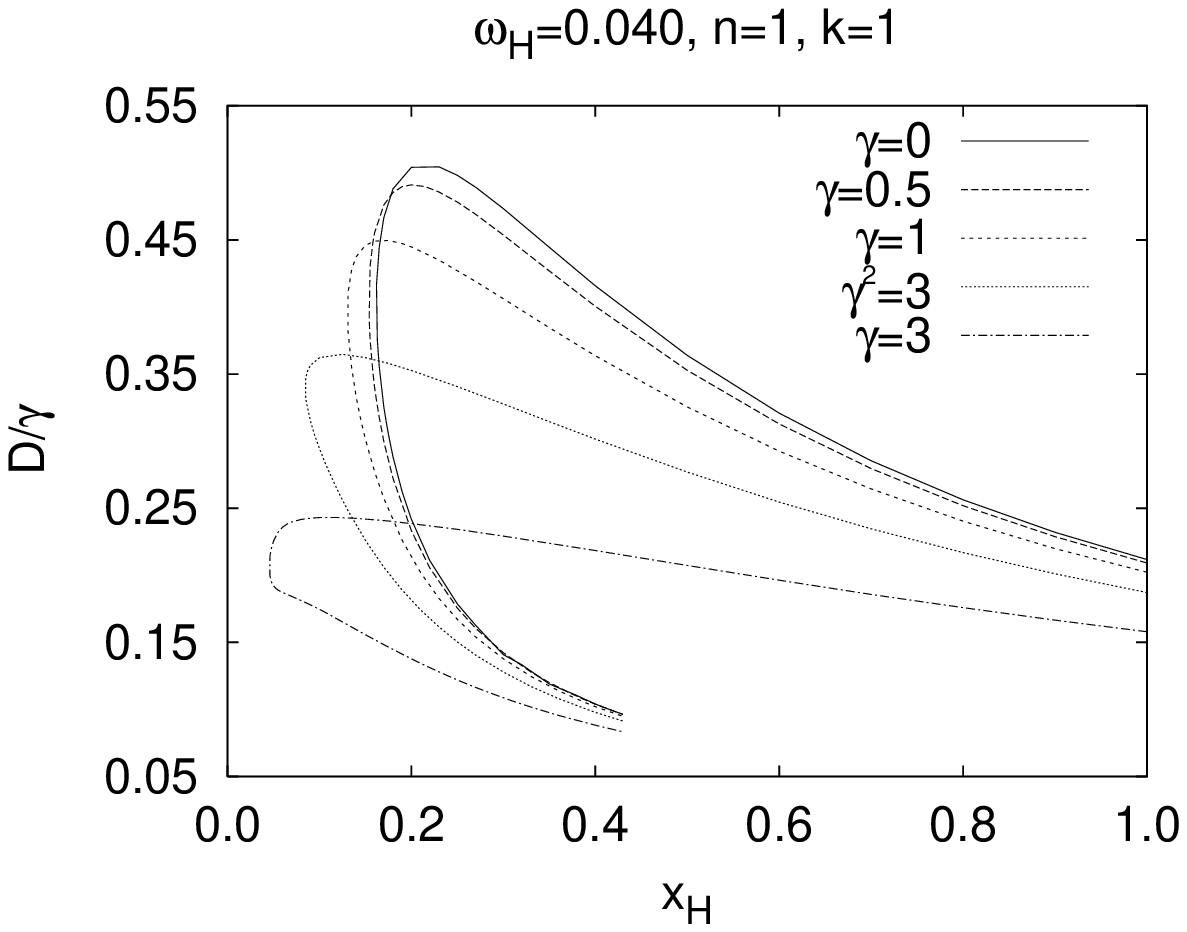}}
\caption{
Same as Fig.~4a for the relative dilaton charge $D/\gamma$.
} \end{figure}

\begin{figure}\centering
{\large Fig. 4d} \vspace{0.0cm}\\
\epsfysize=8cm
\mbox{\epsffile{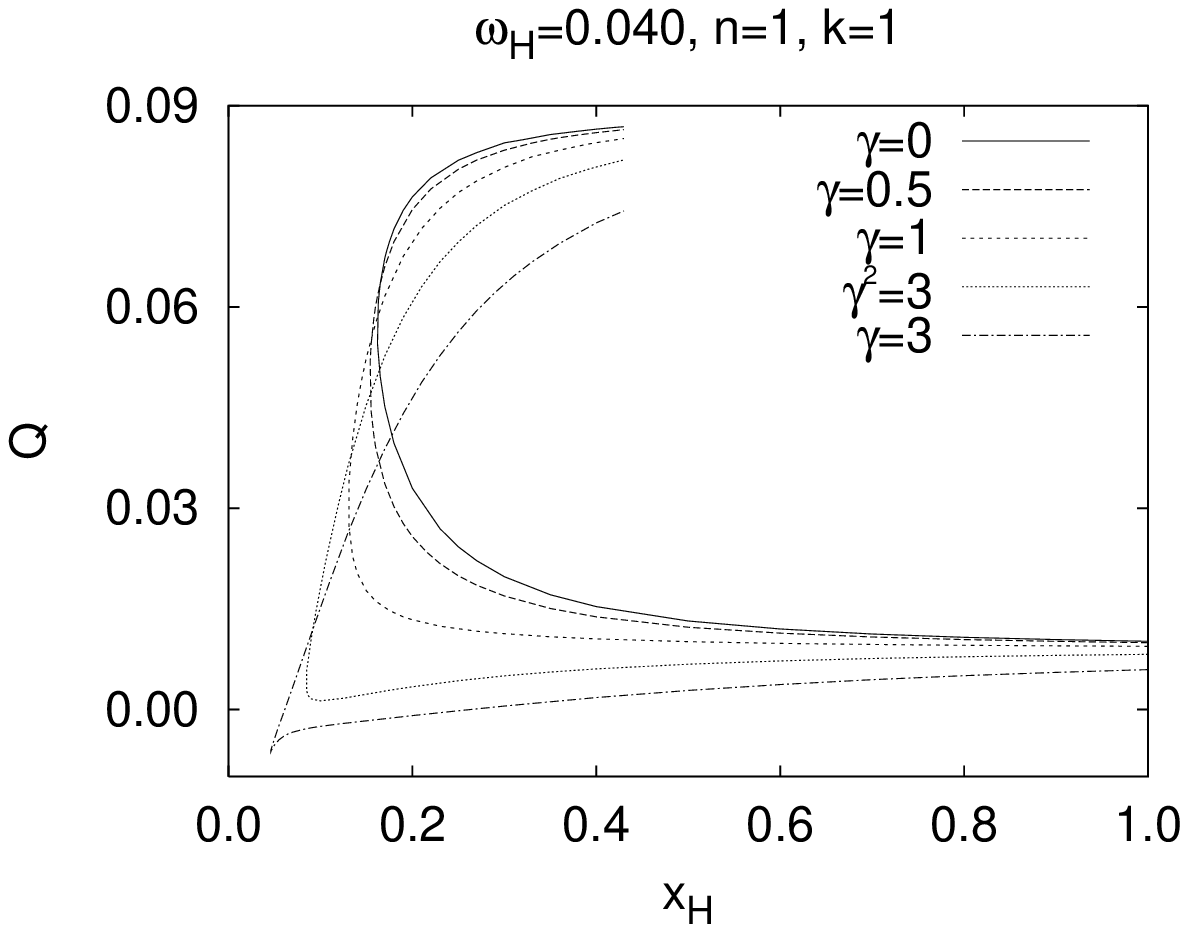}}
\caption{
Same as Fig.~4a for the electric charge $Q$.
} \end{figure}
\end{fixy}

   \begin{fixy} {0}
\begin{figure}\centering
{\large Fig. 5a} \vspace{0.0cm}\\
\epsfysize=8cm
\mbox{\epsffile{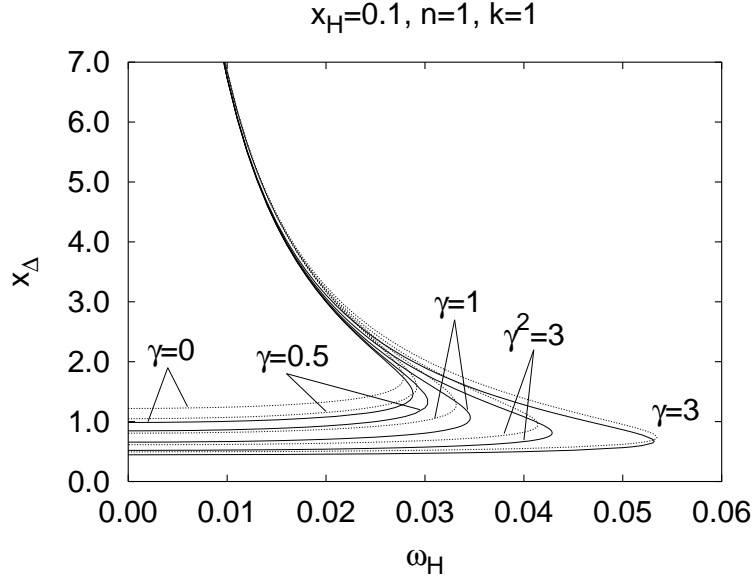}}
\caption{
The area parameter $x_\Delta$ is shown as a function of $\omega_{\rm H}$ 
for EYMD black holes with
winding number $n=1$, node number $k=1$, horizon radius $x_{\rm H}=0.1$, 
for dilaton coupling constant $\gamma=0$, 0.5, 1, $\sqrt{3}$, and 3 
(solid lines). 
The area parameter of the corresponding EMD solutions with
electric charge $Q=0$ and magnetic charge $P=1$ is also shown 
(dotted lines).
} \end{figure}

\begin{figure}\centering
{\large Fig. 5b} \vspace{0.0cm}\\
\epsfysize=8cm
\mbox{\epsffile{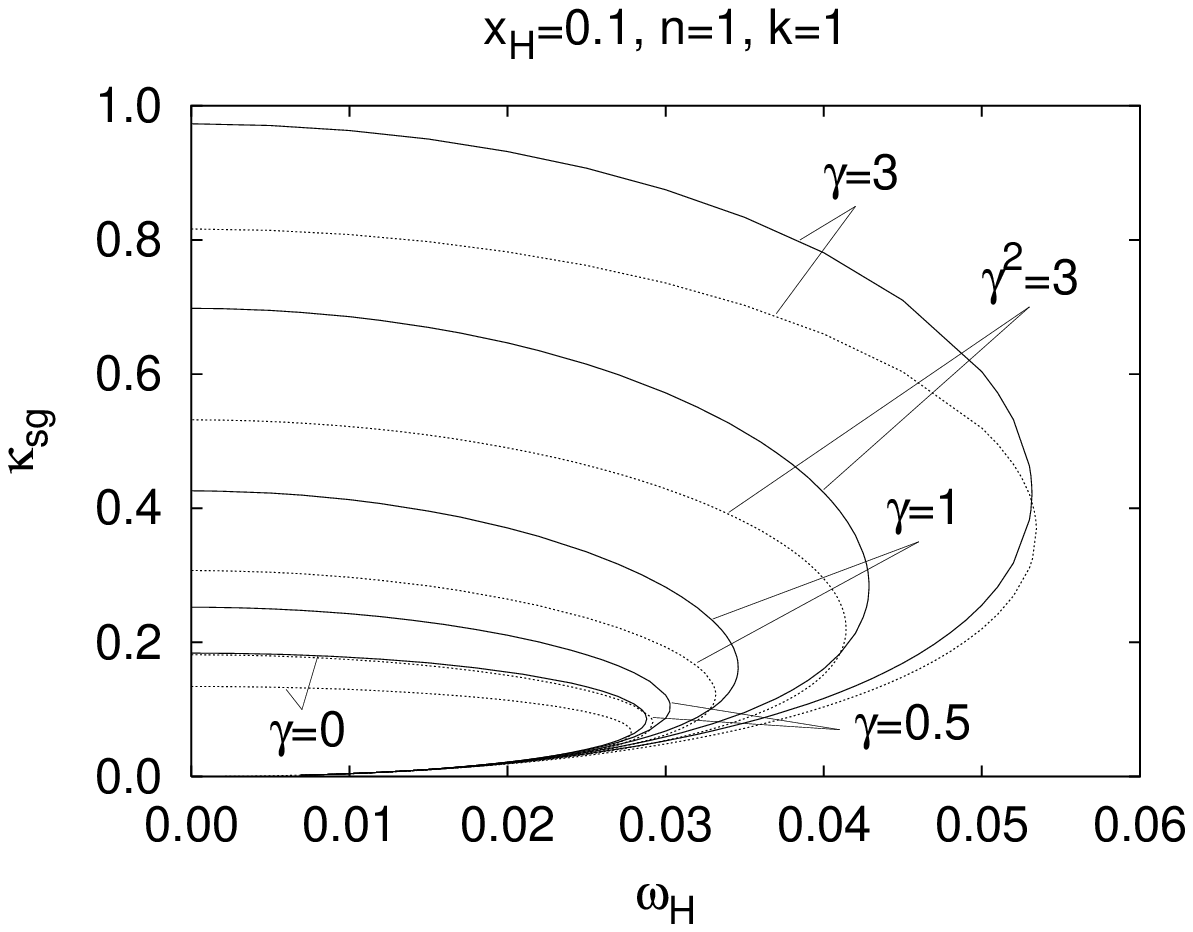}}
\caption{
Same as Fig.~5a for the surface gravity $\kappa_{\rm sg}$. 
} \end{figure}

\clearpage
\newpage

\begin{figure}\centering
{\large Fig. 5c} \vspace{0.0cm}\\
\epsfysize=8cm
\mbox{\epsffile{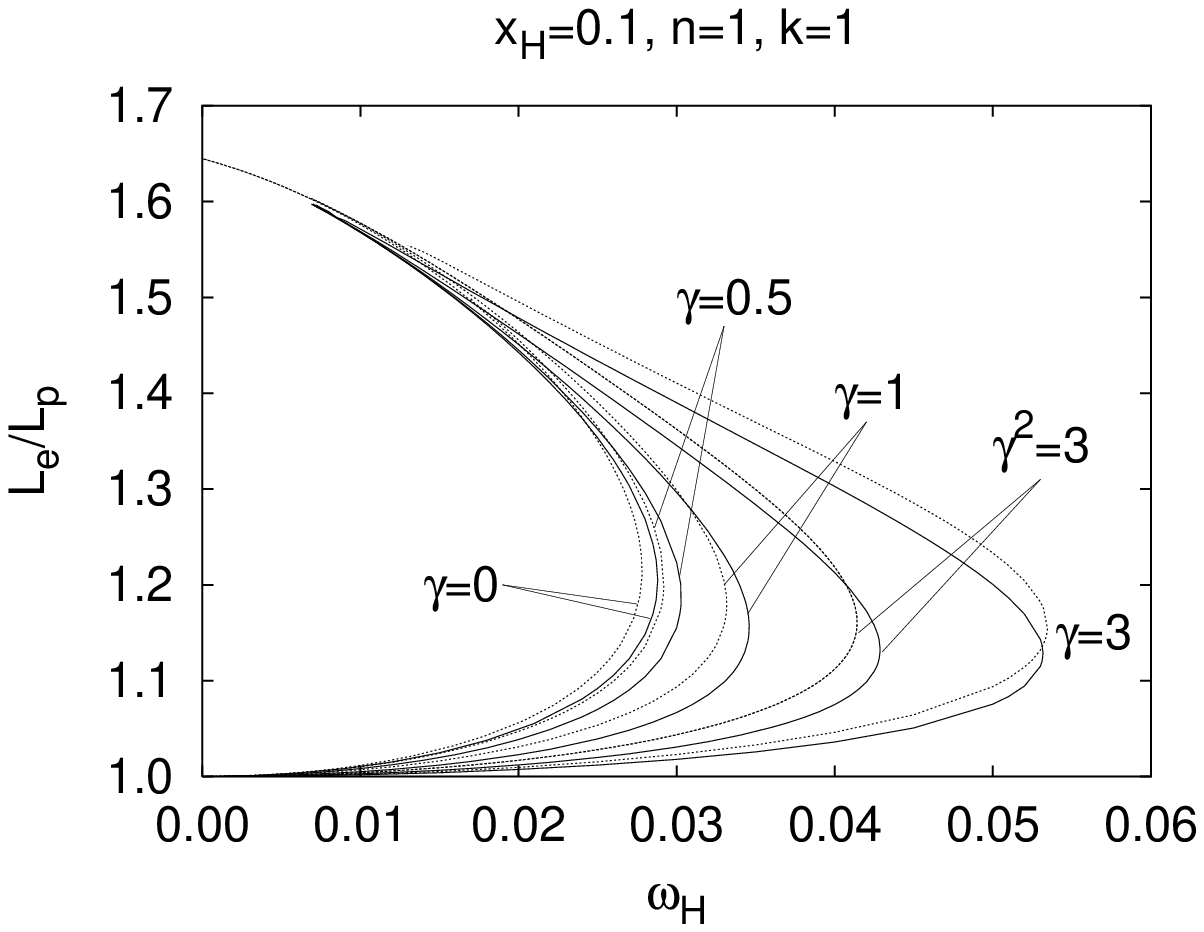}}
\caption{
Same as Fig.~5a for the ratio of circumferences $L_e/L_p$.
} \end{figure}

\begin{figure}\centering
{\large Fig. 5d} \vspace{0.0cm}\\
\epsfysize=8cm
\mbox{\epsffile{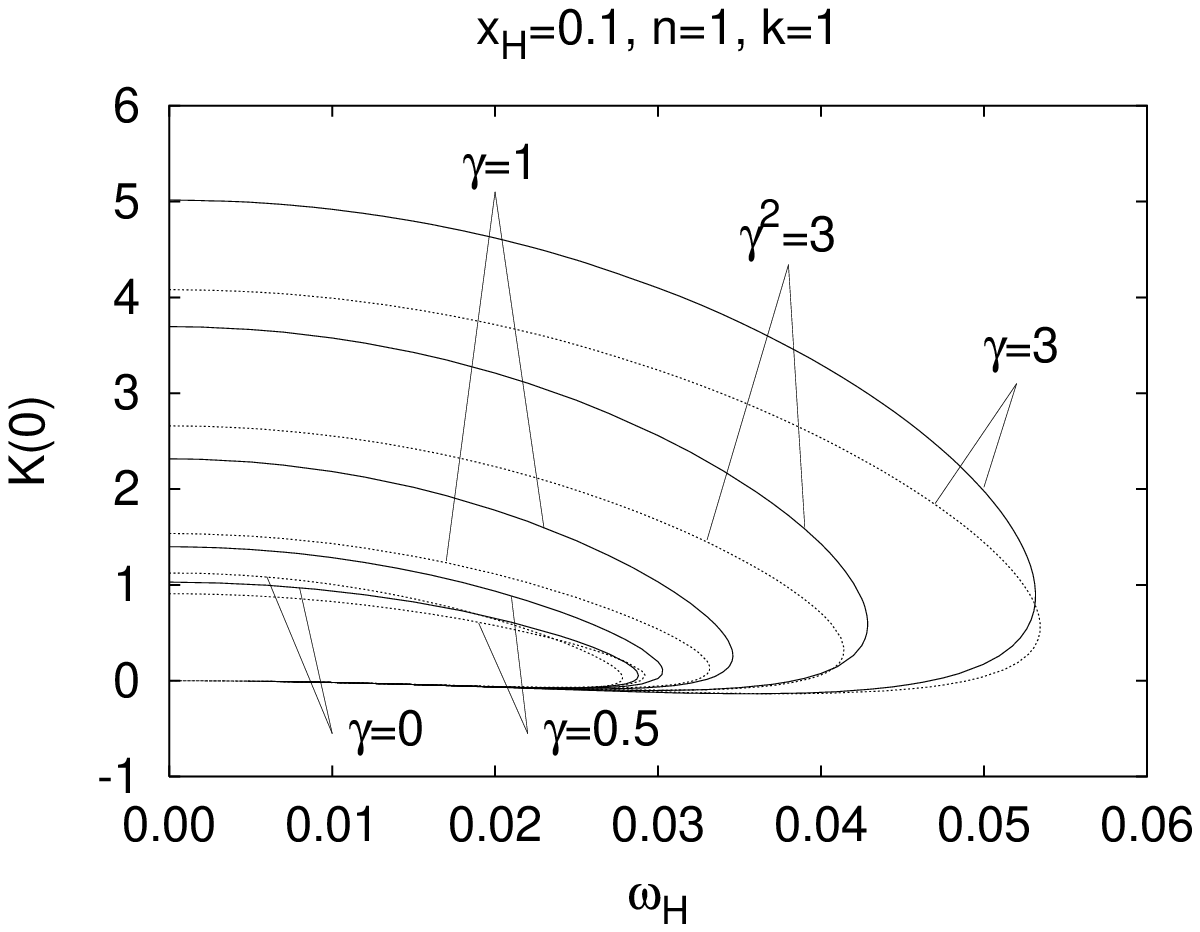}}
\caption{
Same as Fig.~5a for the Gaussian curvature at the poles $K(0)$.  
} \end{figure}
\end{fixy}

\clearpage
\newpage

   \begin{fixy} {0}
\begin{figure}\centering
{\large Fig. 6a} \vspace{0.0cm}\\
\epsfysize=8cm
\mbox{\epsffile{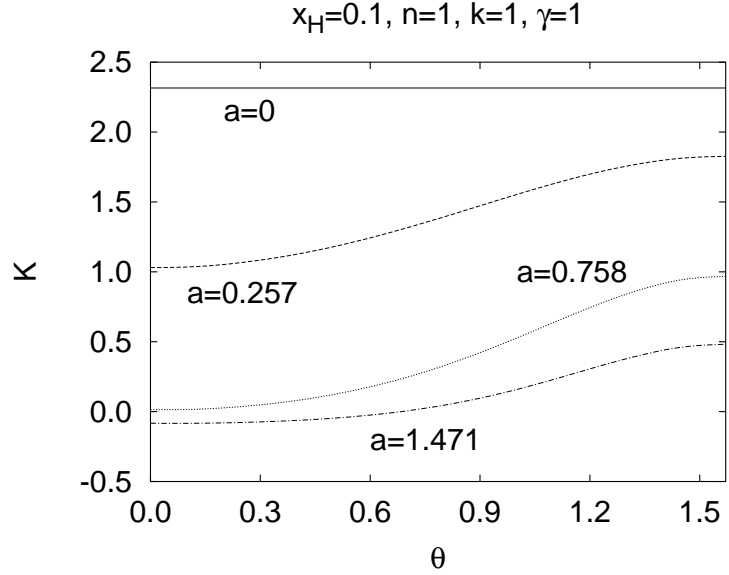}}
\caption{
The Gaussian curvature at the horizon $K(\theta)$
is shown as a function of the angle $\theta$
for EYMD black holes with
winding number $n=1$, node number $k=1$, horizon radius $x_{\rm H}=0.1$,
dilaton coupling constant $\gamma=1$,
for the values of the angular momentum per unit mass
$a=0$, $0.257$, 0.758, and 1.471.
} \end{figure}

\begin{figure}\centering
{\large Fig. 6b} \vspace{0.0cm}\\
\epsfysize=8cm
\mbox{\epsffile{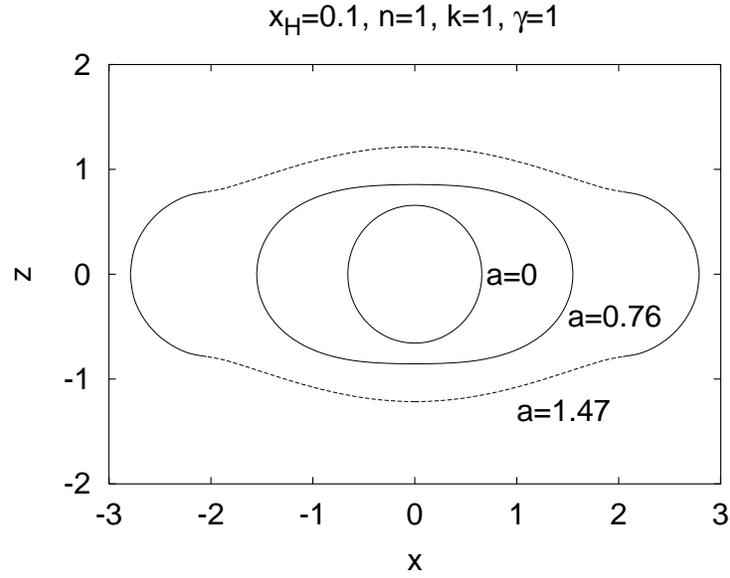}}
\caption{
The shape of the horizon is shown for EYMD black holes with
winding number $n=1$, node number $k=1$, horizon radius $x_{\rm H}=0.1$,
dilaton coupling constant $\gamma=1$,
for the values of the angular momentum per unit mass
$a=0$, 0.76, and 1.47.
Solid lines indicate embedding in Euclidean space,
dashed lines in pseudo-Euclidean space.
} \end{figure}
\end{fixy}

\clearpage
\newpage

   \begin{fixy} {0}
\begin{figure}\centering
{\large Fig. 7a} \vspace{0.0cm}\\
\epsfysize=8cm
\mbox{\epsffile{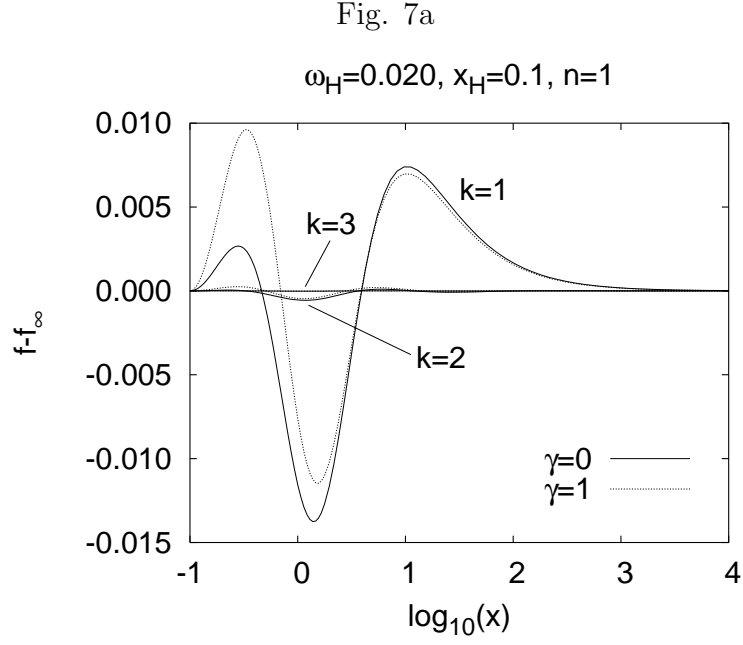}}
\caption{
The difference of the metric function $f$ and the metric function $f_\infty$ of the limiting solutions is shown as a function of the dimensionless coordinate $x$ for $\theta=0$ for the EYMD black hole solutions with $\gamma=0$ (solid lines) and $\gamma=1$ (dotted lines), winding number $n=1$, $\omega_{\rm  H}=0.020$ (lower branch), horizon radius $x_{\rm H}=0.1$, and node numbers $k=1-3$.    
} \end{figure}

\begin{figure}\centering
{\large Fig. 7b} \vspace{0.0cm}\\
\epsfysize=8cm
\mbox{\epsffile{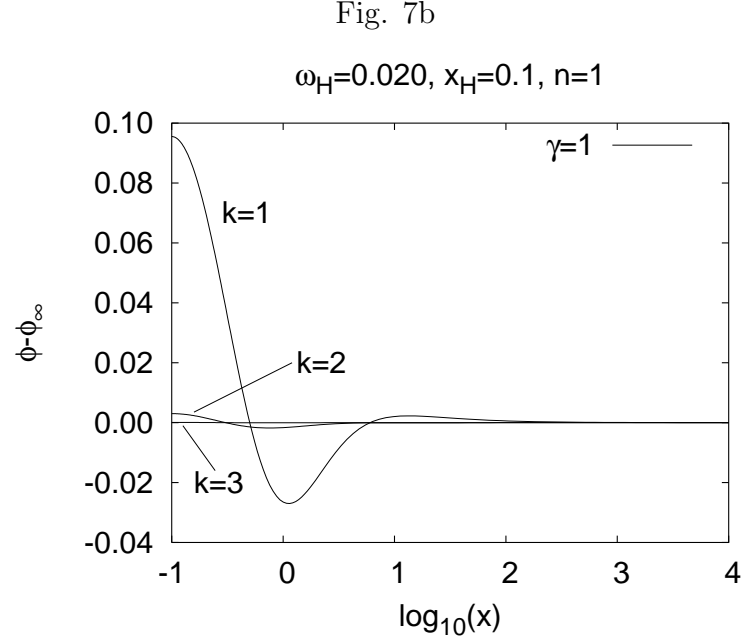}}
\caption{
The same as Fig.~7a for the dilaton function $\phi$.
} \end{figure}

\clearpage
\newpage

\begin{figure}\centering
{\large Fig. 7c} \vspace{0.0cm}\\
\epsfysize=8cm
\mbox{\epsffile{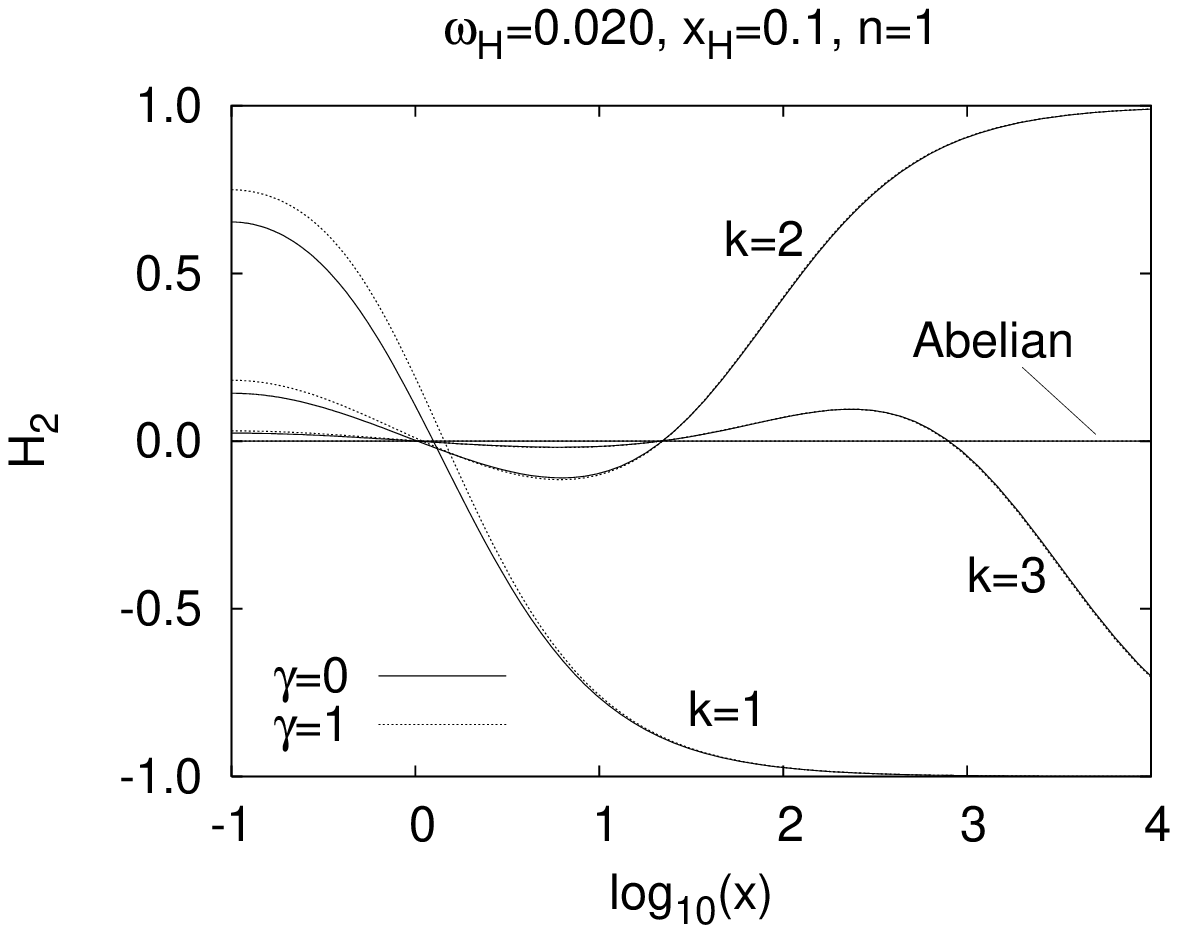}}
\caption{
The same as Fig.~7a for the function $H_2$. ($H_{2\infty}=0$).
} \end{figure}

\begin{figure}\centering
{\large Fig. 7d} \vspace{0.0cm}\\
\epsfysize=8cm
\mbox{\epsffile{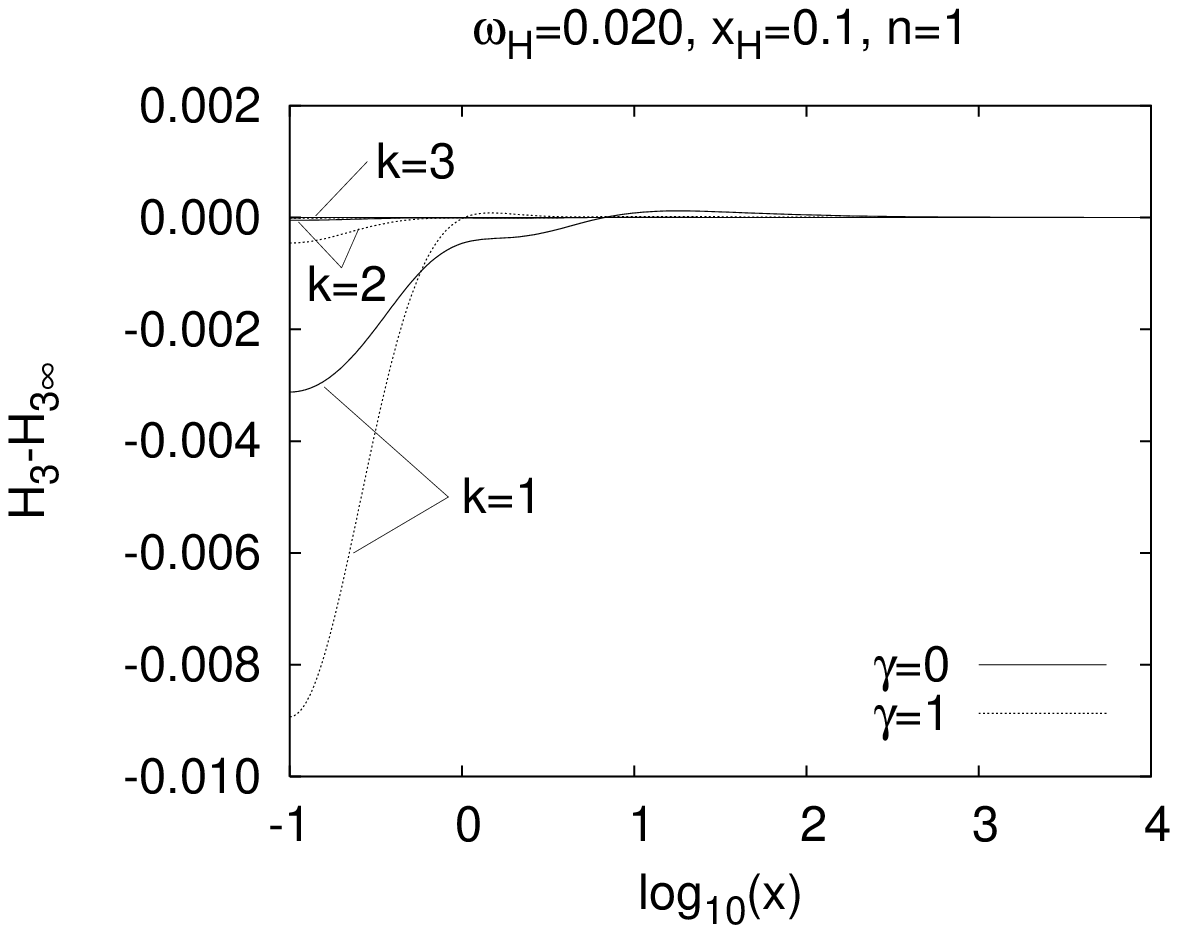}}
\caption{
The same as Fig.~7a for the function $H_3$, but $\theta = \pi/4$.
} \end{figure}

\begin{figure}\centering
{\large Fig. 7e} \vspace{0.0cm}\\
\epsfysize=8cm
\mbox{\epsffile{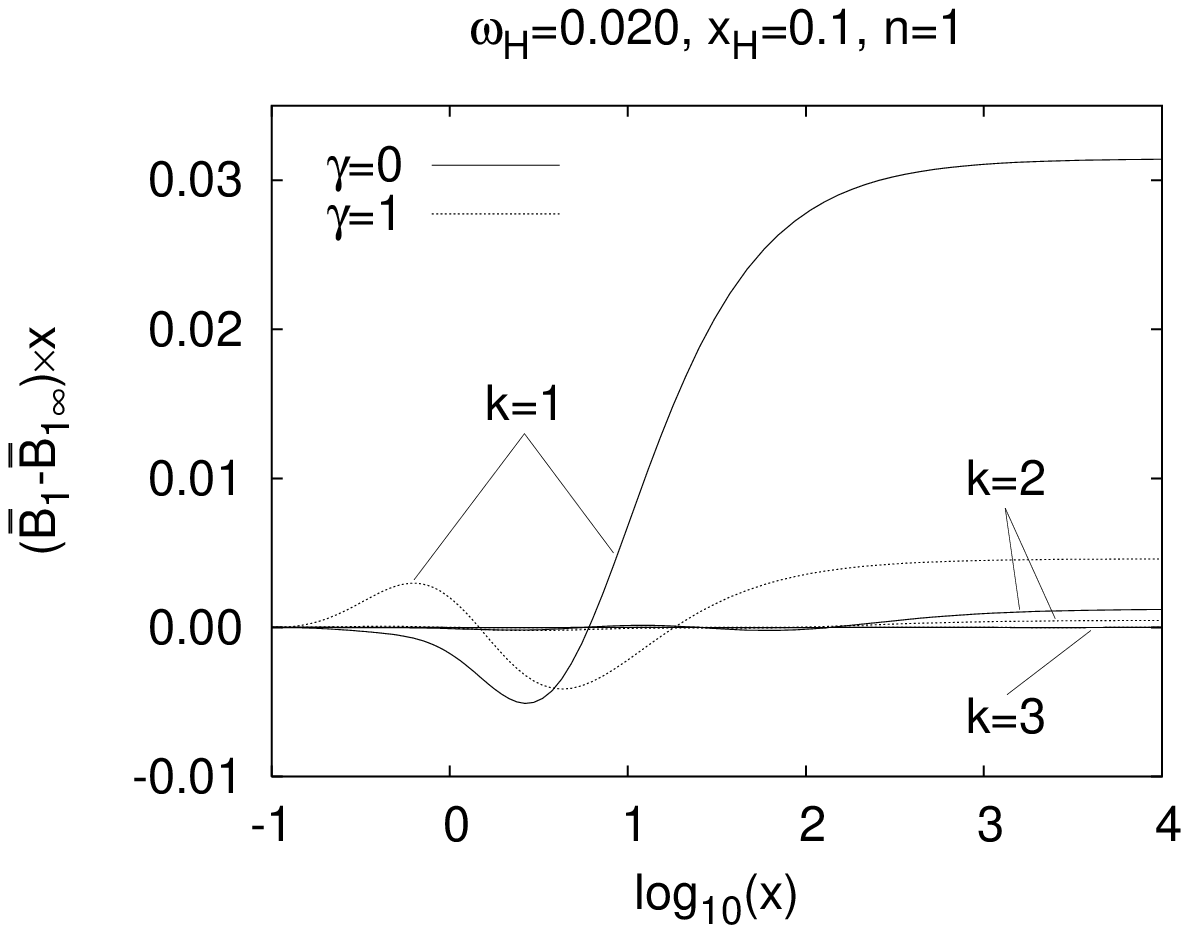}}
\caption{
The same as Fig.~7a for the function $x \bar{B}_1$.
} \end{figure}

\end{fixy}

\clearpage
\newpage

   \begin{fixy} {0}
\begin{figure}\centering
{\large Fig. 8a} \vspace{0.0cm}\\
\epsfysize=8cm
\mbox{\epsffile{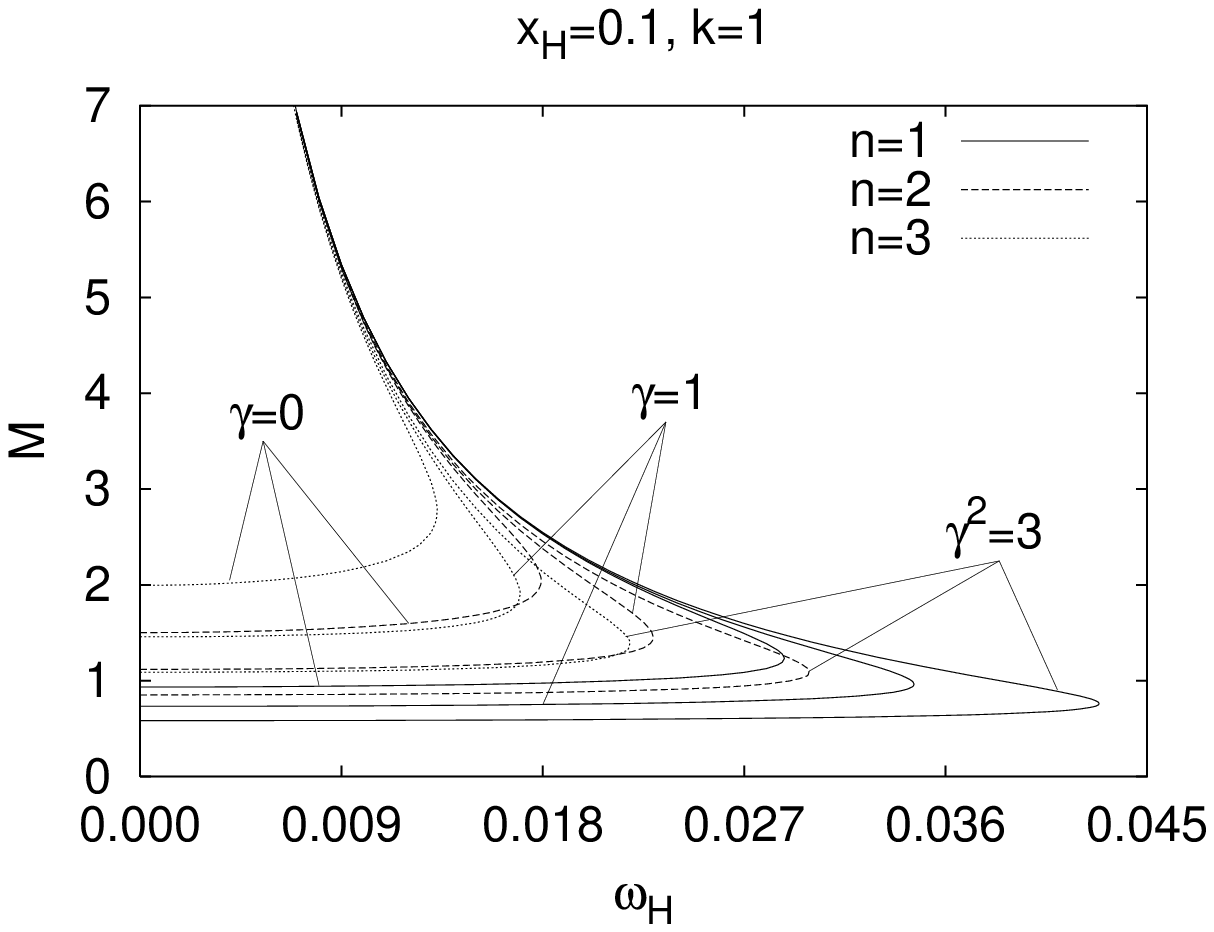}}
\caption{
The dimensionless mass $M$ is shown as a function of $\omega_{\rm H}$ for EYMD black holes with $n=1-3$, $k=1$, $x_{\rm H}=0.1$, and $\gamma=0$, $1$, and $\sqrt{3}$.
} \end{figure}

\begin{figure}\centering
{\large Fig. 8b} \vspace{0.0cm}\\
\epsfysize=8cm
\mbox{\epsffile{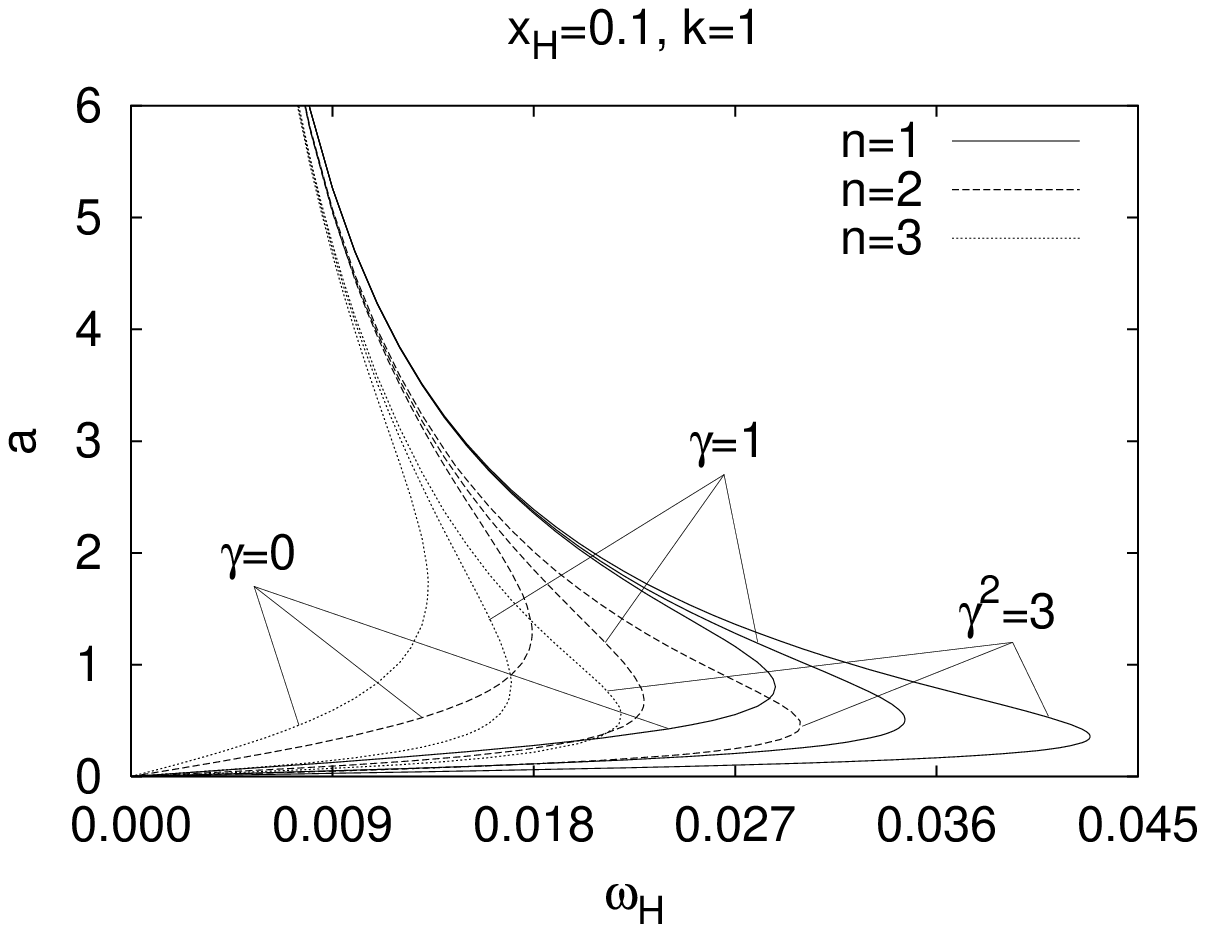}}
\caption{
Same as Fig.~8a for the angular momentum per unit mass $a$.
} \end{figure}

\clearpage
\newpage

\begin{figure}\centering
{\large Fig. 8c} \vspace{0.0cm}\\
\epsfysize=8cm
\mbox{\epsffile{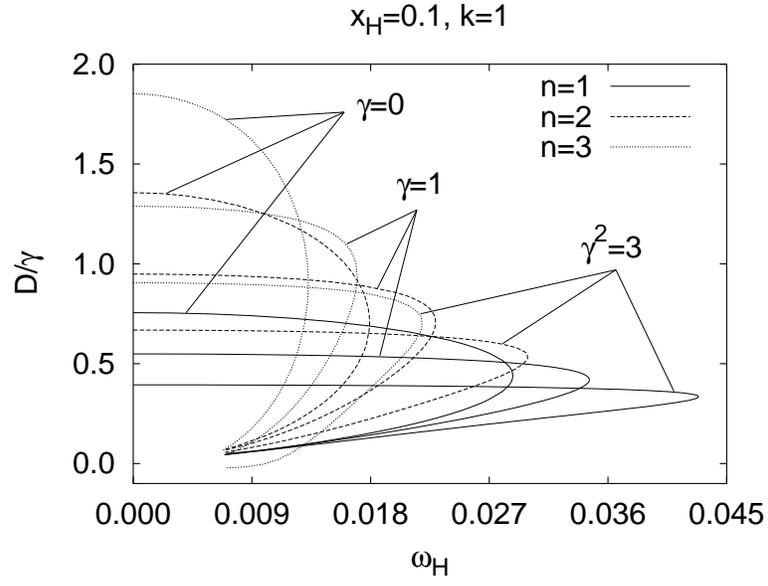}}
\caption{
Same as Fig.~8a for the relative charge $D/\gamma$.
} \end{figure}

\begin{figure}\centering
{\large Fig. 8d} \vspace{0.0cm}\\
\epsfysize=8cm
\mbox{\epsffile{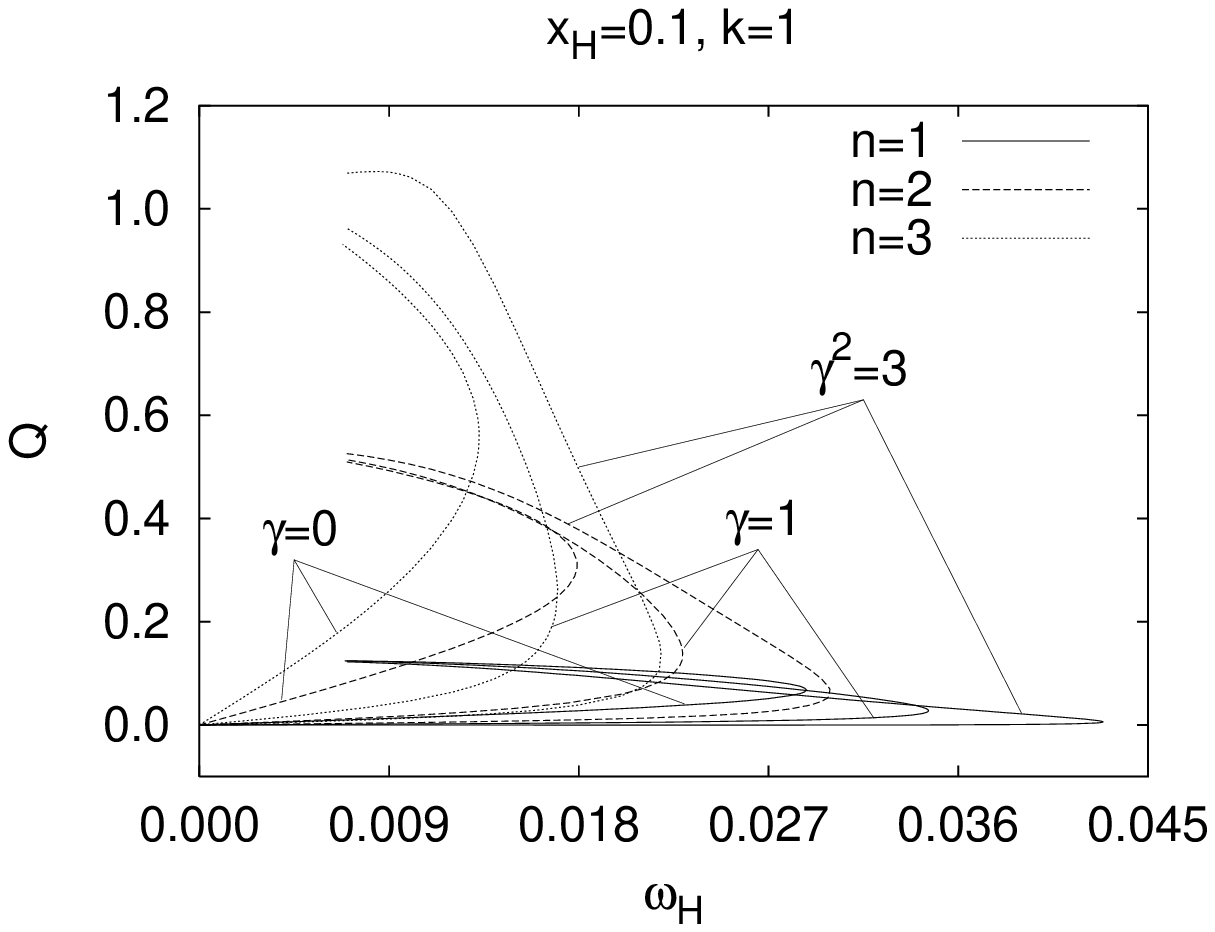}}
\caption{
Same as Fig.~8a for the electric charge $Q$.
} \end{figure}
\end{fixy}

\clearpage
\newpage

   \begin{fixy} {0}
\begin{figure}\centering
{\large Fig. 9a} \vspace{0.0cm}\\
\epsfysize=8cm
\mbox{\epsffile{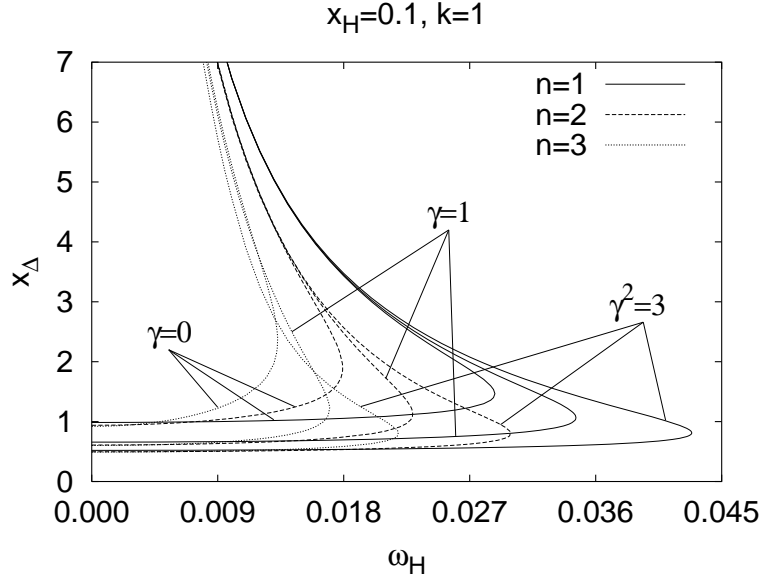}}
\caption{
The area parameter $x_\Delta$ is shown as a function of $\omega_{\rm H}$ for EYMD black holes with $n=1-3$, $k=1$, $x_{\rm H}=0.1$, and $\gamma=0$, $1$, and $\sqrt{3}$.
} \end{figure}

\begin{figure}\centering
{\large Fig. 9b} \vspace{0.0cm}\\
\epsfysize=8cm
\mbox{\epsffile{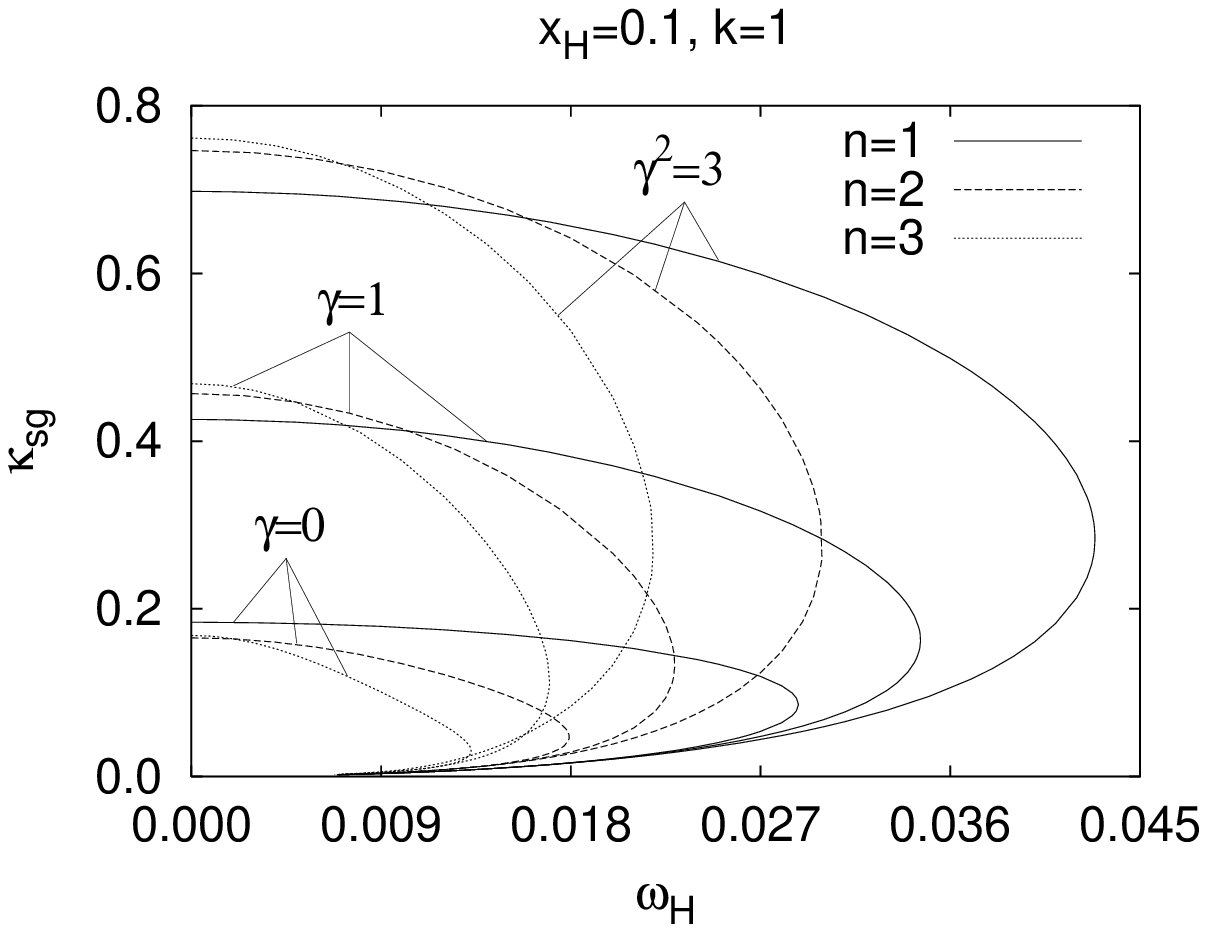}}
\caption{
Same as Fig.~9a for the surface gravity $\kappa_{\rm sg}$.
} \end{figure}

\clearpage
\newpage

\begin{figure}\centering
{\large Fig. 9c} \vspace{0.0cm}\\
\epsfysize=8cm
\mbox{\epsffile{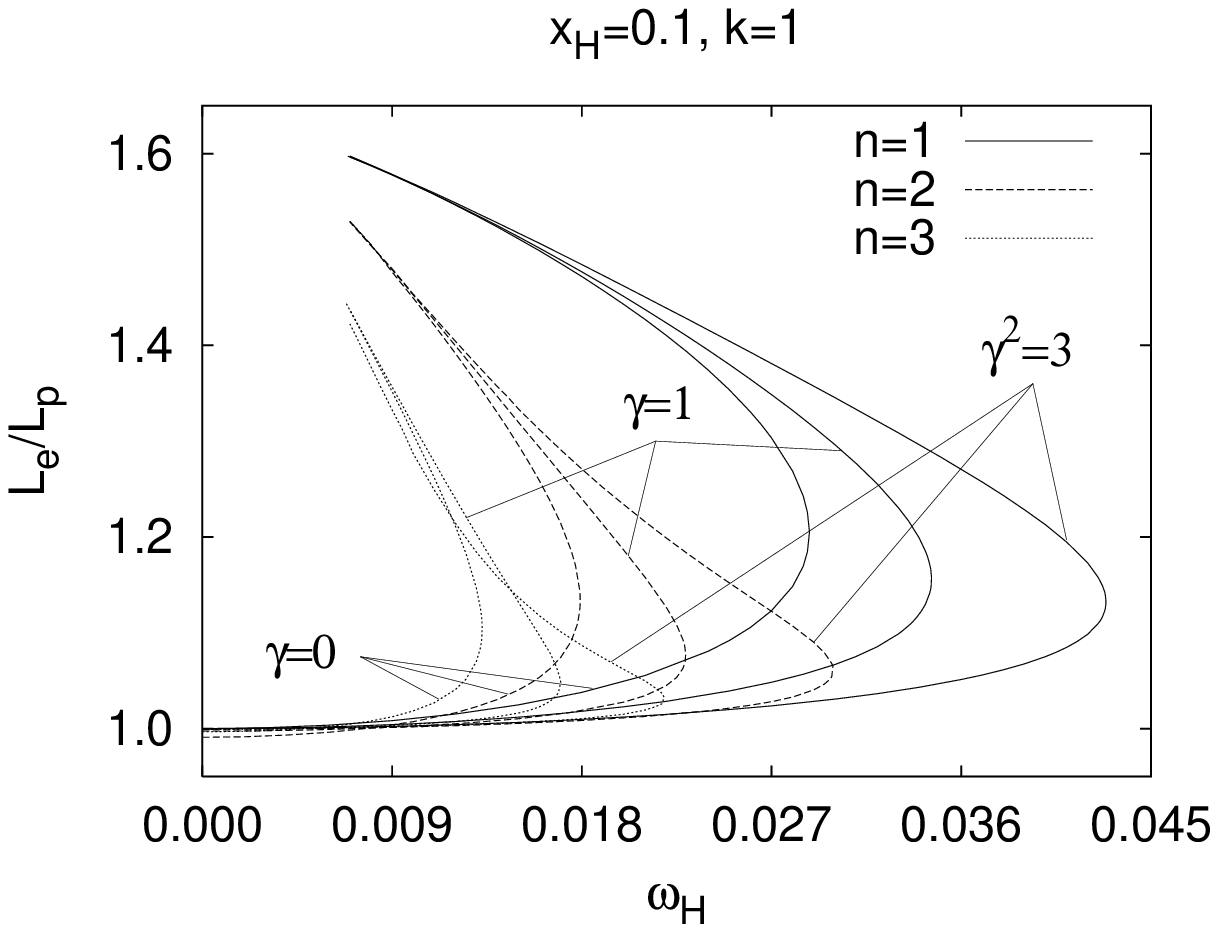}}
\caption{
Same as Fig.~9a for the deformation parameter $L_e/L_p$.
} \end{figure}

\begin{figure}\centering
{\large Fig. 9d} \vspace{0.0cm}\\
\epsfysize=8cm
\mbox{\epsffile{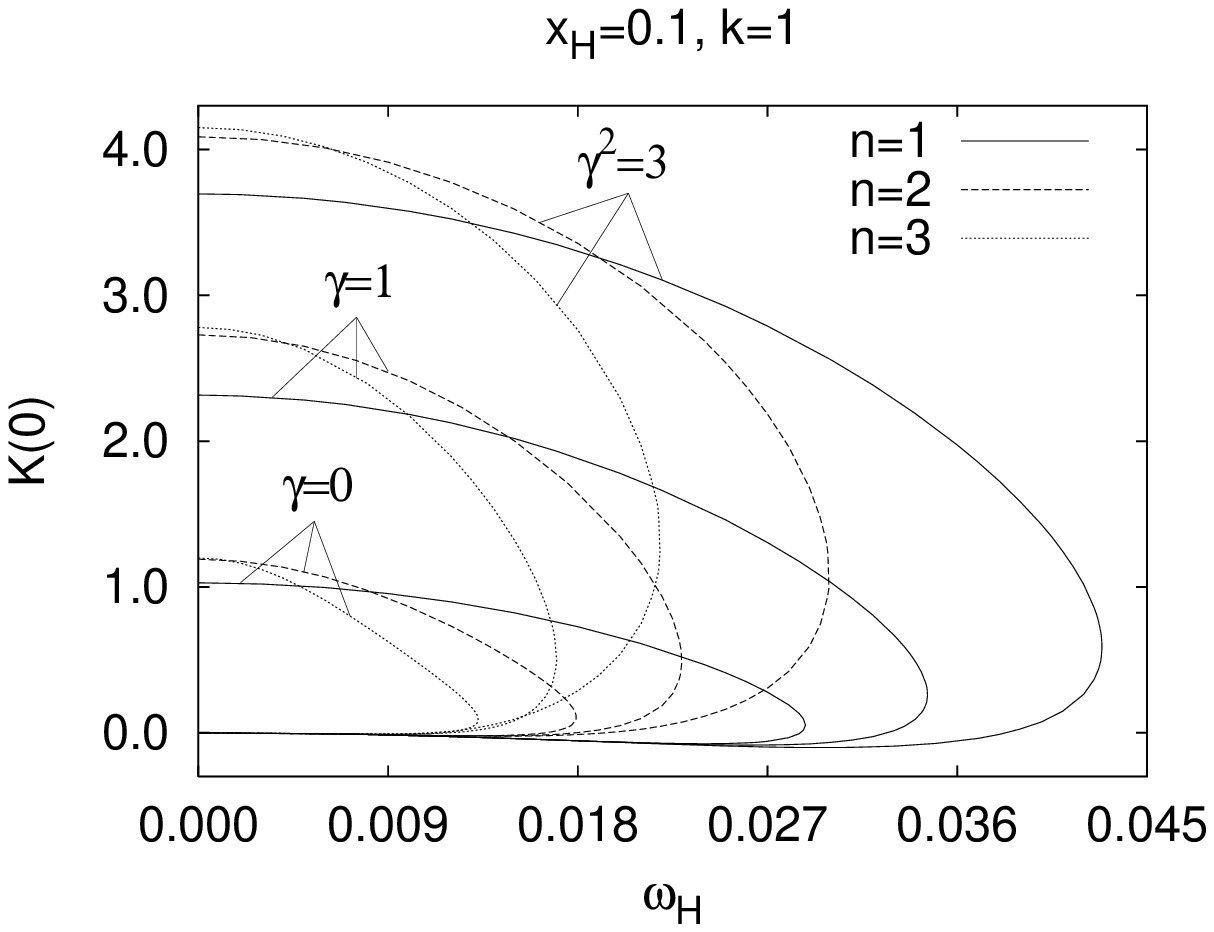}}
\caption{
Same as Fig.~9a for the Gaussian curvature at the pole $K(0)$. 
} \end{figure}
\end{fixy}

\clearpage
\newpage

   \begin{fixy} {0}
\begin{figure}\centering
{\large Fig. 10a} \vspace{0.0cm}\\
\epsfysize=8cm
\mbox{\epsffile{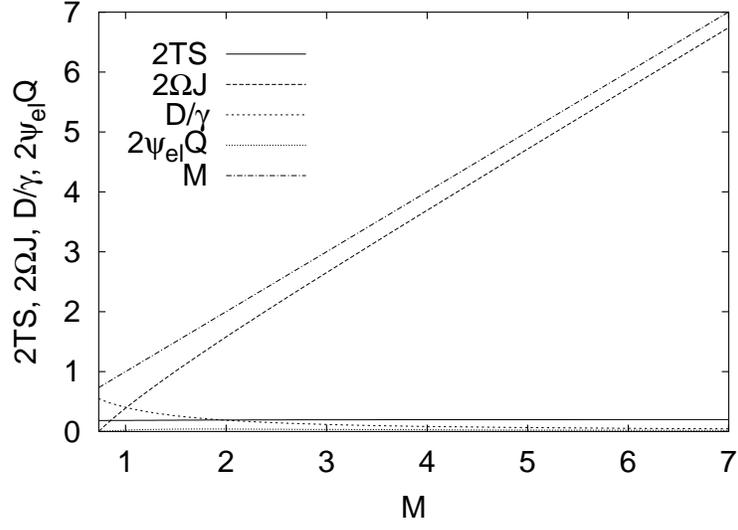}}
\caption{
Numerical check of the mass formula, Eq.~(\ref{namass}), for $n=1$, $k=1$, $x_{\rm H}=0.1$, $\gamma=1$ black hole solutions. The terms in Eq.~(\ref{namass}) are plotted versus $M$, their sum coincides with $M$, within the numerical accuracy.  
} \end{figure}

\begin{figure}\centering
{\large Fig. 10b} \vspace{0.0cm}\\
\epsfysize=8cm
\mbox{\epsffile{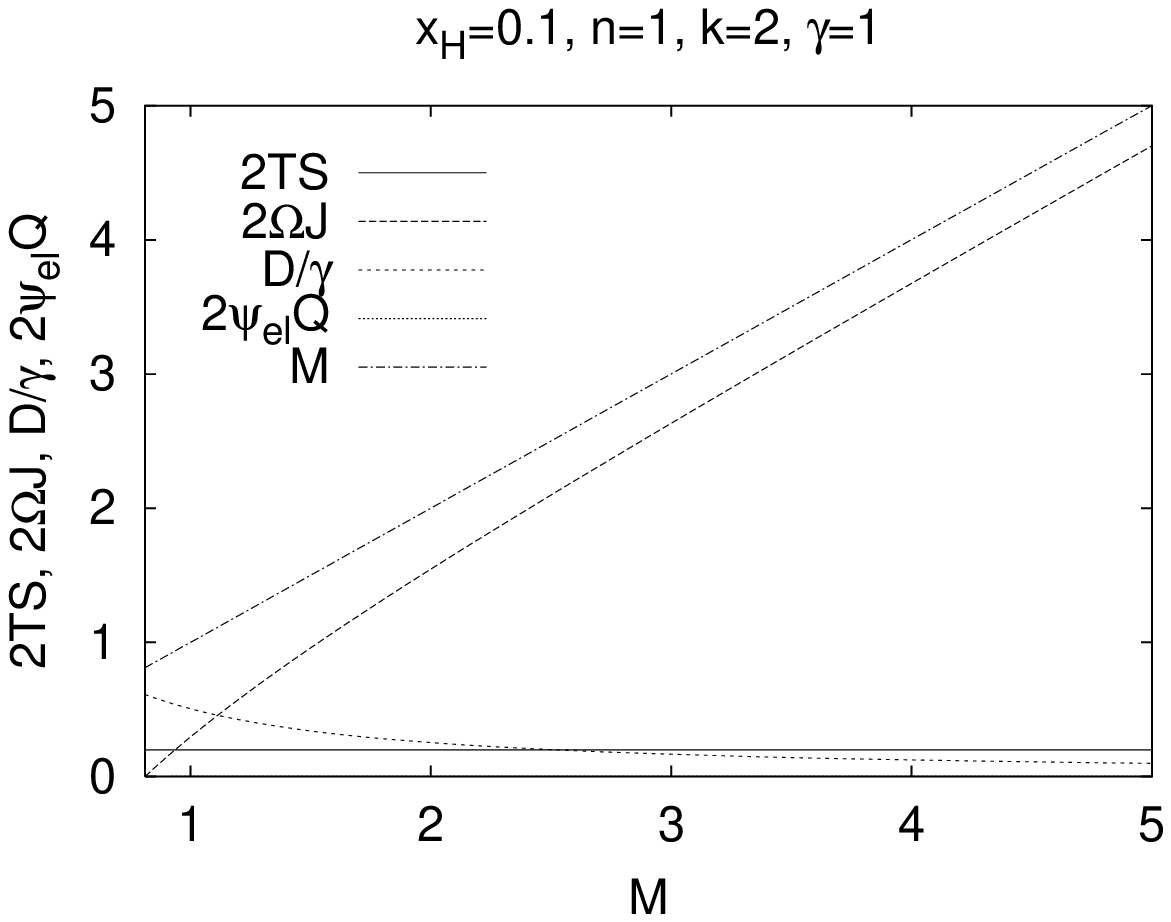}}
\caption{
Same as Fig.~10a for $n=1$ and $k=2$.
} \end{figure}

\clearpage
\newpage

\begin{figure}\centering
{\large Fig. 10c} \vspace{0.0cm}\\
\epsfysize=8cm
\mbox{\epsffile{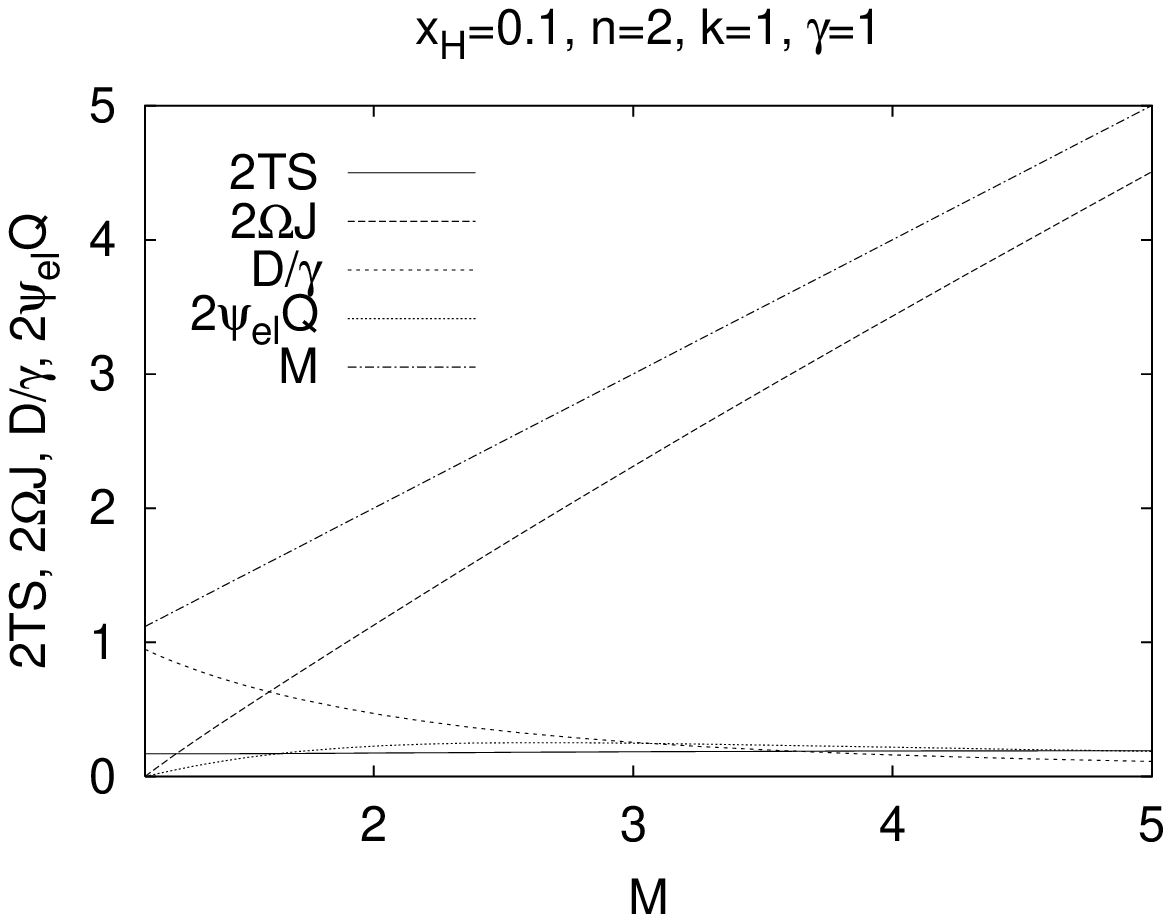}}
\caption{
Same as Fig.~10a for $n=2$ and $k=1$.
} \end{figure}
\end{fixy}

   \begin{fixy} {-1}
\begin{figure}\centering
{\large Fig. 11} \vspace{0.0cm}\\
\epsfysize=8cm
\mbox{\epsffile{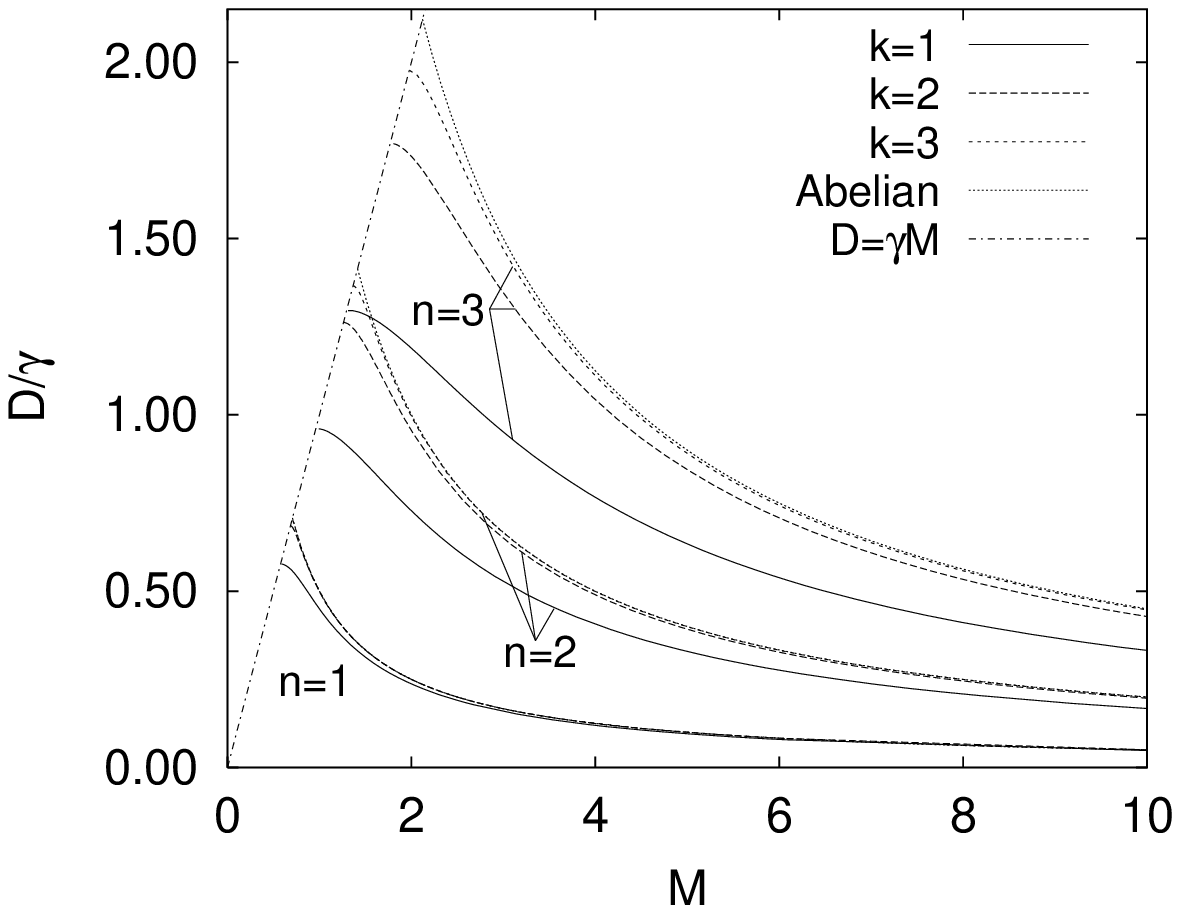}}
\caption{
The relative dilaton charge $D/\gamma$ is shown as a function of the mass for static non-Abelian solutions ($\gamma=1$) and embedded Abelian solutions with $Q=0$ and $P=n$. 
} \end{figure}
\end{fixy}

\clearpage
\newpage

  \begin{fixy} {0}
\begin{figure}\centering
{\large Fig. 12a} \vspace{0.0cm}\\
\epsfysize=8cm
\mbox{\epsffile{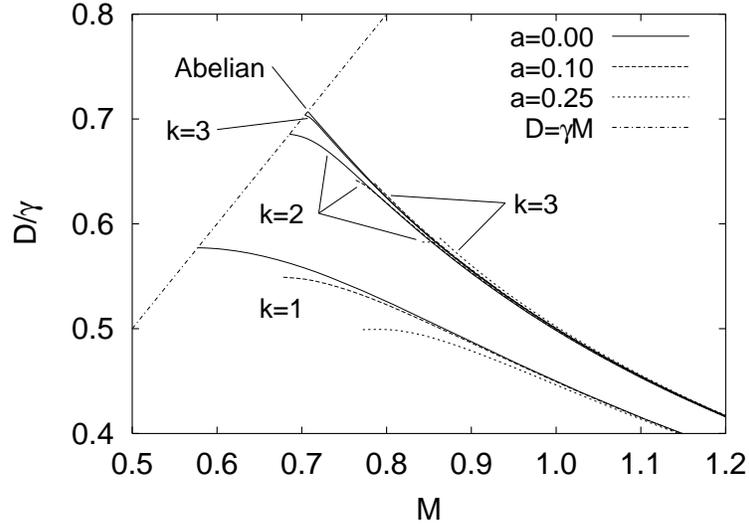}}
\caption{
The relative dilaton charge $D/\gamma$ is shown as a function of the mass $M$
for non-Abelian black holes with $n=1$, $k=1-3$ ($\gamma=1$)
and specific angular momentum $a=0.25$, 0.1 and 0.
Also shown are the embedded Abelian solutions 
with $Q=0$ and $P=1$.
} \end{figure}

\begin{figure}\centering
{\large Fig. 12b} \vspace{0.0cm}\\
\epsfysize=8cm
\mbox{\epsffile{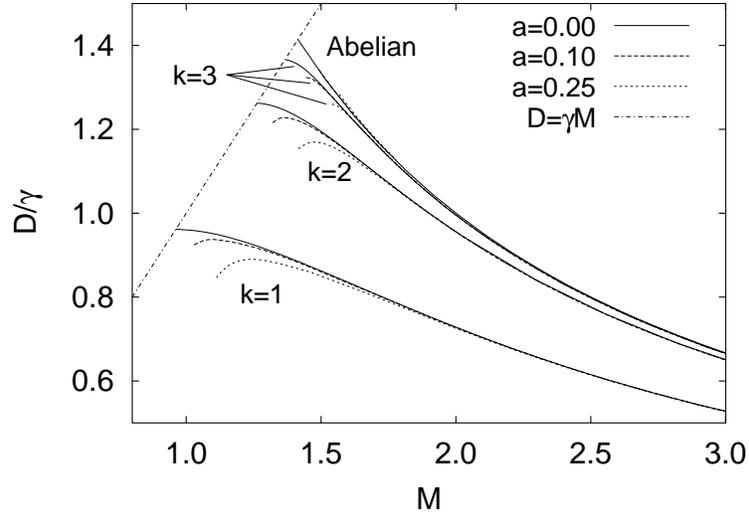}}
\caption{
Same as Fig.~12a for $n=2$ and $P=2$.
} \end{figure}

\clearpage
\newpage

\begin{figure}\centering
{\large Fig. 12c} \vspace{0.0cm}\\
\epsfysize=8cm
\mbox{\epsffile{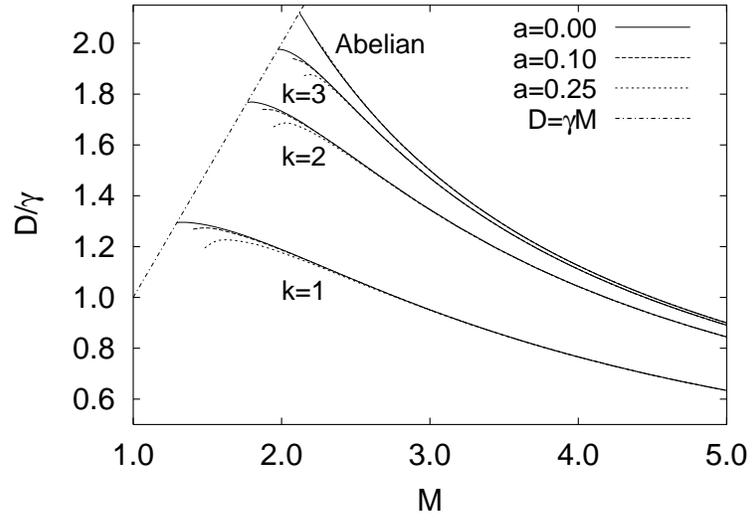}}
\caption{
Same as Fig.~12a for $n=3$ and $P=3$.
} \end{figure}

\end{fixy}

\clearpage
\newpage

   \begin{fixy} {0}
\begin{figure}\centering
{\large Fig. 13a} \vspace{0.0cm}\\
\epsfysize=8cm
\mbox{\epsffile{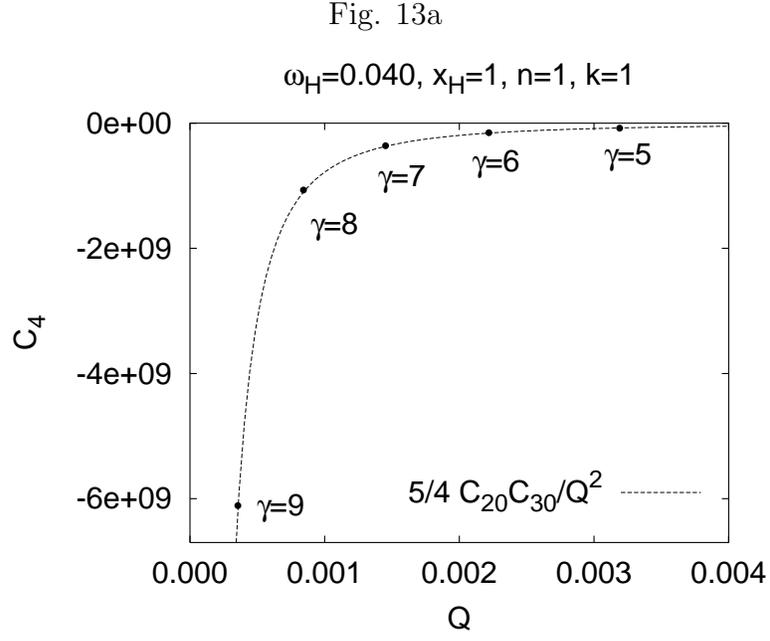}}
\caption{
The dependence of the asymptotic coefficient 
$C_4$ on $Q$ is shown for the black hole solutions 
with $n=1$, $k=1$, $x_{\rm H}=1$, and $\omega_{\rm H}=0.040$.
The dashed line corresponds to the leading term in the expansion of $C_4$. 
The dots correspond to numerical values of $C_4$,
extracted for several values of $\gamma$.
} \end{figure}

\begin{figure}\centering
{\large Fig. 13b} \vspace{0.0cm}\\
\epsfysize=8cm
\mbox{\epsffile{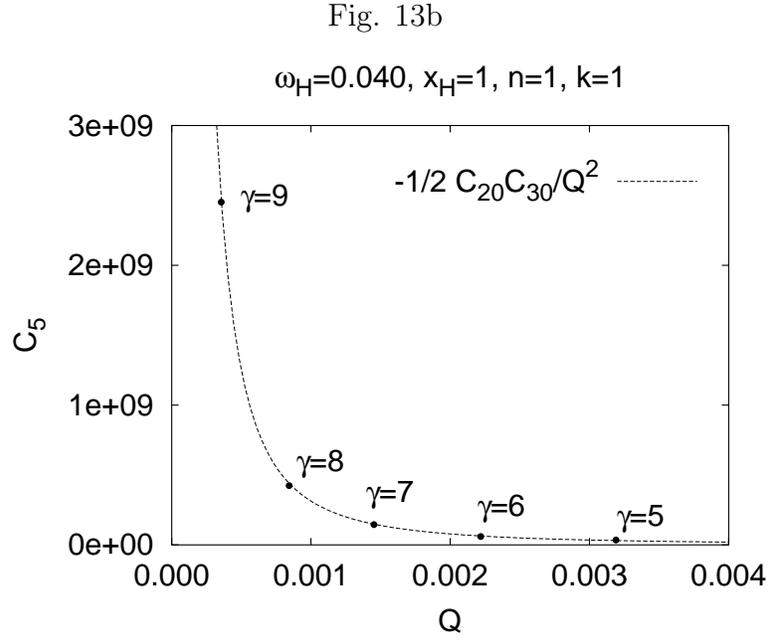}}
\caption{
Same as Fig.~13a for $C_5$.
} \end{figure}
\end{fixy}


\begin{thebibliography}{000}
\bibitem{foot3}
 The perturbative calculations of \cite{pert1} indicate a sign
 change of $Q$ for $n=1$ $k=2$ black hole solutions, however,
 the non-perturbative calculations do not show such a change of sign.
\bibitem{foot1}
 This is to be expected in EYM theory,
 since globally regular rotating solutions would have to satisfy
 a different set of boundary conditions at infinity \cite{pert2}.
 Non-perturbative numerical analysis, however, suggests \cite{radu2}
 that no such globally regular rotating solutions exist.
\end{thebibliography}
\end{document}